\documentclass[a4paper,11pt]{article}
\pdfoutput=1 % if your are submitting a p.d.flatex (i.e. if you have
\usepackage{jinstpub} % for details on the use of the package, please
\usepackage{threeparttable}
\usepackage{realboxes}
\usepackage[utf8]{inputenc}
\usepackage{dirtree}
\usepackage{rotating}
\usepackage[version=3]{mhchem}
\usepackage{makecell}
 \usepackage{lineno}
\usepackage{footnote}
\usepackage{subfigure}
\usepackage{arydshln} 
\usepackage{listings}

\usepackage{siunitx}
\usepackage{color}

\usepackage{realboxes}
\usepackage{dirtree}
\usepackage{rotating}
\usepackage[version=3]{mhchem}
\usepackage{makecell}
\usepackage{footnote}
\usepackage{float}
\usepackage{graphicx}
\usepackage{siunitx}
\usepackage{array}
\usepackage{booktabs}

\usepackage{multirow}

\usepackage{indentfirst}
\title{Reconstruction Algorithm for a Novel Cherenkov Scintillation Detector}

\author{ Wentai Luo$ ^{a}$  Qian Liu$ ^{a}  $  Yangheng Zheng$ ^{a} $ Zhe Wang$ ^{b,c,d}$ Shaomin Chen$ ^{b,c,d} $}

\affiliation{%
$^a$School of Physical Sciences, University of Chinese Academy of Sciences, \\Beijing, China
}%
\affiliation{%
$^b$Department of Engineering Physics, Tsinghua University, \\Beijing, China
}%
\affiliation{%
$^c$Center for High Energy Physics, Tsinghua University, \\Beijing, China
}%
\affiliation{%
$^d$Key Laboratory of Particle \& Radiation Imaging (Tsinghua University), Ministry of Education, \\Beijing, China
}%

\collaboration{}
\emailAdd{luowentai15@mails.ucas.ac.cn}

\abstract{
For future MeV-scale neutrino experiments, a Cherenkov scintillation detector, CSD, is of particular interest for its capability to reconstruct both energy and direction for charged particles.
A type of new target material, slow liquid scintillator, SlowLS, which can be used to separate Cherenkov and scintillation lights, is one of the options for the neutrino detectors. 
A multi-hundred ton spherical CSD is simulated using a Geant4-based Monte Carlo software, which handles the detailed the micro processes of MeV particles and optical photons and the functions for photomultiplier, PMT, and readout electronics. 
Twelve SlowLS samples are simulated and studied to cover a wide range of scintillation light yields and scintillation emission time constants.
Based on the detailed knowledge of the signal processes, simplified functions are constructed to predict the charge and time signals on the PMTs to fulfill an efficient reconstruction for the energy, direction, and position of charged particles.
The performance of the SlowLS reconstruction, including the resulting energy, angular, and position resolution, and particle identification capability, is presented for these samples.
The dependence of the performance on the scintillation light yield and emission time constants is understood. 
This study will be a guideline for future MeV-scale neutrino CSD
design and SlowLS development for the interested physics goals.
}


\arxivnumber{2209.13772} % only if you have one

\collaboration[c]{}

\begin{document}
\maketitle
\flushbottom
%\linenumbers

\section{Introduction}
Liquid neutrino detectors typically come in two different varieties. One is based on liquid scintillator, and typical experiments are KamLAND, Borexino, Daya Bay, and RENO~\citep{KamLAND, Borexino2002, Dayabaytheta13, RENOLS}. The other uses water, and the experiments are Super-Kamiokande, SNO, and IMB~\citep{SK1998, SNOAngle, IMB1992}. 
Water Cherenkov detectors, despite the good directional performance, have a low light yield and poor energy resolution and are not very effective in detecting neutrinos below \SI{4}{MeV}~\citep{Solar2016Super-K4}. 
In contrast, the liquid scintillator neutrino detector
has a high light yield and, correspondingly, a good energy resolution
and can detect neutrinos below \SI{1}{MeV}, but almost no signature of direction can be seen~\citep{Borexino2022CID}. High energy resolution and directionality are two key ingredients in studying neutrino physics.

The Cherenkov scintillation detector, CSD, is a novel type of detector that can reconstruct the number of Cherenkov photons and scintillation photons. It not only has good position resolution and energy resolution but also has a good  angular resolution. A few slow liquid scintillators, SlowLS, candidates (water-based or oil-based) are being studied as the detection medium of the CSD~\citep{SLSYeMinfang, SLS2015, SLSLiMohan, SLSGuoZiyi, StevenBiller2020SLS, WbLS2020Yeh, StevenBiller2022SLS, land2021mev, kaptanoglu2019cherenkov, CSSeparation2017, 0vbb2014Winslowa}, whose main feature is the separation of scintillation light from Cherenkov light through the time profile.
The emission time profile of organic solution scintillators shows a rising time $\tau_r$ and a decay time $\tau_d$, resulting in the normalized pulse shape of the scintillation light~\citep{SLSLiMohan, SLSGuoZiyi}:
\begin{equation}
n_{\mathrm{s}}(t)=\frac{\tau_r+\tau_d}{\tau_d^2}(1-e^{-t/\tau_r})\cdot e^{-t/\tau_d}.
\label{Eq: SLSTime}
\end{equation}

For the SlowLS, $\tau_r$ is usually greater than \SI{1}{ns}, $\tau_d$ is more than \SI{20}{ns}, and the scintillation light yields range from a few hundred photons to a few thousand photons per MeV.
Cherenkov light of charged particles can be reconstructed as a prompt component and scintillation light mainly as a delayed component. Directional information is obtained from Cherenkov light and energy information can be estimated from both scintillation and Cherenkov lights.

Based on the properties of the SlowLS, the CSD will be of great benefit in many studies. The directional information provides the essential solar angle measurement and suppresses isotropic radioactive backgrounds in solar neutrino experiments~\cite{Solar2006Super-K1, Solar2008Super-K2, Solar2011Super-K3, Solar2016Super-K4, Solar2001SNO, Solar2014Borexino, Direction2013SNO3}. 
In addition, determining the numbers of Cherenkov photons and scintillation photons also helps to do particle identification, PID, for electrons, gamma, protons, and alpha. 
In neutrinoless double beta, $0 \nu \beta \beta$, decay experiments, good directional information and PID would reduce the solar neutrino's background and radioactive background~\citep{KamLAND-Zen2022, gando2012measurement, KamLANDZen2020, gando2016search, andreotti2011130te, albert2014improved, caccianiga2000neutrinoless}. 
Supernova neutrinos research will also be beneficial for the PID information~\citep{SuperNovaWeiHanyu}.

China's Jinping underground laboratory has a \SI{2400}{m} rock overburden. A CSD analysis is based on the design of a multi-hundred-ton Jinping CSD proposal, which will measure MeV-scale neutrinos, such as solar-, geo-, supernova neutrinos, and $0 \nu \beta \beta$ decay. Other detectors with comparable structures can also use the approaches described in this article. 

In this paper, a CSD reconstruction algorithm and the reconstruction performance of the CSD with full detector simulation study are presented.  
The results can be used as a guideline for the new detector design or for a new SlowLS development for a specific physics goal.
Section~\ref{Sec: SimulationTotal} describes the details of the simulation, including the detector geometry, the setup of the simulation, and the properties of twelve SlowLS samples. Section~\ref{Sec: Waveform} focuses on the analysis of the simulated output waveforms. To reconstruct the energy, position, and direction, and do PID for the events, first, in section~\ref{Sec: SignalProcess}, we decompose and analyze the signal analysis and make the corresponding simplification. Additionally the study of the direct photons and indirect photons is included in this section. Next, in section~\ref{Sec: Model}, we construct the prediction models of the charge and time profiles and compare them with the full simulation. 
Then, reconstruction is done with the prediction models using a maximal likelihood method, and the performance of the CSD detector reconstruction is shown in section~\ref{Sec: Recon}. Section~\ref{Sec: Conclusion} is the conclusion.

\section{Detector Setup and Simulation}\label{Sec: SimulationTotal}

The geometry of the proposed multi-hundred-ton detector, the settings of the full simulation, and the chosen twelve SlowLS samples are all described in this section.

\subsection{Detector Setup} 

\begin{figure}[!htbp]
\centering
\includegraphics[scale=0.48]{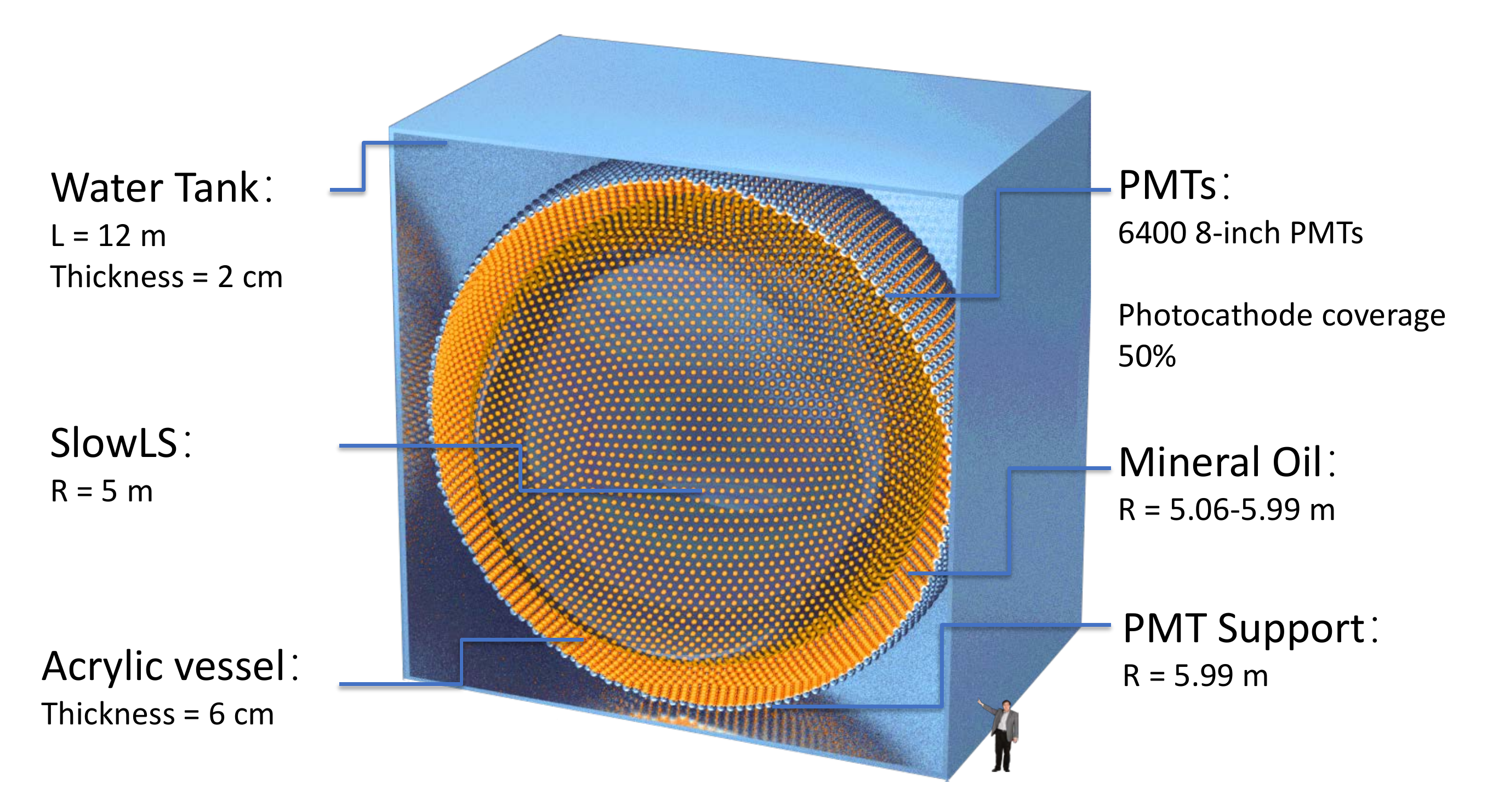}
\caption{Structure of the multi-hundred-ton detector and its geometry setup.}
\label{500tdetector}
\end{figure}

 A multi-hundred-ton neutrino detector will be built, which is comparable to the Borexino, Kamland, and SNO+ detectors; thus, the following geometry is set up for simulation, which is depicted in figure~\ref{500tdetector}. Five hundred metric tons of the SlowLS are contained in a 10-m diameter transparent acrylic vessel. The acrylic vessel is constructed from 6-cm-thick transparent acrylic and is suspended at the center of the CSD. Scintillation and Cherenkov light produced by neutrino interactions and radioactive backgrounds is detected by an array of 6400 Hammamatsu model 8-inch photomultipliers (photocathode coverage to nearly 50\%), PMTs, supported by a stainless steel geodesic sphere. The setup of the 8-inch PMT is as follows. 34\% quantum efficiency at nearly \SI{390}{nm}, 100\% collection efficiency, \SI{2}{kHz} dark noise rate, \SI{20}{ns} transmission time, TT, and \SI{1}{ns} transmission time spread (full width at half maximum), TTS~\citep{HighQEByHama}. Additionally, over 1,200 tons of mineral oil and water shield the SlowLS from external radioactive backgrounds: 300 tons of mineral oil of the acrylic vessel and the PMT support sphere and 800 tons of water between the PMT support sphere and a water tank.

\subsection{Simulation} \label{Sec: Simulation}
\begin{figure}[!htbp]
\centering
\includegraphics[width=0.6\textwidth]{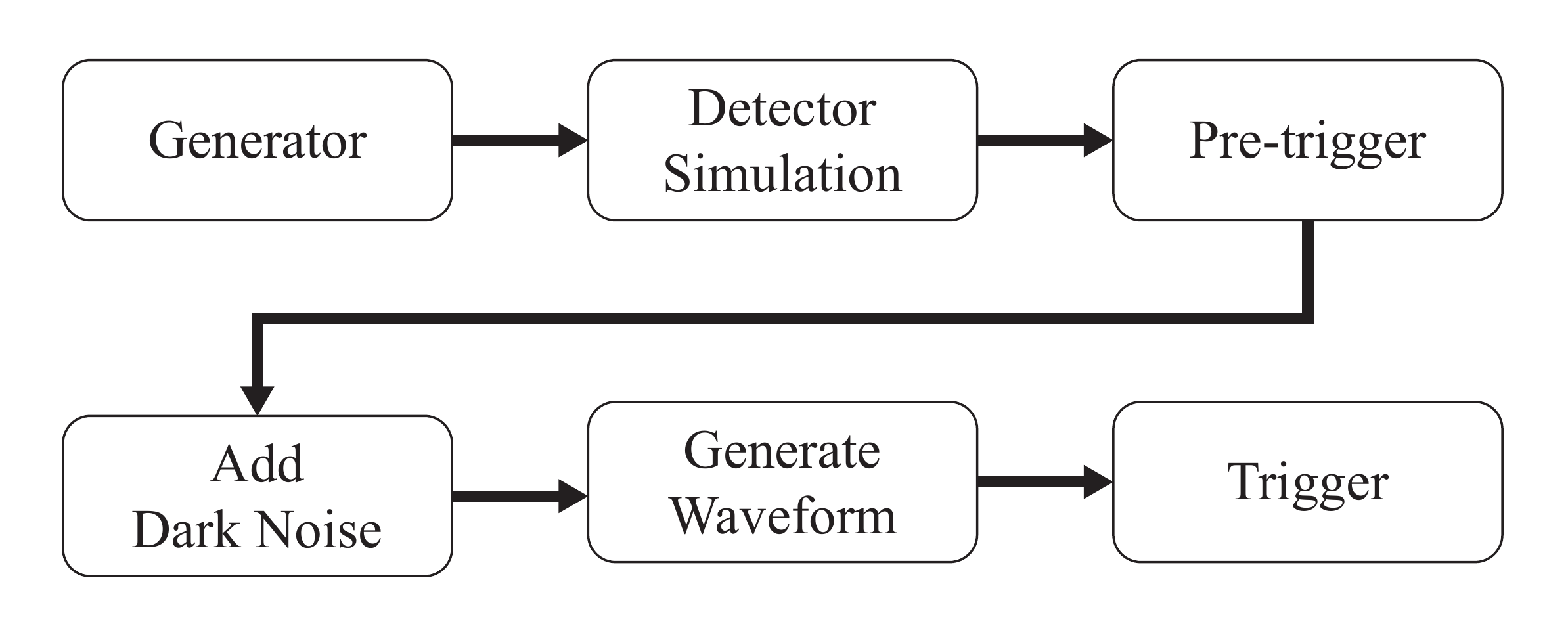}
\caption{Simulation flow chart of the proposed MeV-scale neutrino experiment. }
\label{fig:ov}
\end{figure}

A Monte-Carlo, MC, simulation package, the Jinping Simulation and Analysis Package, JSAP, is used to simulate the physical process in the CSD. The JSAP is based on Geant4~\citep{Geant4} with many customized features. The main components are shown in a simulation flow chart in figure~\ref{fig:ov} and expanded below.

\paragraph{Generator}
The generator creates a primary particle and generates its position, time, and momentum. The generator for neutrino interactions only begins with secondary particles such as recoiled electrons. 

\begin{figure}[!htbp]
\centering
\includegraphics[width= .60\textwidth]{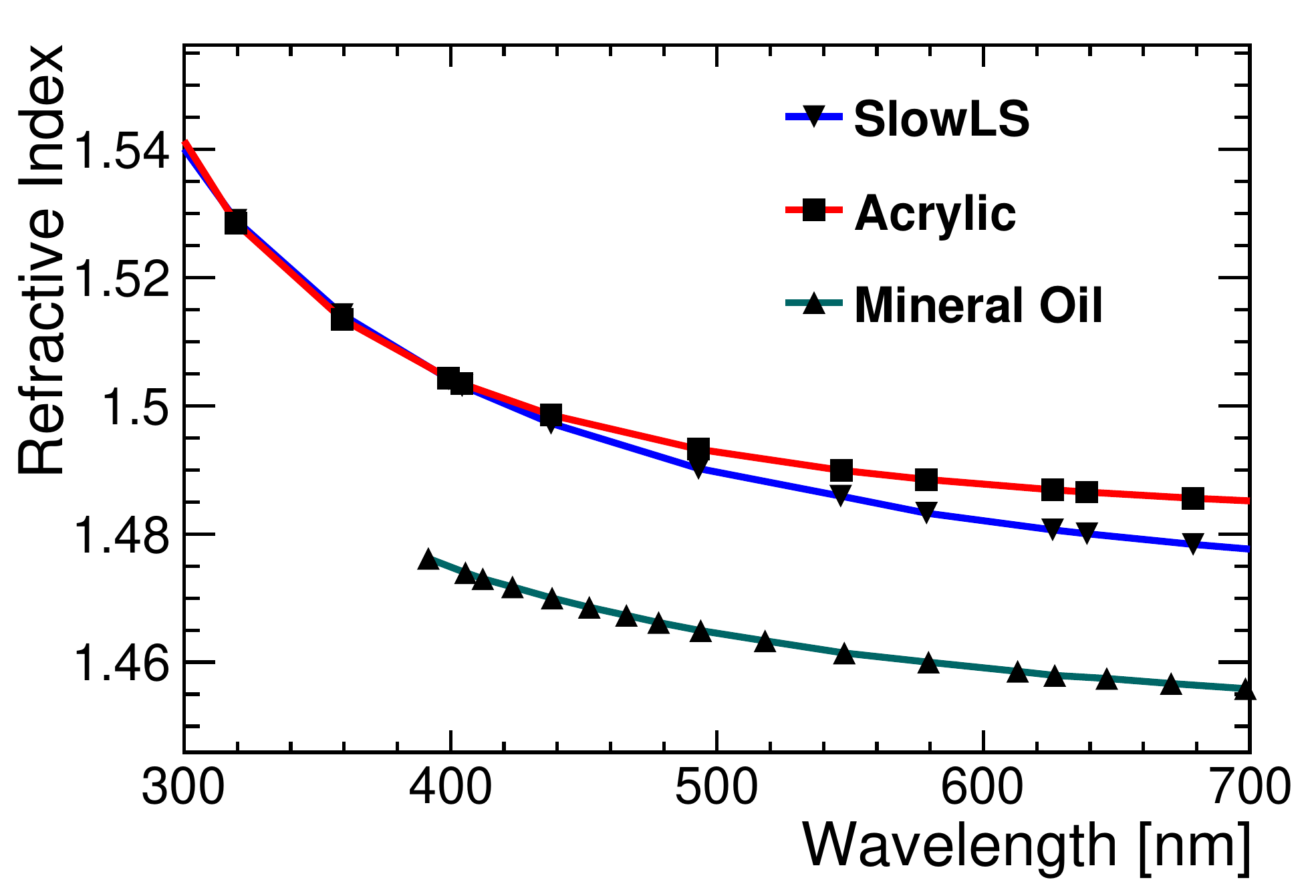}
\caption{A refractive index spectrum of the SlowLS, acrylic and mineral oil in simulation~\citep{RindexReno, LABRindex}.}
\label{Fig: Rindex}
\end{figure}

\begin{figure}[!htbp]
\centering
\subfigure{\includegraphics[scale=0.37]{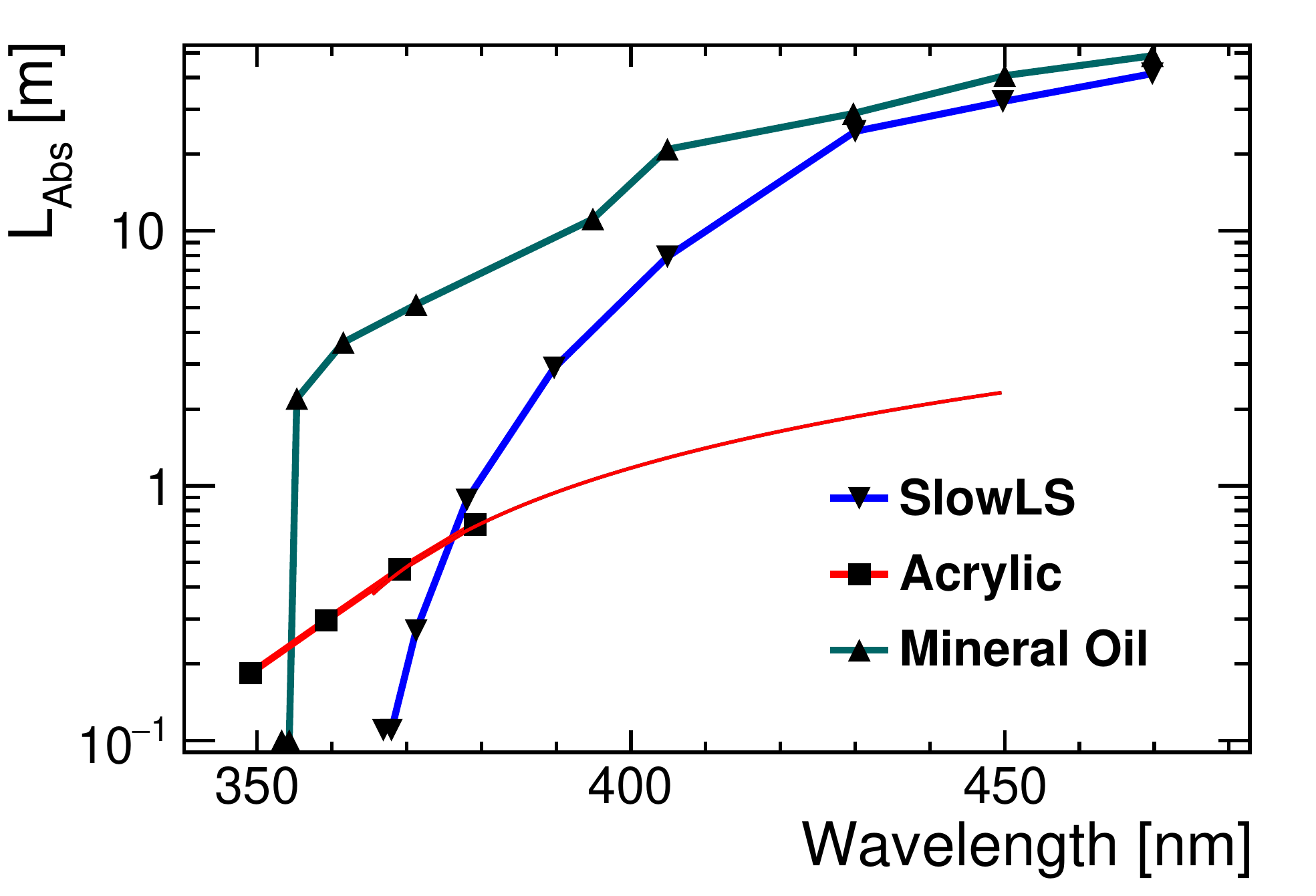}}
\subfigure{\includegraphics[scale=0.37]{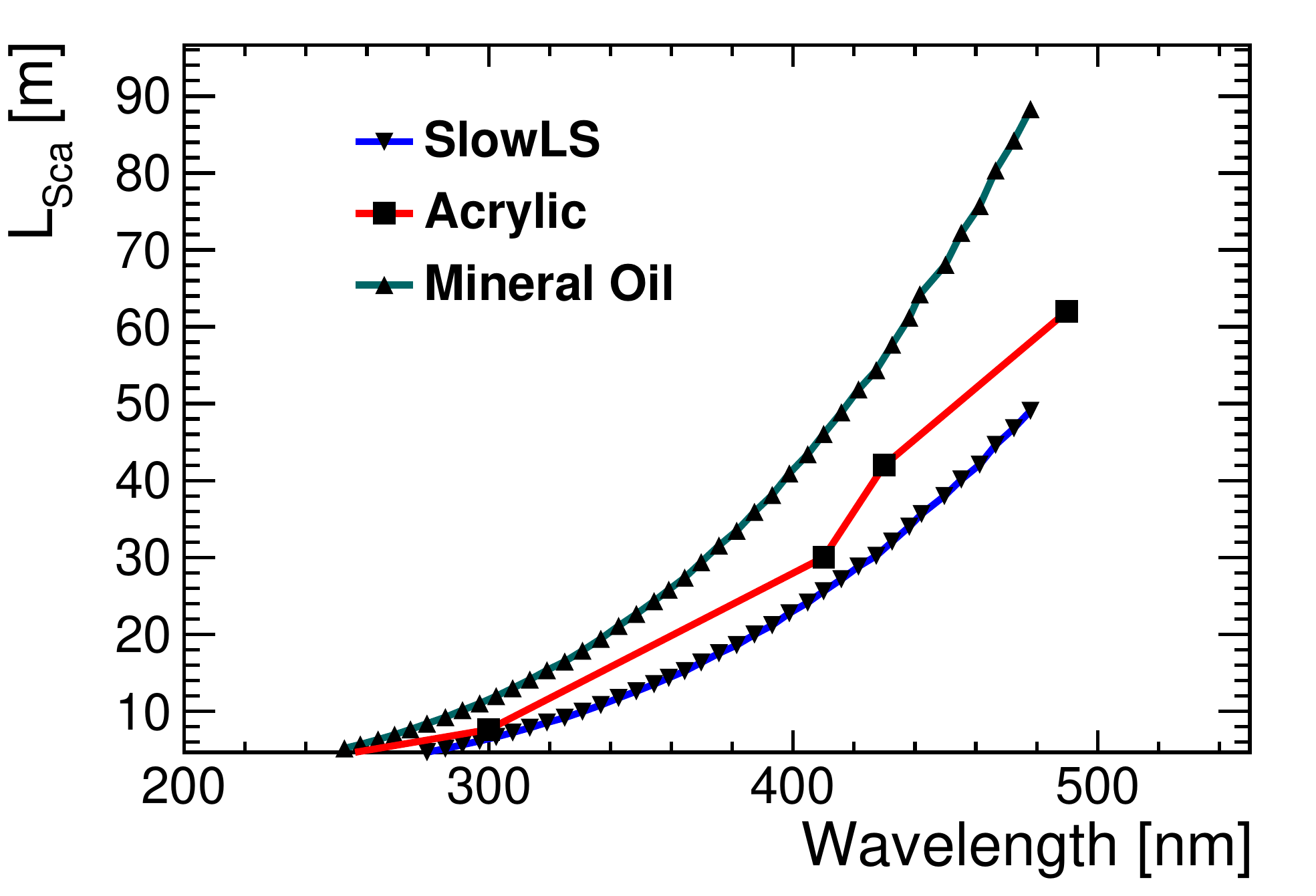}}
\caption{Left: The absorption length spectrum of transparent materials in simulation. Right: The Rayleigh scattering length spectrum of transparent materials in simulation~\citep{AcrylicDayabay, LABAbs, RayleighZhouXiang, MORayMiniBoone}.}
\label{ABSRaySim}
\end{figure}

\begin{figure}[!htbp]
\centering
\includegraphics[width= .60\textwidth]{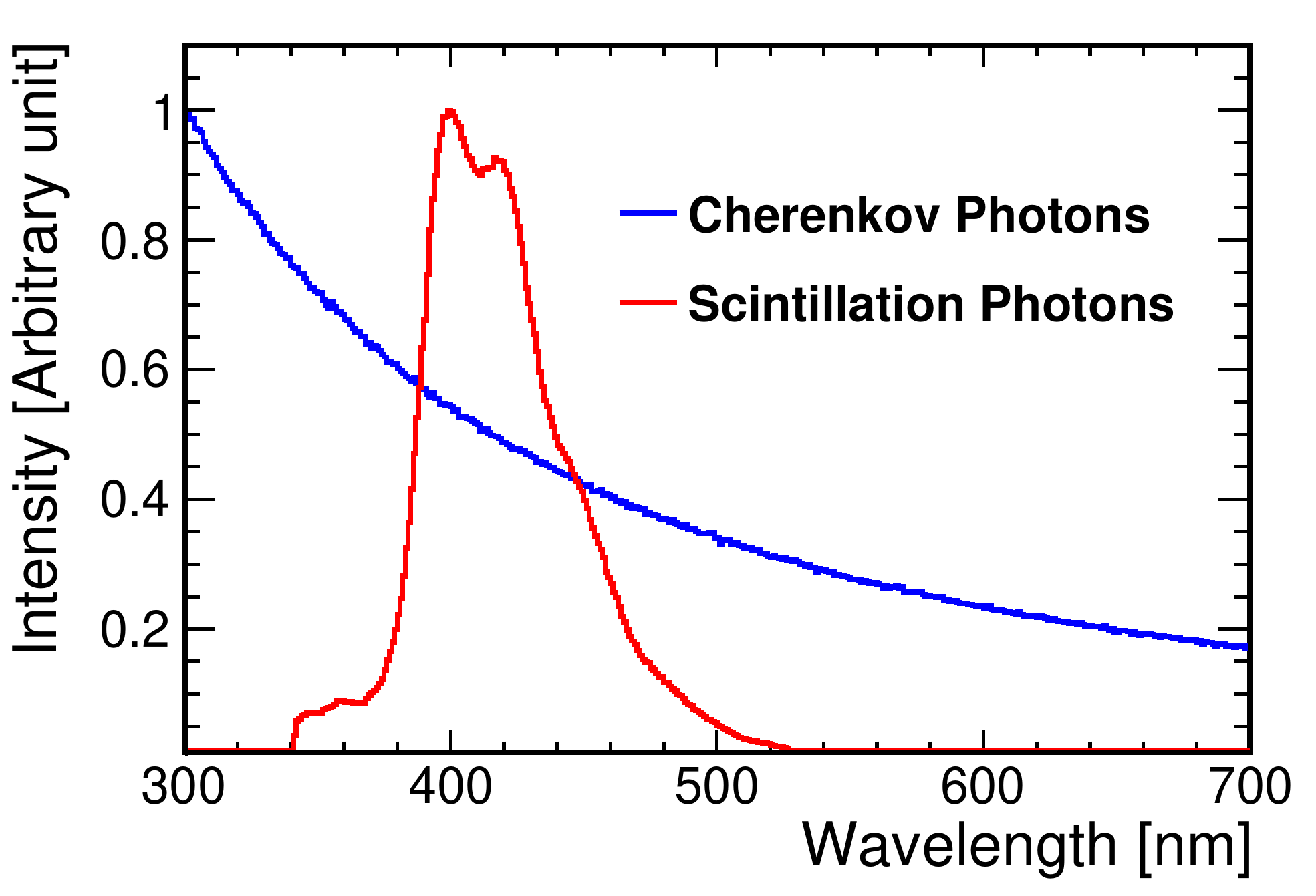}
\caption{The wavelength spectrum of scintillation and Cherenkov photons in simulation~\citep{SLSGuoZiyi, Jackson1999}.}
\label{CSPhotonSpec}
\end{figure}

\paragraph{Detector simulation}
The software simulates the detector’s response using information from the physics generator as well as the detector’s geometry and materials. In the CSD, photons will pass through substances such as the SlowLS, acrylic, and mineral oil. For the above substances, the JSAP can define the propagation characteristics of the photons in the substance, which are included:

\begin{itemize}
\item The spectra of the refractive index, absorption length, and Rayleigh scattering length are shown in figure~\ref{Fig: Rindex}~\citep{RindexReno, LABRindex} and figure~\ref{ABSRaySim}~\citep{AcrylicDayabay, LABAbs, RayleighZhouXiang, MORayMiniBoone}.
\item The scintillation emission spectrum is shown in figure~\ref{CSPhotonSpec}~\citep{SLSGuoZiyi, Jackson1999}.
\item The scintillation light yield, rise time and decay time will be discussed in table~\ref{SimSetup}.
\item Birks' constant $k_\mathrm{B}$ in this work is \SI{0.016}{cm/MeV}~\citep{Birksconstant2011}.
\end{itemize}

The output is the information about a photoelectron, PE, hit. The PE hit is a sampling based on quantum efficiency that determines whether an optical photon that hits a PMT photocathode may be converted to an electron. 

\paragraph{Dark noise and Pre-trigger}
On the PMT photocathode, thermal electron emission produces dark noise, which results in an equivalent PE signal. 
The pre-trigger determines whether the total count rate of the physically caused PEs and dark noise signals reaches the minimum number of hit PMTs within the time window (\SI{500}{ns}).
If the total count rate cannot reach the minimum number of hit PMTs, no more simulations will be performed, which saves simulation time.

\paragraph{Waveform generation}
The input is the PE hit time and dark noise hit time. The PEs are convoluted with the single PE response of the 8-inch PMT~\citep{HighQEByHama}. A flash analog-to-digital converter is simulated in the JSAP's readout electronics system with a dynamic range of \SI{1}{V}, a precision of \SI{10}{bits}, and a sampling rate of \SI{1}{GHz}. The electronics readout system then combines these PEs with a single PE waveform to form the output waveforms, which is the same as the actual experimental data. The gain of each PMT is considered. The single PE waveform is fit using the empirical formula as follows:
\begin{equation}
\mathrm{PE_{single}}(t)= A_0\cdot \frac{1}{\tau_{\mathrm{PE}}} \mathrm{exp}(-\frac{t}{\tau_{\mathrm{PE}}})\theta(t)\otimes \frac{1}{\sigma_{\mathrm{PE}} \sqrt{2\pi}}\exp{(-\frac{t^2}{2\sigma_{\mathrm{PE}}^2})},
\label{Eq: SingPEWave}
\end{equation}

\noindent where $A_0$ is the single PE charge, $\theta(t)$ is a step function, and $\sigma_{\mathrm{PE}}$ and $\tau_{\mathrm{PE}}$ are the shape parameters.

\begin{figure}[!htbp]
\centering
\includegraphics[width= .60\textwidth]{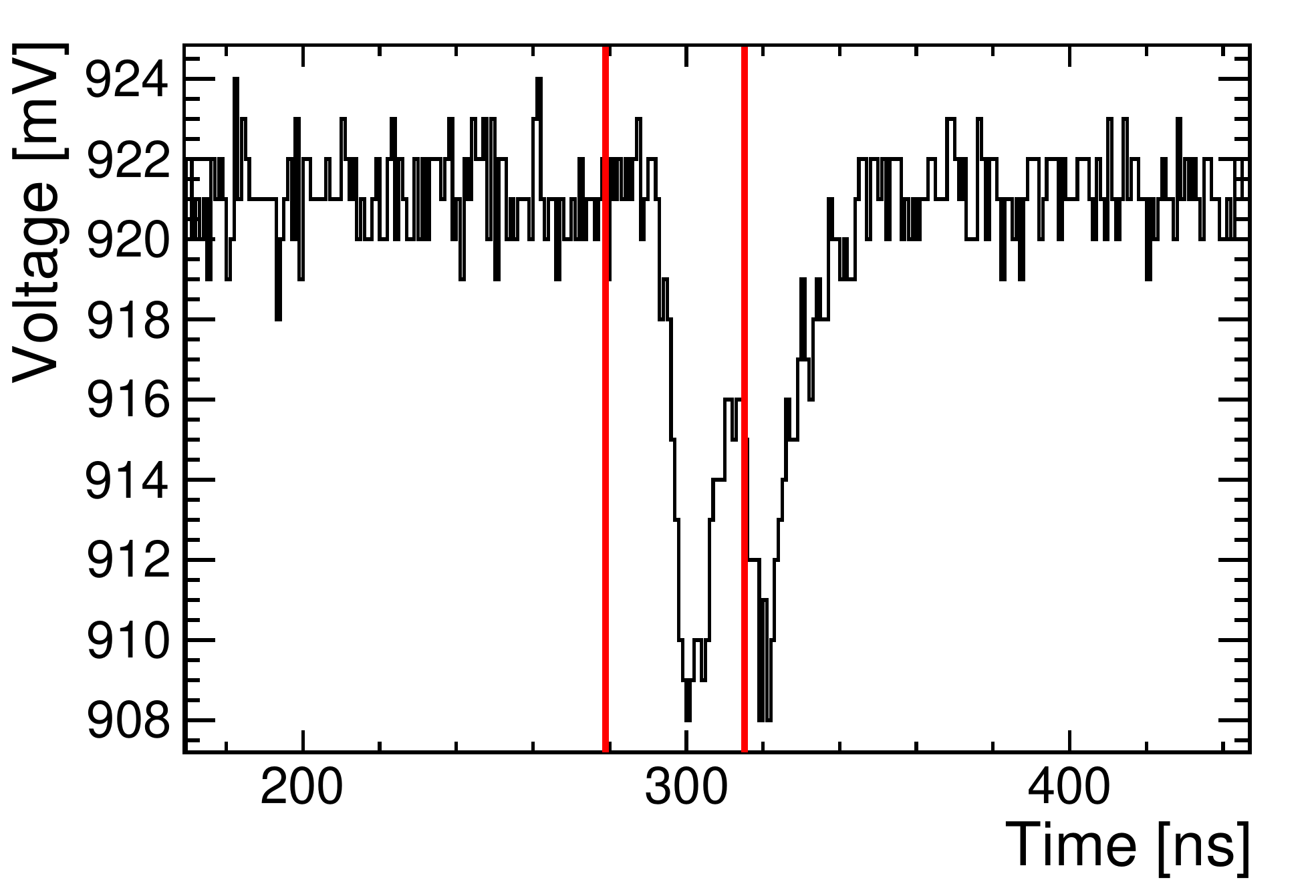}
\caption{A waveform of the simulation output. The PE time is indicated by the red line.}
\label{fig:wav}
\end{figure}

Moreover, a baseline fluctuation is considered to be \SI{1}{mV}, and a waveform is depicted in figure~\ref{fig:wav}.

\paragraph{Trigger}

This step synchronizes and assimilates the information from the raw waveform and checks the trigger conditions. A threshold of the PMTs is \SI{5}{mV}, and the trigger condition of the event is that at least 15 hit PMTs within \SI{500}{ns}.
\subsection{Twelve SlowLS Samples}
In this paper, we attempt to separate Cherenkov light from scintillation light, which is related to two keys of the SlowLS: the scintillation  light yield and scintillation time profile. 
The current study of the SlowLS indicates that the higher the light yield is, the smaller the time constants of the time profile for the oil-based liquid scintillator~\citep{SLSGuoZiyi}. For water-based liquid scintillators, the higher the light yield is, the larger the time constants of the time profile~\citep{WbLS2020Yeh}. To find the potential SlowLS in the CSD, the four typical light yields and three typical time profiles are selected, as shown in table~\ref{SimSetup}. 

The four typical scintillation light yields are 200, 350, 3750 and \SI{5000}{photons/MeV}. The low light yield SlowLS samples correspond to the water-based liquid scintillator, which usually has a low light yield of a few hundred photons per MeV~\citep{WbLS2020Yeh}; the high light yield SlowLS samples correspond to the oil-based liquid scintillator, which is usually as high as several thousand photons per MeV~\citep{SLSGuoZiyi}. The rise times, $\tau_r$ of the three typical time profiles are 7.7, 1.2 and \SI{0.5}{ns} and the decay times, $\tau_d$ are 37, 27, and \SI{10}{ns}. The two slower time profiles are derived from existing measurements of the SlowLS~\citep{SLSGuoZiyi,SLSLiMohan}, and the fastest time profile is used to explore the time profile of the potential SlowLS.

Table~\ref{SimSetup} lists four PE yields, $E_{\mathrm{scale}}$, and their corresponding scintillation light yields. Although the number of PEs is related to the attenuation length spectrum of the medium, the PMT photocathode coverage, and the detection efficiency of the PMT, similar PE yields can be obtained by slightly varying the spectra of these parameters in the simulation. In addition, a signal process will be discussed in detail in later sections.

Table~\ref{SimSetup} also lists a Cherenkov-to-scintillation ratio, CSR, within the first \SI{10}{ns}. The times of the hit PEs are arranged from smallest to largest, and the number of PEs generated by Cherenkov and scintillation light is counted to obtain the CSR. Combined with the time response of the detector, the CSR within the first \SI{10}{ns} (equal to 5 times the time resolution) can better characterize the distinction between scintillation and Cherenkov light. As a result, we use the CSR in the first \SI{10}{ns} of the detector center's \SI{5}{MeV} electrons as a criterion to judge the distinction between Cherenkov and scintillation light.

For classification purposes, we named the twelve different samples such as $\mathrm{LS_{\mathrm{376~PE}}^{\mathrm{7.7~ns}}}$ or $\mathrm{LS_{\mathrm{35~PE}}^{\mathrm{0.5~ns}}}$, with PE yield (\SI{376}{PE/MeV} or \SI{35}{PE/MeV}) in the lower corner and $\tau_r$ (\SI{7.7}{ns} or \SI{0.5}{ns}) in the upper corner. In section~\ref{Sec: Recon}, the twelve different samples of the SlowLS are reconstructed in the CSD.

\begingroup
\setlength{\tabcolsep}{18pt} % Default value: 6pt
\renewcommand{\arraystretch}{1.2} % Default value: 1
\begin{table}[!htbp]
\centering
\begin{tabular}{@{}ccccc@{}}
\toprule
\multirow{2}{*}{\begin{tabular}[c]{@{}c@{}}Scintillation \\light yield\\ (photons/MeV)\end{tabular}} 
& \multirow{2}{*}{\begin{tabular}[c]{@{}c@{}}PE yield  \\$E_{\mathrm{scale}}$\\(PE/MeV)\end{tabular}} 
&\multicolumn{3}{c}{CSR in the first \SI{10}{ns}} \\ \cmidrule(l){3-5} 
 
&  

& \begin{tabular}[c]{@{}c@{}}$\tau_r=7.7~\mathrm{ns}$\\ $\tau_d=37~\mathrm{ns}$\end{tabular} 
& \begin{tabular}[c]{@{}c@{}}$\tau_r=1.2~\mathrm{ns}$\\ $\tau_d=27~\mathrm{ns}$\end{tabular} 
& \begin{tabular}[c]{@{}c@{}}$\tau_r=0.5~\mathrm{ns}$\\ $\tau_d=10~\mathrm{ns}$\end{tabular} \\ \midrule

         200& 24 &\begin{tabular}[c]{@{}l@{}} 1:~0.1 \end{tabular} &\begin{tabular}[c]{@{}l@{}} 1:~0.3 \end{tabular} & \begin{tabular}[c]{@{}l@{}} 1:~0.7 \end{tabular} \\ 
        
        350 &35 & \begin{tabular}[c]{@{}l@{}} 1:~0.2 \end{tabular} &\begin{tabular}[c]{@{}l@{}} 1:~0.6\end{tabular} & \begin{tabular}[c]{@{}l@{}} 1:~1.3\end{tabular} \\

         3750 &285&\begin{tabular}[c]{@{}l@{}} 1:~2.4\end{tabular} &\begin{tabular}[c]{@{}l@{}} 1:~6.1\end{tabular} & \begin{tabular}[c]{@{}l@{}} 1:~14\end{tabular} \\ 
         
        5000 &376 & \begin{tabular}[c]{@{}l@{}} 1:~3.2\end{tabular} &\begin{tabular}[c]{@{}l@{}} 1:~8.2\end{tabular} & \begin{tabular}[c]{@{}l@{}} 1:~19\end{tabular}  \\ \bottomrule
\end{tabular}
\caption{Twelve samples of three scintillation time profiles and four scintillation light yields are shown in the table. The PE yields of them are obtained by the full MC simulation. The CSR in the first \SI{10}{ns} is obtained by simulating the \SI{5}{MeV} electrons at the detector center for these samples.}
\label{SimSetup}
\end{table}
\endgroup

\section{Readout and Waveform}\label{Sec: Waveform}

The simulated output waveform needs to be analyzed to obtain the number of the observed PEs by each PMT and the time of each PE. This section focuses on the waveform analysis method and the performance.

\subsection{Waveform Analysis}

The time and number of PEs are obtained by fitting the $i$th waveform with the constructed likelihood function. The time window of the waveform is \SI{500}{ns}, and the trigger time starts from \SI{125}{ns}, so the waveform is fit within a time window of 100-\SI{500}{ns}. As a result, a likelihood function of waveform analysis is shown as follows:
\begin{equation}
\mathcal{L}(n_{i}^{\mathrm{Obs}},t_{ij},R_{ij}) = f_{\mathrm{Poisson}}(n_{i}^{\mathrm{Obs}};n_{\mathrm{average}}) \cdot \prod^{500}_{k=100}\mathrm{Gauss}([V_{ik}-\sum_{j=1}^{n_{i}^{\mathrm{Obs}}} R_{ij} \mathrm{PE_{single}}(t-t_{ij})],\sigma_{\mathrm{base}}),
\label{Eq: LikelihoodWave}
\end{equation}

\noindent where $n_{i}^{\mathrm{Obs}}$ is the number of PEs of the $i$th waveform, $t_{ij}$ is the start time of the $j$th PE, and $R_{ij}$ is the ratio of the charge of the $j$th PE to the average charge. $V_{ik}$ is the amplitude of the $k$th bin, and $\mathrm{PE_{single}}$ is the single PE waveform shown in eq.~(\ref{Eq: SingPEWave}), and $\sigma_{\mathrm{base}}$ is the standard deviation of the baseline, which is \SI{1}{mV}. $f_{\mathrm{Poisson}}$ is the Poisson distribution with the average number of PEs, $n_{\mathrm{average}}$ is defined as,

\begin{equation}
n_{\mathrm{average}}=\displaystyle {\sum_{i=1}^{N_{\mathrm{hitPMT}}}{q_i}} \frac{1}{q_{\mathrm{single}} N_{\mathrm{PMT}}},
\label{Eq: Navaerage}
\end{equation}

\noindent where $N_{\mathrm{hitPMT}}$ is the number of the hit PMT, $q_i$ is the charge of the $i$th waveform, $q_{\mathrm{single}}$ is the mean value of the single PE charge obtained from the calibration, and $N_{\mathrm{PMT}}$ is the total number of PMTs.

\subsection{Performance of Waveform Analysis}
\begin{figure}[!htbp]
\centering
\subfigure{\includegraphics[scale=0.37]{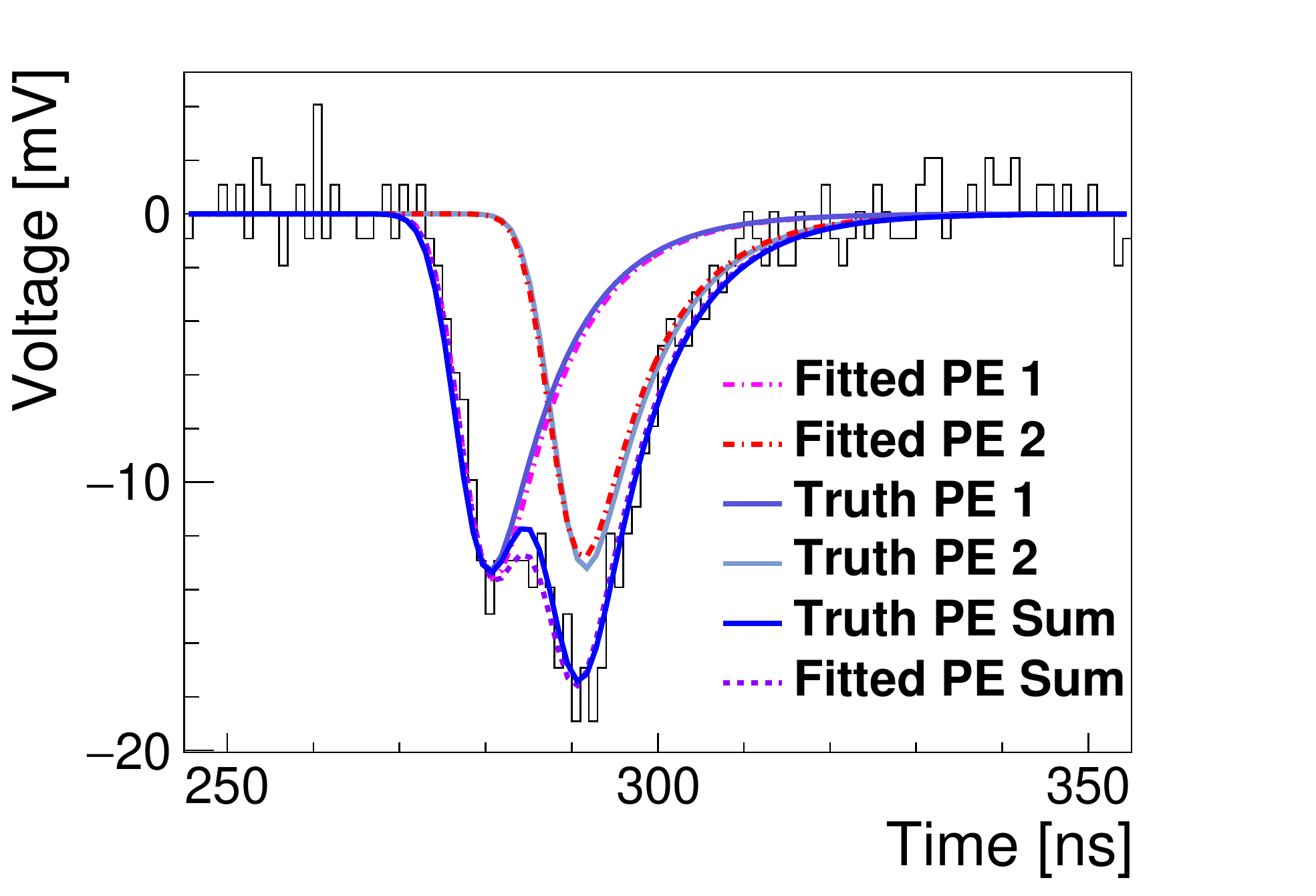}}
\subfigure{\includegraphics[scale=0.37]{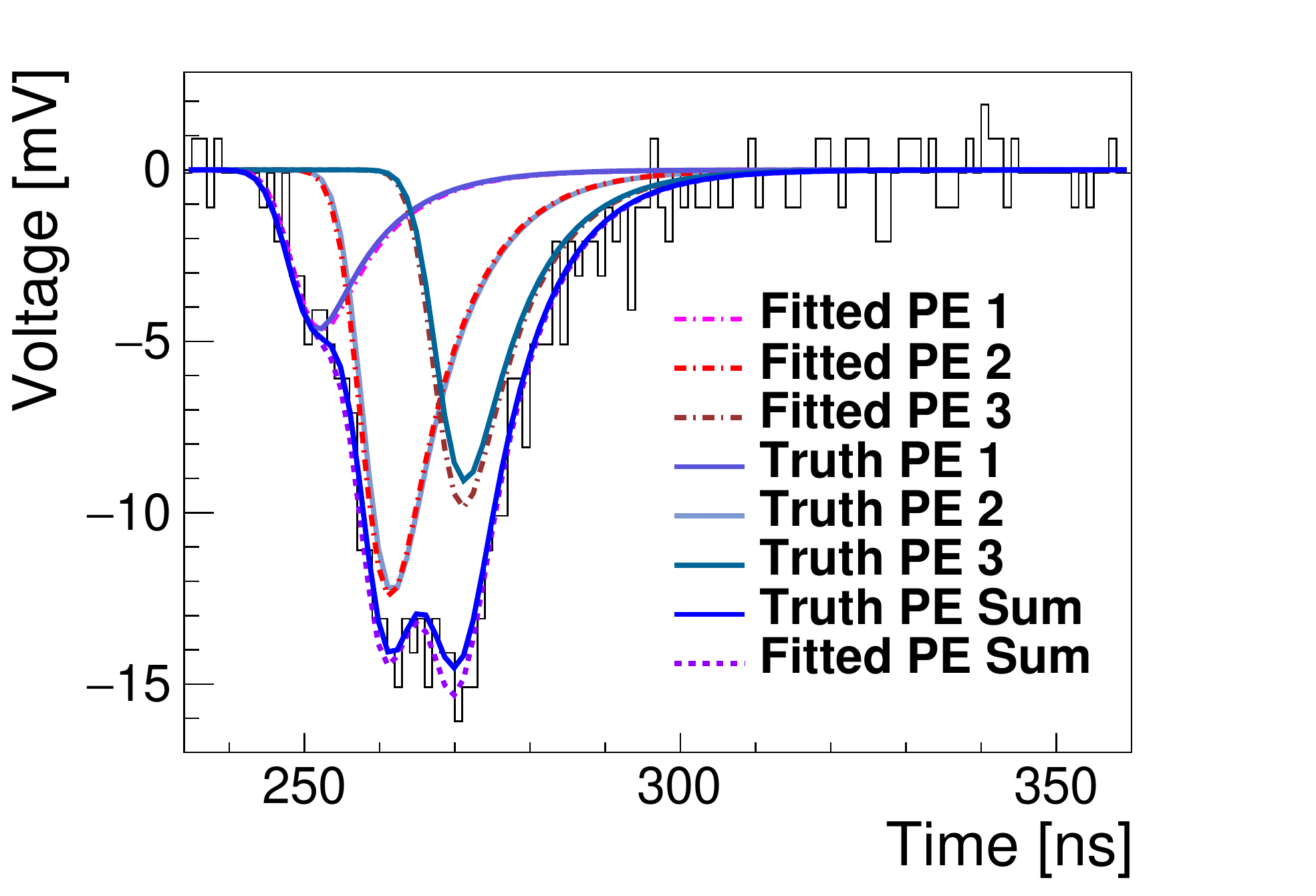}}
\caption{Two examples of the fit waveforms.}
\label{FitWaveform}
\end{figure}

\begin{figure}[!htbp]
\centering
\subfigure{\includegraphics[scale=0.37]{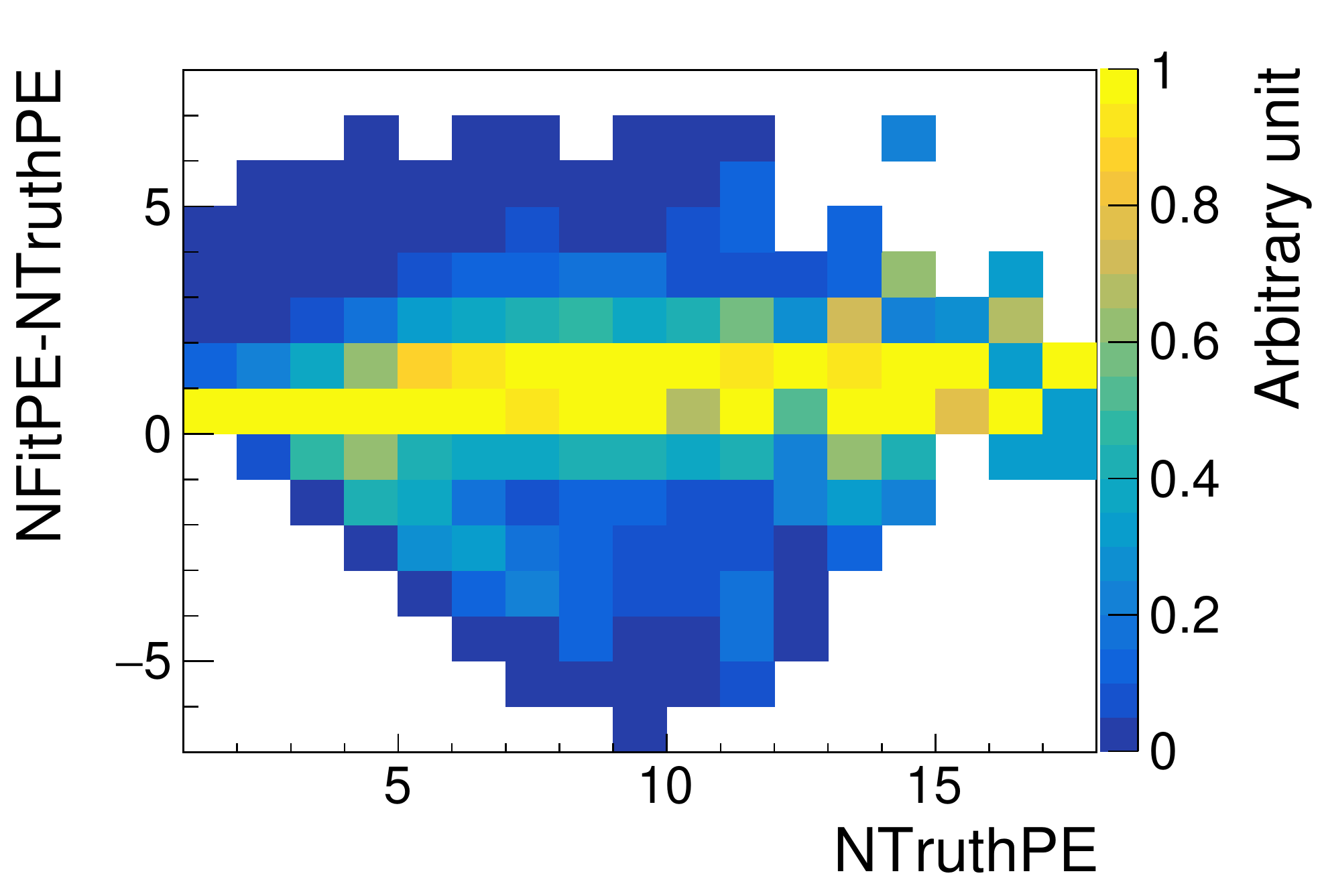}}
\subfigure{\includegraphics[scale=0.37]{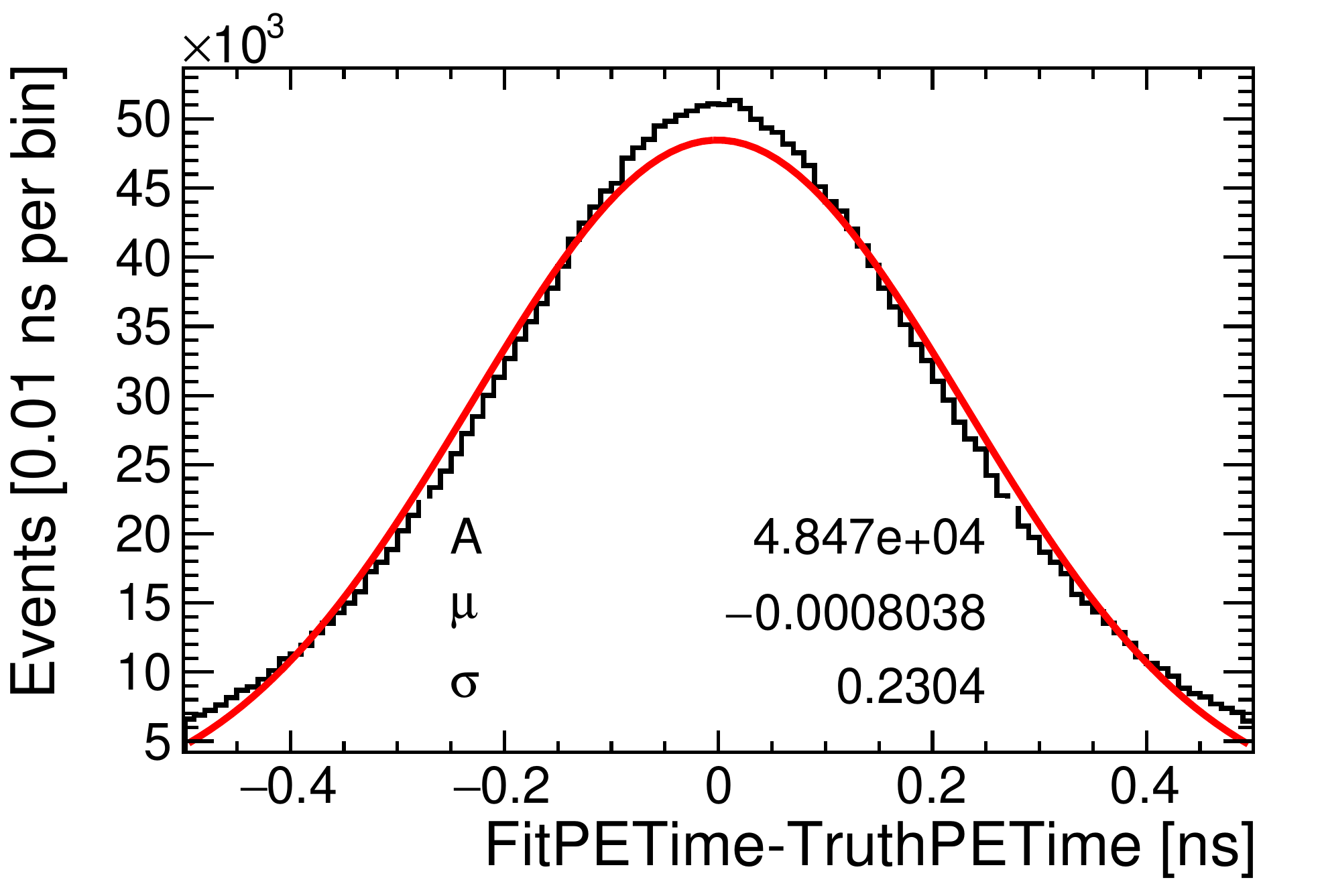}}
\caption{Left: The variations in the number of PEs for each waveform. Right: The difference between the true and fit PE time.}
\label{fig: PE Compare}
\end{figure}

$\mathrm{LS_{\mathrm{35~PE}}^{\mathrm{0.5~ns}}}$ is used as an example with a scintillation light yield of \SI{350}{photons/MeV}. The \SI{5}{MeV} kinetic energy electrons in the SlowLS are simulated. With the use of the maximum likelihood function method, the waveform is analysed. The PE time and the number of PEs on each PMT are eventually produced by maximization of the likelihood function. The results, as shown in figure~\ref{FitWaveform}. Moreover, the results also compare the true PE with the fit PE are shown in figure~\ref{fig: PE Compare}.

As seen in the left panel of figure~\ref{fig: PE Compare}, in most of the analysis results of the waveforms, the number of fit PEs and the number of true PEs are basically the same. In addition, the fit PE time is compared with the true PE time as shown in the right panel of figure~\ref{fig: PE Compare}. The distribution of the PE time difference is obtained by subtracting the fit PE time from the true PE time. We fit the distribution with a Gaussian function with a fit standard deviation of \SI{0.23}{ns}. Moreover, for more than 93\% of the PEs, the time difference between the fit result and the true value is less than \SI{0.5}{ns}.

\section{Signal Process and Simplification}\label{Sec: SignalProcess}
The number of observed PEs on each PMT and the PE times are obtained from the waveform analysis. Further reconstruction of energy, position, direction, and PID requires the prediction of the number of PEs and PE time on each PMT, so the signal process needs to be carefully studied.

The process from a particle's energy deposition to PEs is very complex. If the signal process was fully considered in the original physics style during reconstruction, the computation time would be astronomical. To save reconstruction time, the corresponding simplified functions are constructed for the signal process. Additionally, the thoughts on direct and indirect photons are covered in detail. 

 \subsection{Photon Emission} \label{PY}
The mechanisms of scintillation and Cherenkov light generation and the Cherenkov light angle distribution are described here, and the light yields of both are calculated.

\subsubsection{Scintillation Light}\label{Sec: YS}
Scintillation light is a type of photon produced by ionization or excitation of charged particles in a medium, and its emission direction is isotropic. In addition, the quenching effect of charged particles is generally expressed by the semiempirical formula Birks' law. Birks' law gives the scintillation light yield $Y_{\mathrm{s}}$ per unit of travel distance~\citep{BirksLaw}:

 \begin{equation}
\frac{\mathrm{d}Y_{\mathrm{s}}}{\mathrm{d}x} = A\frac{\mathrm{d}E/\mathrm{d}x}{1+k_{\mathrm{B}} \mathrm{d}E/\mathrm{d}x},
\label{Eq: YS}
\end{equation}

\noindent where the energy loss density for charged particle production of scintillation light is $\mathrm{d}E/\mathrm{d}x$, $k_{\mathrm{B}}$ is Birks' constant, as introduced in section~\ref{Sec: Simulation} and $A$ is a normalization constant. 

The quenching effect for 1 to \SI{9}{MeV} electrons is nearly 0.88 in the simulation. As a result, for \SI{2}{MeV} kinetic energy electrons, when the scintillation light yield is \SI{5000}{photons/MeV}, the total number of scintillation photons is $Y_{\mathrm{s}}(E)=0.88Y_{\mathrm{s}}E=8800~\mathrm{photons}$.
Moreover, the wavelength spectrum of the scintillation light for simulation is shown in figure~\ref{CSPhotonSpec}. 

\subsubsection{Cherenkov Light}\label{Sec: YC}
Cherenkov photons are produced in a cone around the trajectory of a charged particle when the charged particle travels faster than the speed of light in a medium.~\citep{pdg}. The cosine of the emission angle, $\theta_{\mathrm{c}}$, can be expressed as:
\begin{equation}
\cos\theta_{\mathrm{c}} = \frac{1}{n(\lambda)\beta},
\end{equation}

\noindent where $n$ is the refractive index determined as shown in figure~\ref{Fig: Rindex}, and $\lambda$ is the wavelength of the photon. The charged particle's relative velocity $\beta$ is shown as

\begin{equation}
\beta=v/c,
\end{equation}
\noindent where $v$ is the velocity of a charged particle in a medium, and $c$ is the speed of light in a vacuum. As a result, the energy threshold for electrons in the SlowLS with $n = 1.5$ is \SI{0.686}{MeV}. In addition, the Cherenkov photon yield $Y_{\mathrm{c}}$ per traveled distance $x$ per unit $\lambda$ is

\begin{equation}
\frac{\mathrm{d}^2Y_{\mathrm{c}}}{\mathrm{d}x\mathrm{d}{\lambda}}=\frac{2\pi a q^2}{\lambda^2}(1-\frac{1}{\beta^2 n^2(\lambda)}),
\label{Eq: YC}
\end{equation}

\noindent where $a$ is the fine structure constant and $q$ is the charge of the particle. Assuming that the refractive index is wavelength independent from 300 to \SI{700}{nm} (the detectable wavelength range), the Cherenkov light yield $Y_{\mathrm{c}}$ for 1 to \SI{9}{MeV} electrons with kinetic energy $E$ is obtained by integrating $\lambda$,

\begin{equation}
Y_{\mathrm{c}}(E)=436E\sin^2{\theta_{\mathrm{c}}},
\label{Eq: YC_Simple}
\end{equation}

\noindent where $\theta_{\mathrm{c}}$ is the initial emission angle of the electron.
Moreover, the wavelength spectrum of the Cherenkov light for simulation is shown in figure~\ref{CSPhotonSpec}. 

\paragraph{Angular Distribution}

\begin{figure}[!htbp]
\centering
\subfigure{\includegraphics[scale=0.37]{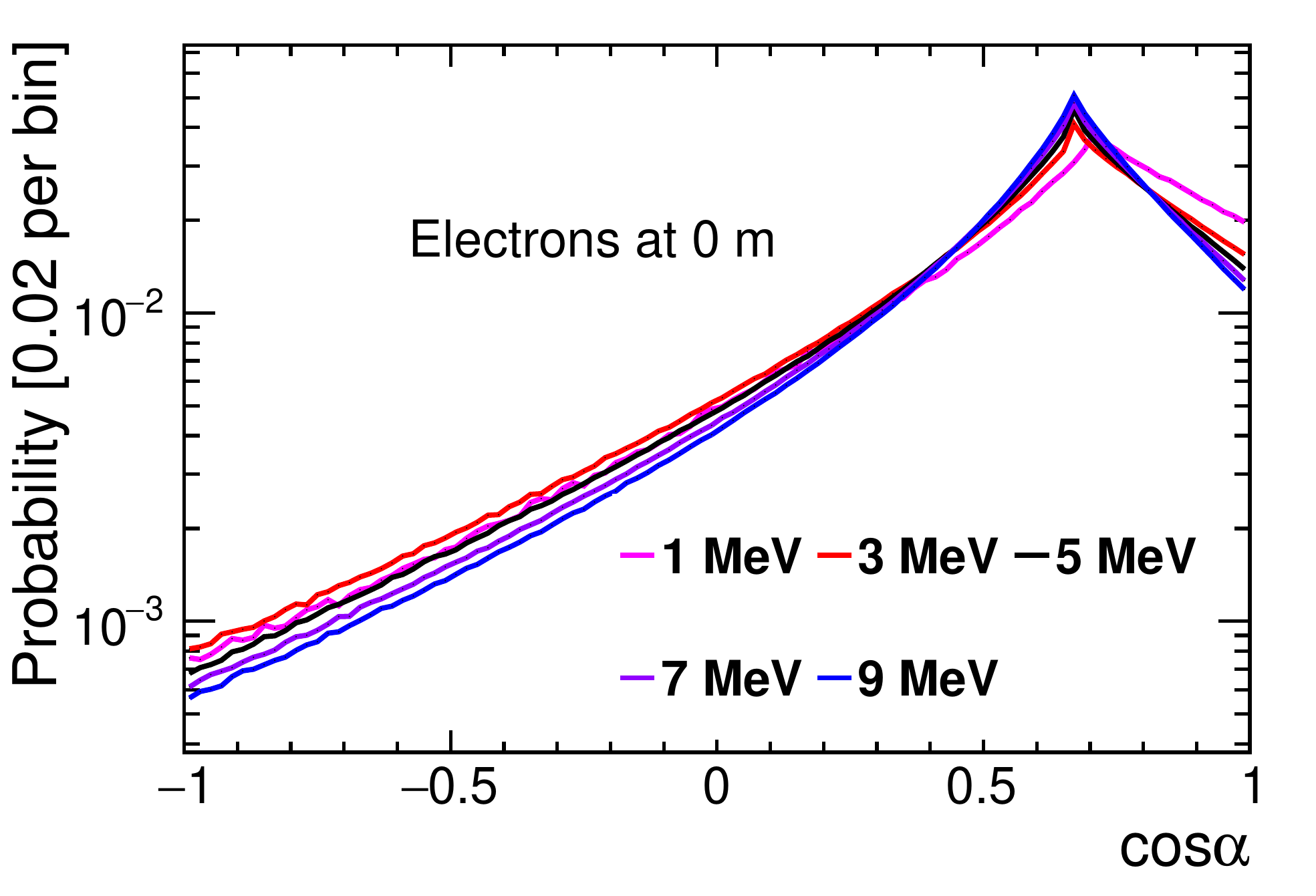}}
\subfigure{\includegraphics[scale=0.37]{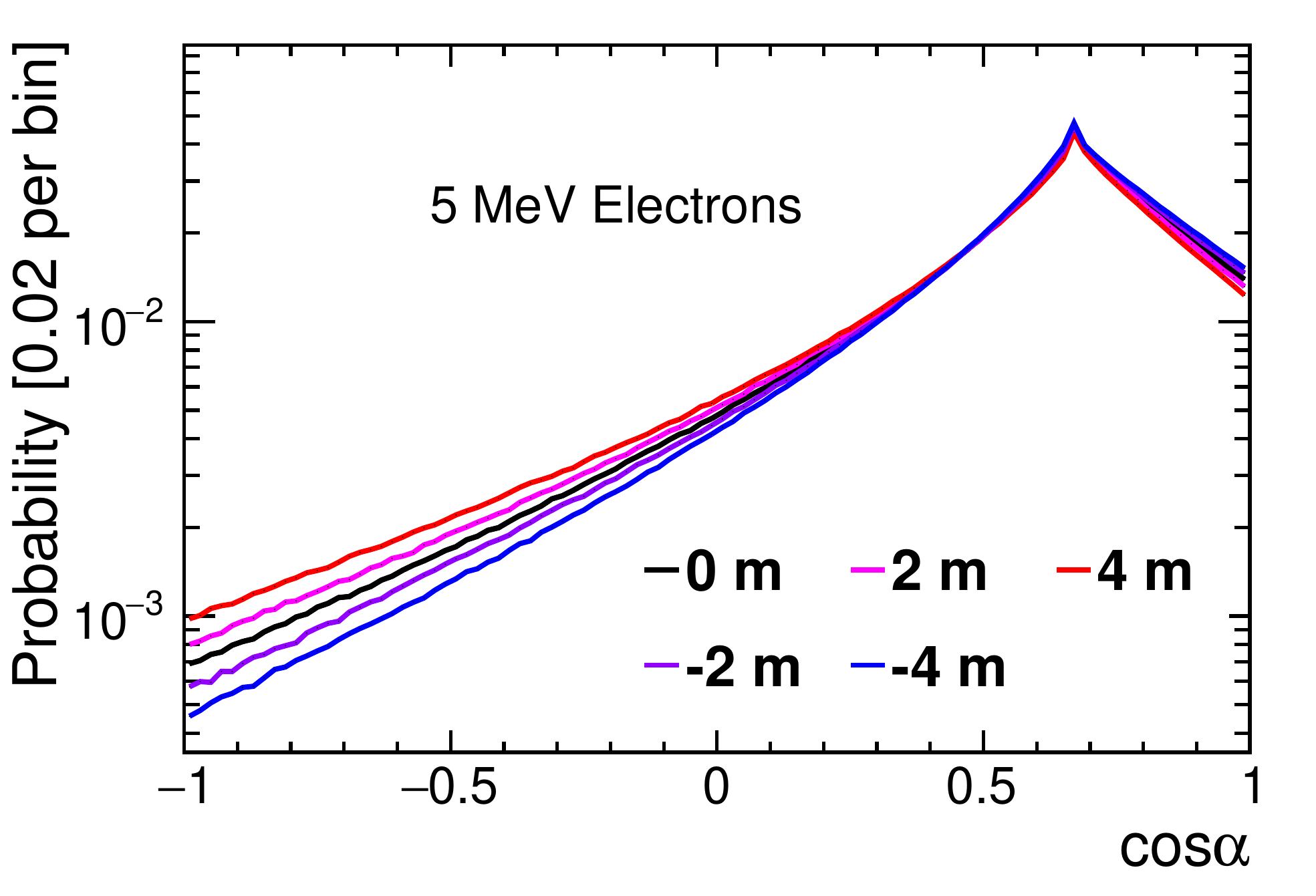}}
\caption{Left: An angular distribution of the Cherenkov light of electrons at various energies (1 to \SI{9}{MeV}). Right: An angular distribution of the Cherenkov light of electrons at different positions. The position on the detector’s X-axis, from \SI{-4}{m} to \SI{4}{m} in O-xyz. It is not significantly different for energies and positions. As a result, the \SI{5}{MeV} Cherenkov photon angular distribution is employed as the p.d.f.~in reconstructing electrons with energies ranging from 1 to \SI{9}{MeV}~\citep{SNOAngle}.}
\label{CPEdistribution}
\end{figure}

In figure~\ref{CPEdistribution}, the angular distributions of the Cherenkov light of electrons at various positions and energies are obtained by full MC simulation, where $\alpha$ is the angle of the hit PMT relative to the electron direction from the vertex.

The left panel of figure~\ref{CPEdistribution} shows that for 1 to \SI{9}{MeV} electrons at the detector center, there is a slight variation in the angular distribution of the Cherenkov light for different energies. 
In the right panel of figure~\ref{CPEdistribution}, a Cherenkov angular distribution of various positions is plotted by simulating \SI{5}{MeV} electrons at the position of \SI{-4}{m} to \SI{4}{m} on the X-axis in the spatial Cartesian coordinate system, O-xyz. It shows that the angular distribution of the Cherenkov light at different positions varies slightly. For the purpose of simplification, we assume that the Cherenkov light angular distribution is independent of energy and position. As a result, the Cherenkov light angular distribution as the probability density function, p.d.f.,~employed in the reconstruction is the one created by the \SI{5}{MeV} electrons at the detector center.

\begin{figure}[!htbp]
\centering
\includegraphics[width= .60\textwidth]{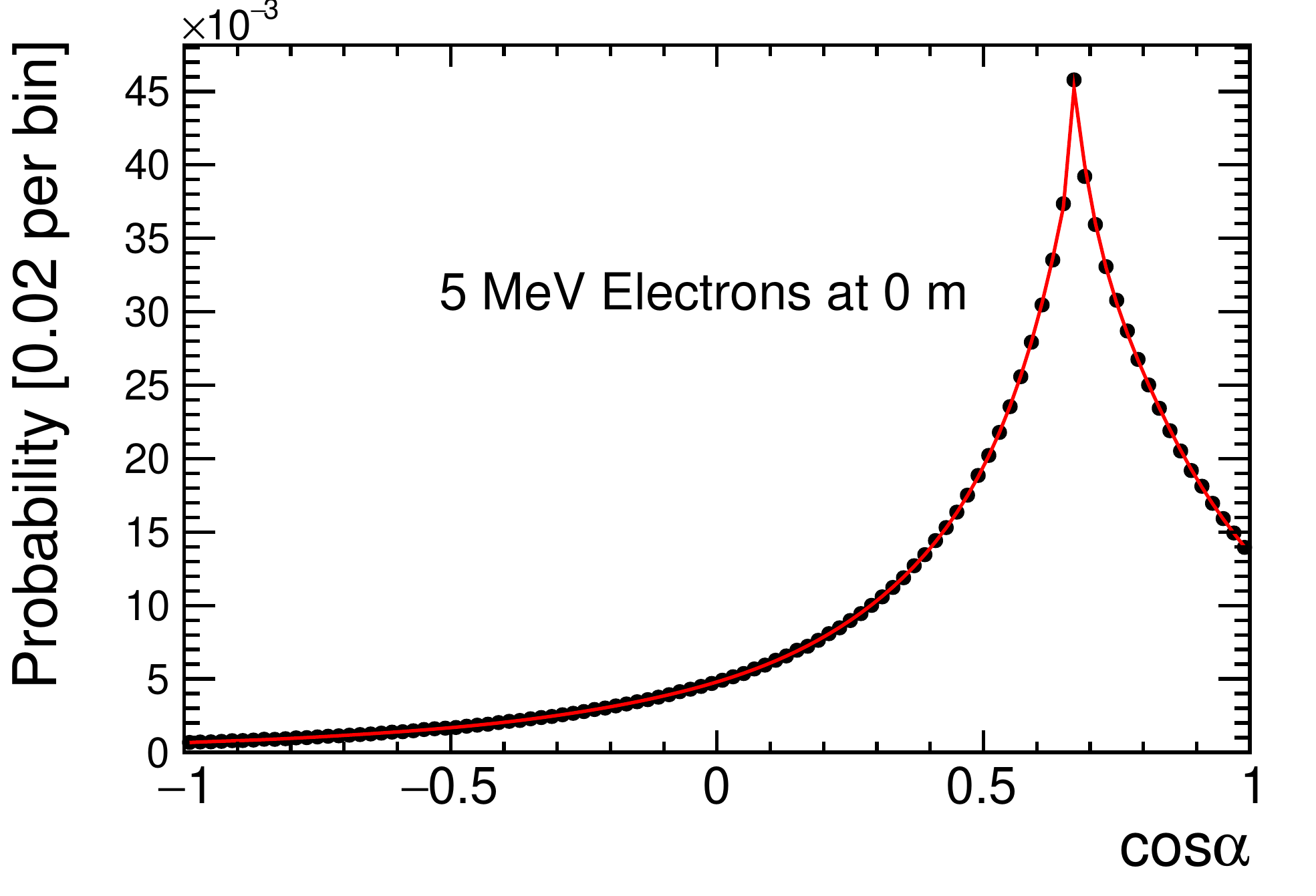}
\caption{A Cherenkov light angular distribution of 5 MeV electrons at the center of the detector.}
\label{Fig: FitCPEAngle}
\end{figure}

The Cherenkov light angular p.d.f.~of \SI{5}{MeV} electrons is fit by a segmented function consisting of a ninth-order polynomial and a sixth-order polynomial, as shown in figure~\ref{Fig: FitCPEAngle}.

\subsection{Optical Transportation}\label{OP}
This subsection focuses on two aspects: one is to obtain the effective group velocity by the full MC simulation in order to calculate the time of flight, and the other is to establish a simplified function for the attenuation of photons in the medium.

\subsubsection{Light Speed}\label{Sec: GroupSpeed}
\begin{figure}[!htbp]
\centering
\includegraphics[width= .60\textwidth]{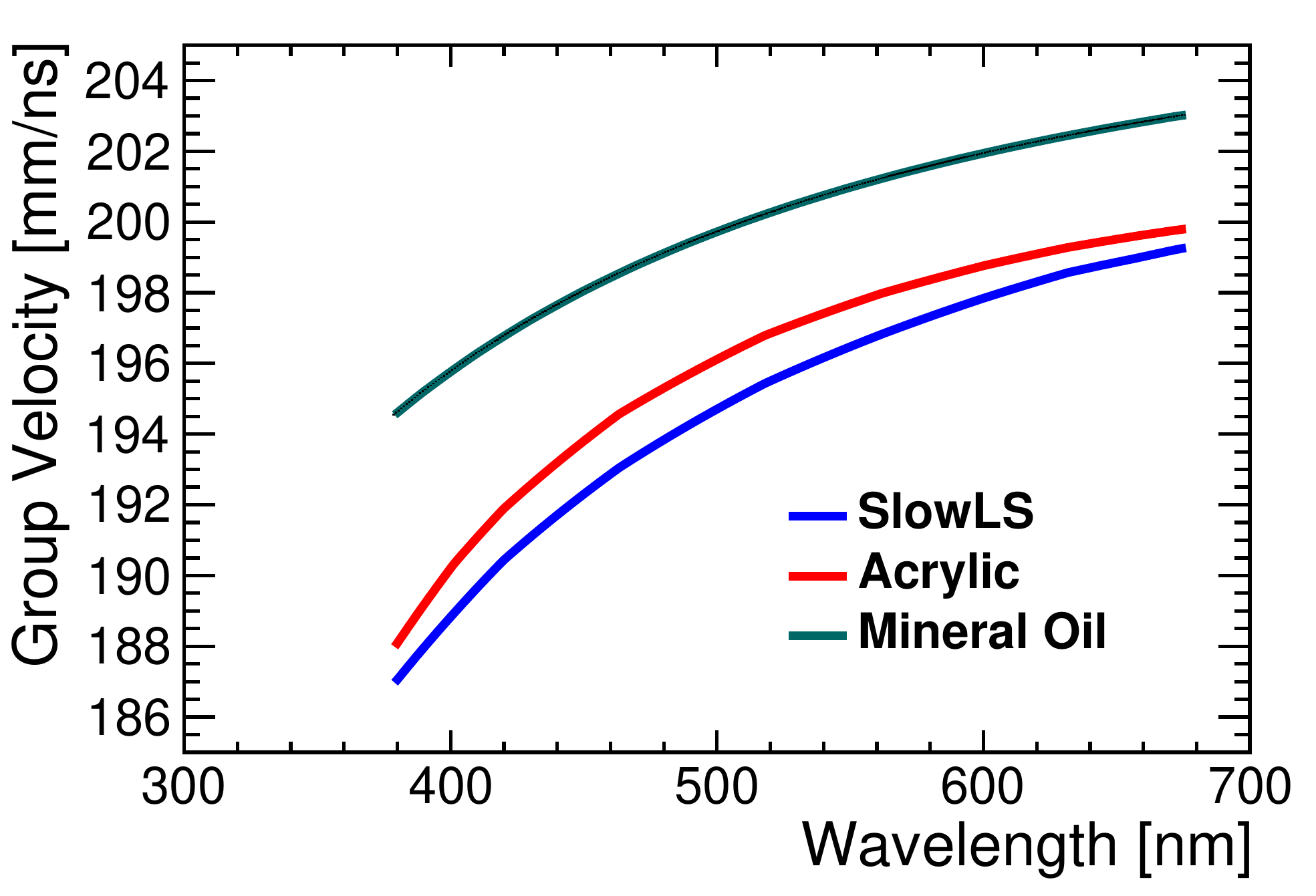}
\caption{A spectrum of the group velocity versus wavelength with different transparent materials in the full MC simulation.}
\label{GroupVSpec}
\end{figure}

Since Cherenkov light and scintillation light are not photons of a single wavelength, the speed of flight is calculated as a group velocity, which is given by~\citep{GroupV}: 
\begin{equation}
v_{\mathrm{g}}(\lambda) = \frac{c}{n(\lambda)-\lambda\frac{\partial n(\lambda)}{\partial \lambda}},
\end{equation}

\noindent where $v_{\mathrm{g}}$ is the group velocity, $n$ is the refractive index determined as shown in figure~\ref{Fig: Rindex}, $c$ is the speed of light in a vacuum, and $\lambda$ is the wavelength of the photon. The group velocity of different wavelengths in different materials is obtained by the full MC simulation, which is shown in figure~\ref{GroupVSpec}.

\begin{figure}[!htbp]
\centering
\subfigure{\includegraphics[scale=0.37]{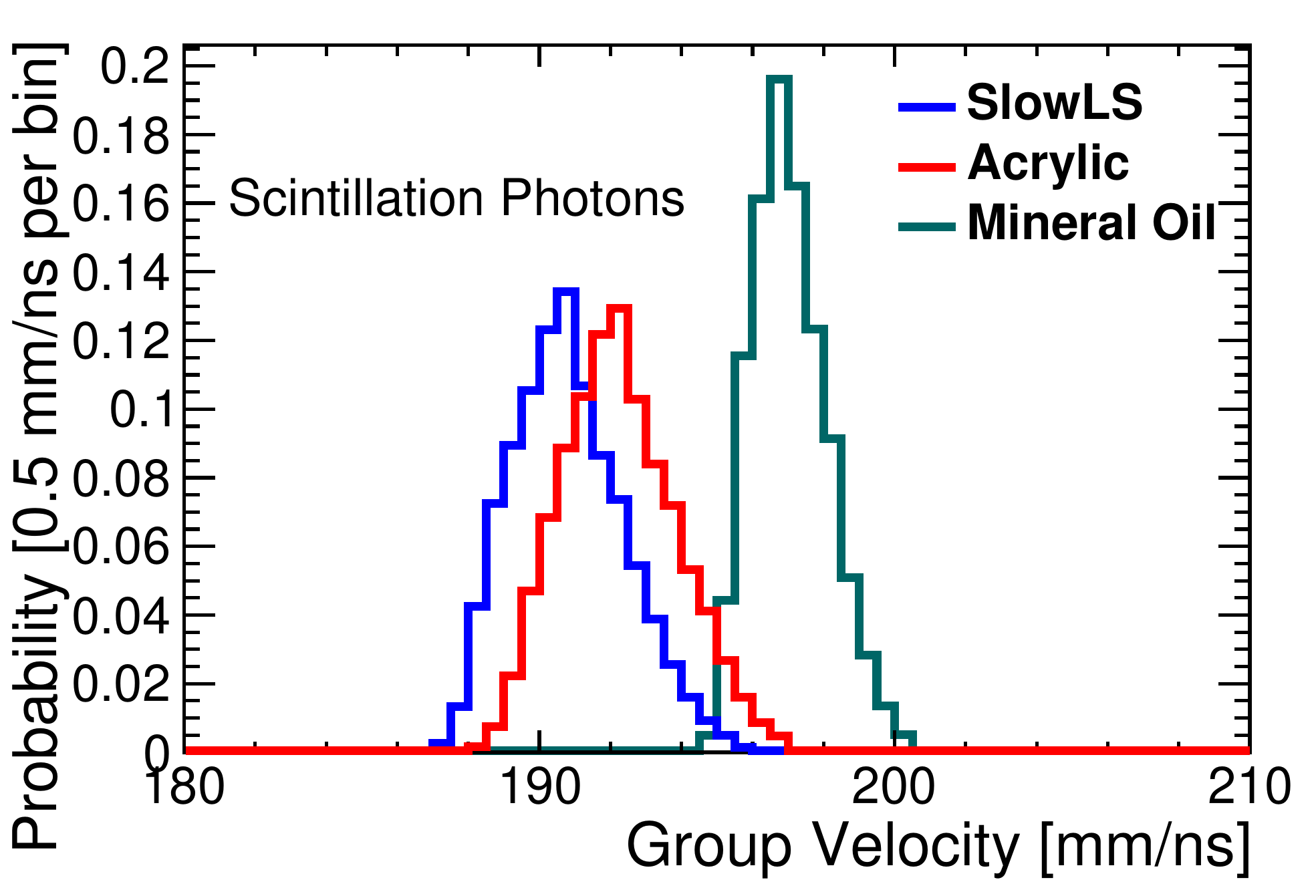}}
\subfigure{\includegraphics[scale=0.37]{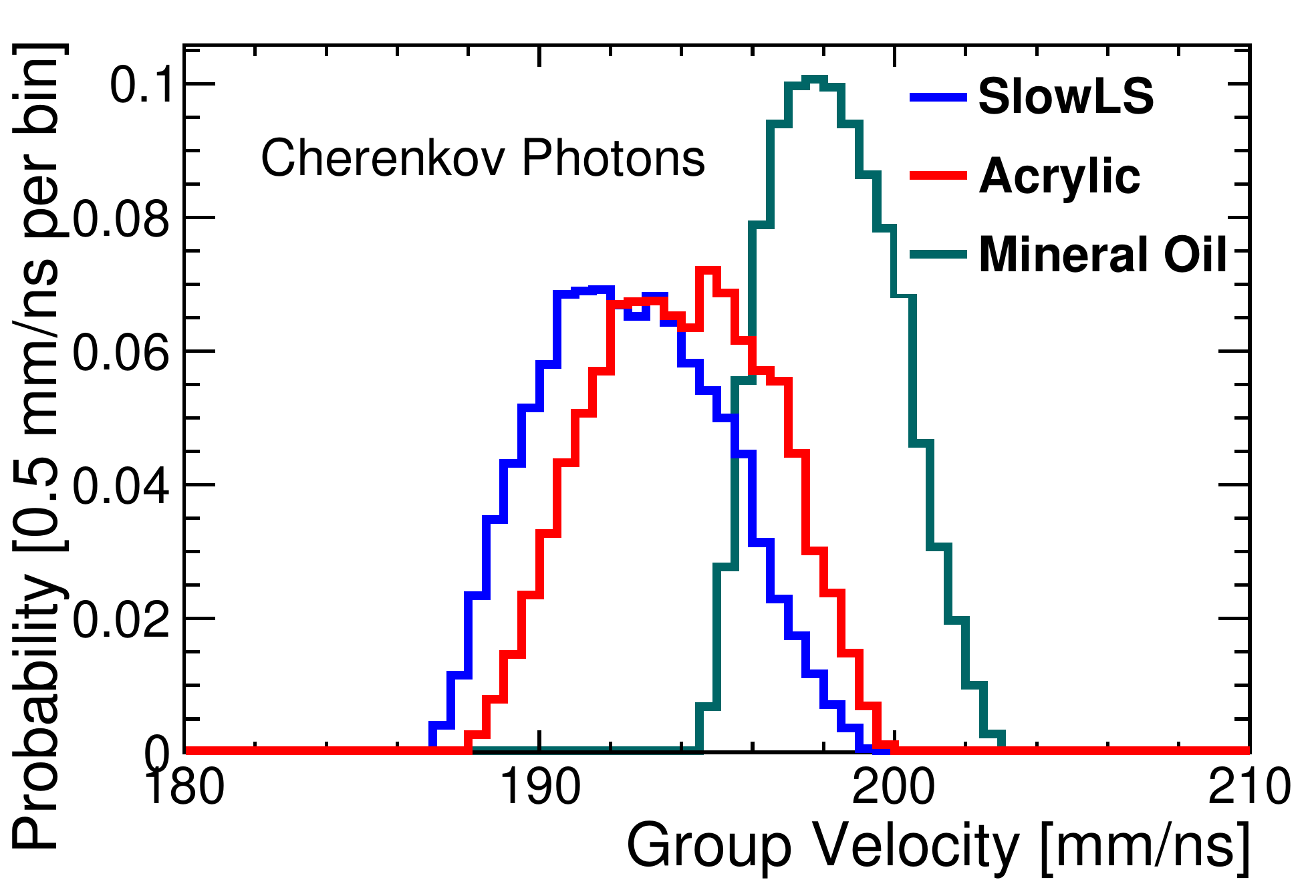}}
\caption{The group velocity distributions of the detected scintillation (left) and Cherenkov photons (right) in the different materials.}
\label{GroupV}
\end{figure}

Because current electronic readout techniques cannot detect the wavelength of each photon, the time of flight, TOF, will be calculated using the average group velocity of the detected photons, i.e., the effective group velocity. Therefore, the group velocities of the detected scintillation and Cherenkov photons are obtained by the full MC simulation, as shown in figure~\ref{GroupV}. 

Figure~\ref{GroupV} shows that the group velocity distribution of the acrylic is slightly larger than that of the SlowLS. In contrast, figure~\ref{Fig: Rindex} illustrates that the group velocity distribution of the acrylic should be somewhat smaller than the distribution of the SlowLS. It is because figure~\ref{GroupV} is only for the detected photons, with a wavelength of around \SI{400}{nm}, giving a slightly smaller refractive index of the acrylic than that of the SlowLS.

As a result, the effective group velocities of scintillation photons in the SlowLS, acrylic, and mineral oil are \SI{190.83}{mm/ns}, \SI{192.30}{mm/ns}, and \SI{197.10}{mm/ns}, respectively. The effective group velocities of Cherenkov photons in the SlowLS, acrylic, and mineral oil are \SI{192.54}{mm/ns}, \SI{194.04}{mm/ns}, and \SI{198.28}{mm/ns}, respectively.

\subsubsection{Absorption and Scattering of Photons}\label{Sec: Abs}

When a photon passes through a medium, there is a probability that the photon will be absorbed or scattered. Compared with the absorption process, the scattering process only changes the direction of the photon, while the absorption process causes the photon to disappear. The remaining number of the photons $I$ obeys Beer’s law~\citep{Beerslaw} and exponentially decreases from the initial number of the photons $I_0$ according to the travel distance of $l$ in a medium with an attenuation length $L(\lambda)$:
\begin{equation}
I = I_0e^{\frac{-l}{L(\lambda)}}.
\end{equation}

The combination of the absorption and scattering processes results in the attenuation of photons. This can be explained using the following formula~\citep{RayleighZhouXiang},
\begin{equation}
\frac{1}{L(\lambda)}=\frac{1}{L_{\mathrm{Abs}}(\lambda)}+\frac{1}{L_{\mathrm{Sca}}(\lambda)},
\end{equation}

\noindent where $L_{\mathrm{Abs}}(\lambda)$ and $L_{\mathrm{Sca}}(\lambda)$ represent the absorption and scattering lengths. The absorption length and scattering length in the SlowLS, acrylic, and mineral oil are shown in figure~\ref{ABSRaySim}.

\begin{figure}[!htbp]
\centering
\subfigure{\includegraphics[scale=0.37]{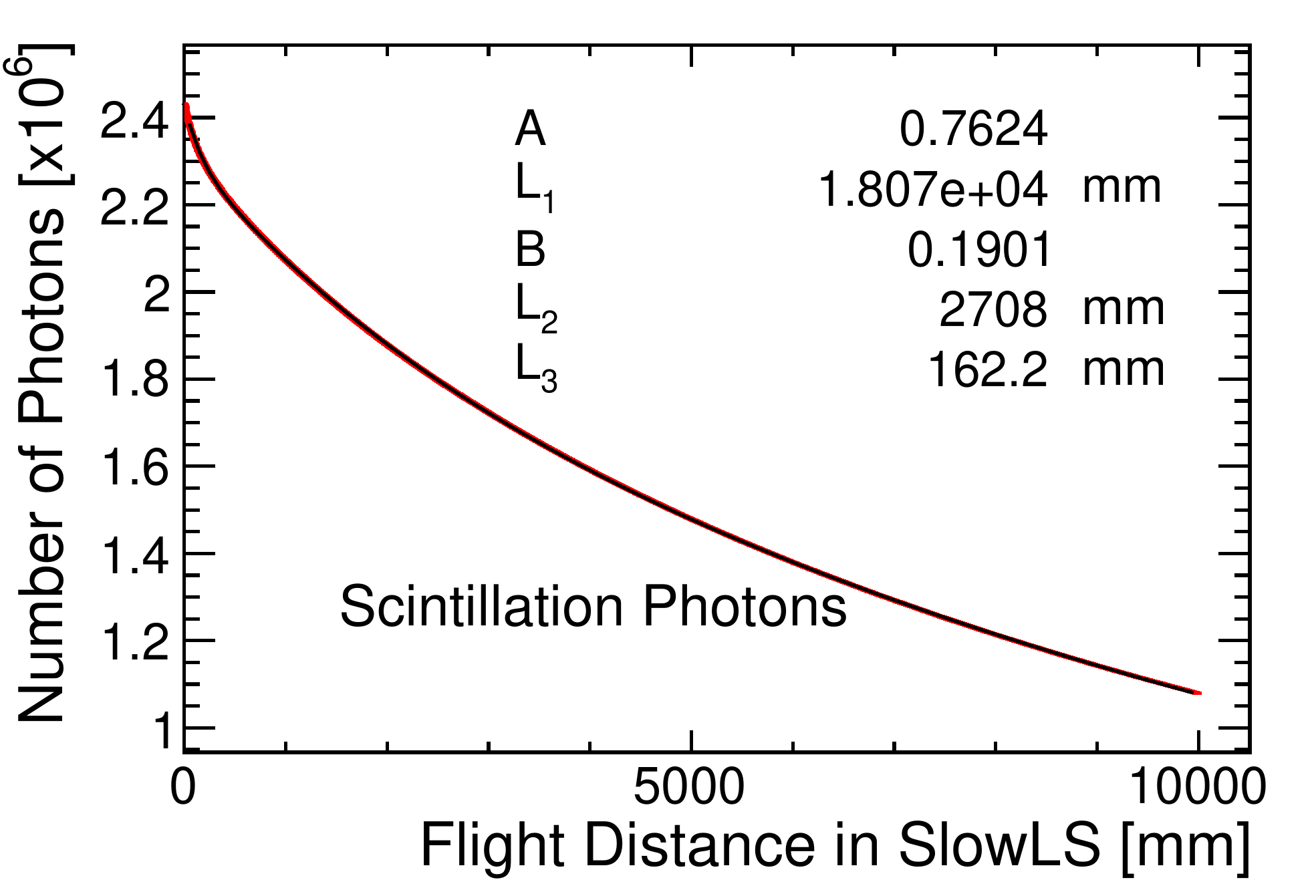}}
\subfigure{\includegraphics[scale=0.37]{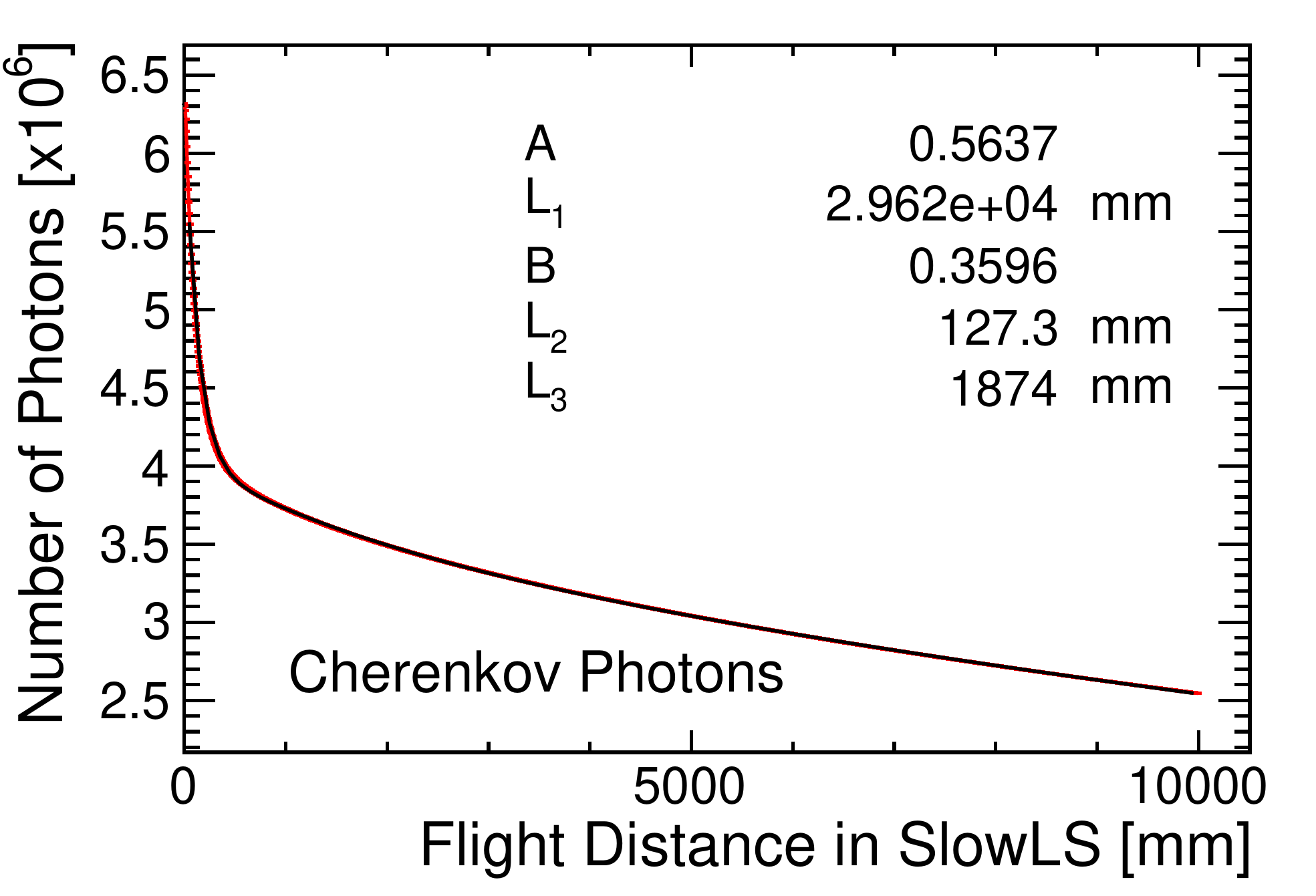}}
\caption{Left: The fit attenuation length and coefficients of scintillation photons in the SlowLS. Right: The fit attenuation length and coefficients of Cherenkov photons in the SlowLS.}
\label{ABSinLS}
\end{figure}

The photon survival number is defined as the number of scintillation or Cherenkov light photons surviving through different lengths of the medium, as counted by the full MC simulation. We take the example of the electrons at the center of the detector. The photon survival distribution is displayed in figure~\ref{ABSinLS}. In SlowLS, the absorption length of short wavelengths is small and that of long wavelengths is large. The photons spectrum of Cherenkov light is concentrated at short wavelengths, so the photon survival spectrum drops sharply in the first \SI{500}{mm}. The spectrum of scintillation light is concentrated near \SI{400}{nm}, but the photon survival spectrum also drops faster in the first \SI{500}{mm}. After fitting tests, a triple exponential function is fit well as follows,
\begin{equation}
I(x) = I_0 [\mathrm{A}\cdot \mathrm{exp}(-x/L_1)+\mathrm{B}\cdot \mathrm{exp}(-x/L_2)+\mathrm{(1-A-B)}\cdot \mathrm{exp}(-x/L_3)],
\label{Eq: ABSCPE}
\end{equation}

\noindent where $x$ is the length through the SlowLS. $\mathrm{A}$ and $\mathrm{B}$ are coefficients. $L_1$, $L_2$, and $L_3$ are the fit attenuation lengths. The fit results in the SlowLS are displayed in figure~\ref{ABSinLS}.  We used the same method to obtain a fitting function for attenuation length in acrylic and mineral oil in order to estimate the number of surviving photons that cross different distances in the three media.

\subsection{Detection}
In this subsection, we construct a formula for the simplified solid angle considering the 3-dimensional, 3D, PMT photocathode geometry and obtain the effective detection efficiency of the PMT.

\subsubsection{Solid Angle}
\begin{figure}[!htbp]
    \centering
   \includegraphics[width= .60\textwidth]{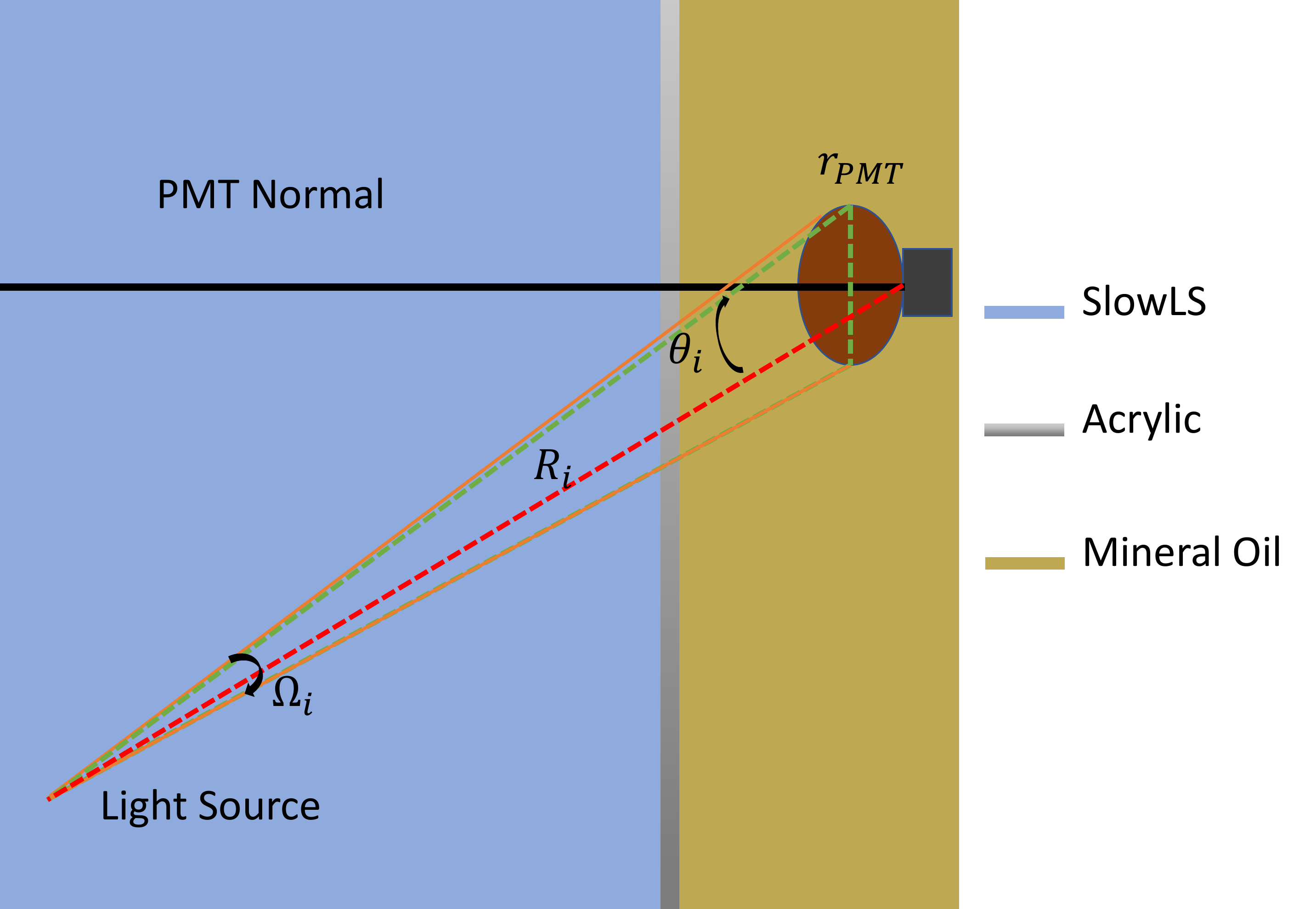}
    \caption{The schematic for the solid angle $\Omega_i$ of the PMT photocathode. The orange solid lines show the true solid angle considering the 3D PMT photocathode geometry, whereas the green dotted lines represent the solid angle determined by the 2D disk model. The length $R_i$ of the PMT to the vertex position is shown as the red dotted line. The PMT normal vector is represented by the black line. The incident angle $\theta_i$ is between the vector of the PMT to the vertex and the PMT normal vector.}
    \label{PMT solid angle}
\end{figure}

The calculation of solid angles, $\Omega_i$, takes into consideration the 3D PMT photocathode geometry. As shown in figure~\ref{PMT solid angle}, the 3D solid angle (orange solid lines) should consider not only the radius of the PMT photocathode, $r_{\mathrm{PMT}}$, but also the geometric influence of the PMT photocathode. 

\begin{figure}[!htbp]
\centering
\subfigure{\includegraphics[width= .49\textwidth]{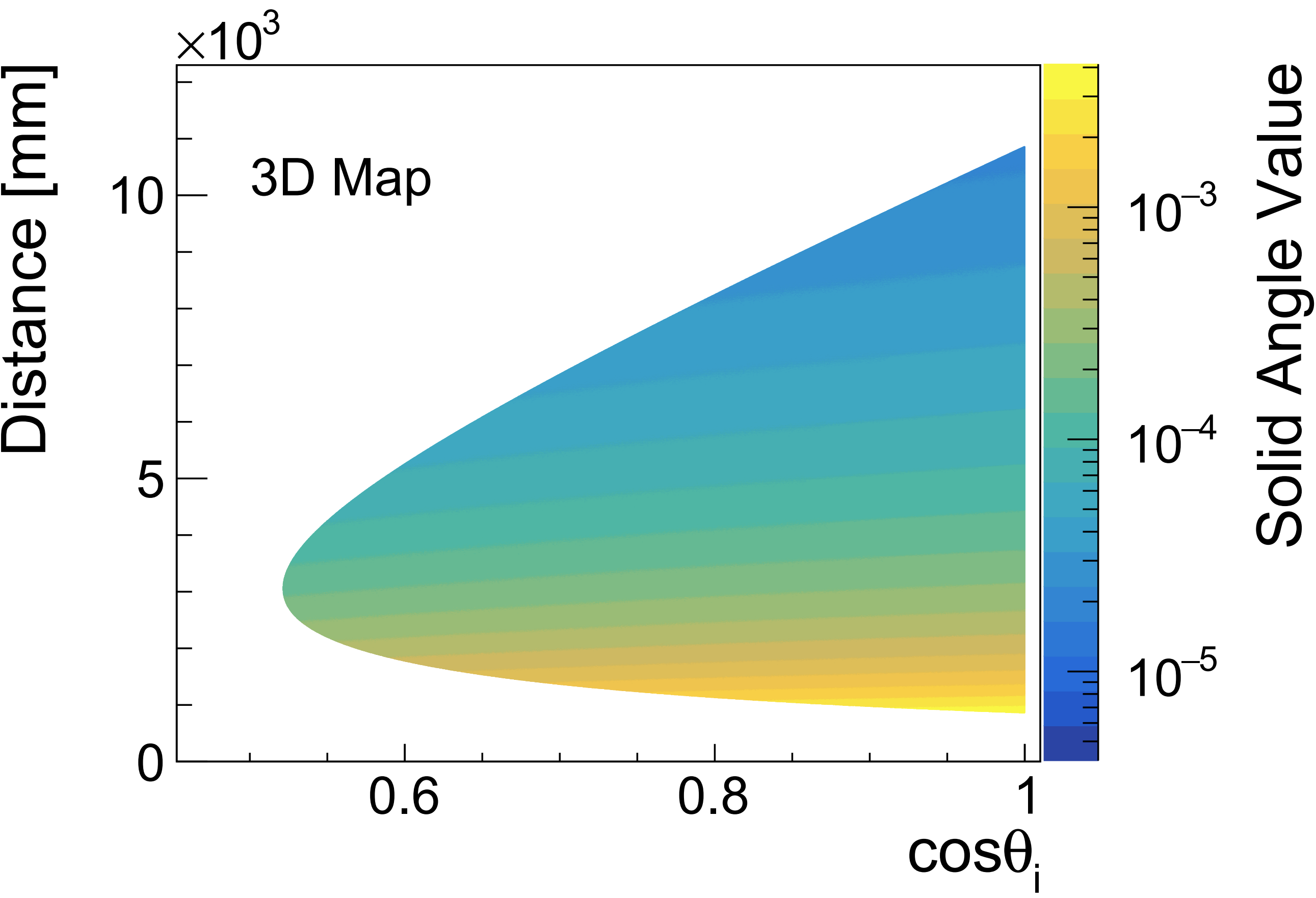}}
\subfigure{\includegraphics[width= .49\textwidth]{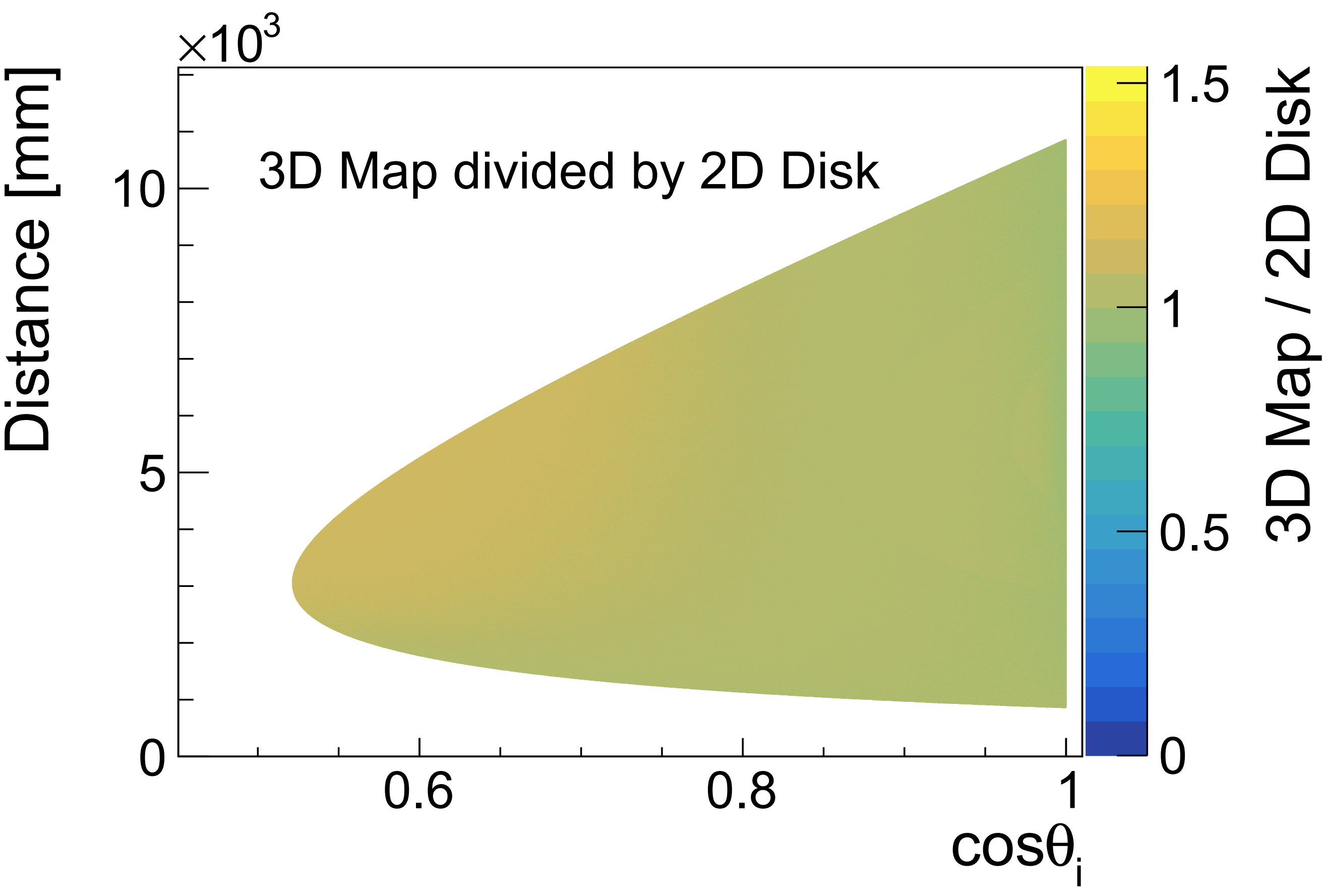}}
\caption{Left: A 3D solid angle map of the 8-inch PMT photocathode is obtained by the MC simulation. The cosine of the incident angle $\theta_i$ is represented on the X-axis. The Y-axis presents the distance $R_i$ from the PMT to the vertex position. The Z-axis represents the value of the solid angle. Right: The value of the 3D divided by the 2D solid angle with $\cos{\theta_i}$ and $R_i$. The Z-axis represents the value of the 3D divided by the 2D solid angle.}
\label{PMT solid angle map}
\end{figure}

A MC simulation is used to obtain the 3D solid angle map of the 8-inch PMT. The left panel of figure~\ref{PMT solid angle map} shows the variation of the 3D solid angle with the incidence angle, $\theta_i$, and distance, $R_i$. The 3D solid angle map needs a simplified function for the reconstruction, and was done as follows.

First, the predicted solid angle is approximated by a 2-dimensional, 2D, disk model. As shown in figure~\ref{PMT solid angle}, when the $R_i$ is significantly more than $r_{\mathrm{PMT}}$ (8-inch), the predicted 2D solid angle by the disk model (green dotted lines) is approximated as follows:

\begin{equation}
\Omega_i=\frac{\pi r_{\mathrm{PMT}}^2}{4\pi R_i^2}\cos{\theta_i}.
\label{Solid Angle}
\end{equation}

As we can see in figure~\ref{PMT solid angle}, the 3D solid angle (orange solid lines ) is larger than the 2D solid angle (green dotted lines), so we next explore the relationship between the 3D solid angle and the approximated 2D solid angle. We divide the 3D solid angle by the 2D solid angle, as shown in the right panel of figure~\ref{PMT solid angle map}. We find that the value of the 3D solid angle divided by the 2D solid angle has an approximately linear relationship with the $\cos{\theta_i}$ for the same $R_i$. 

 \begin{figure}[!htbp]
    \centering
   \includegraphics[width= .60\textwidth]{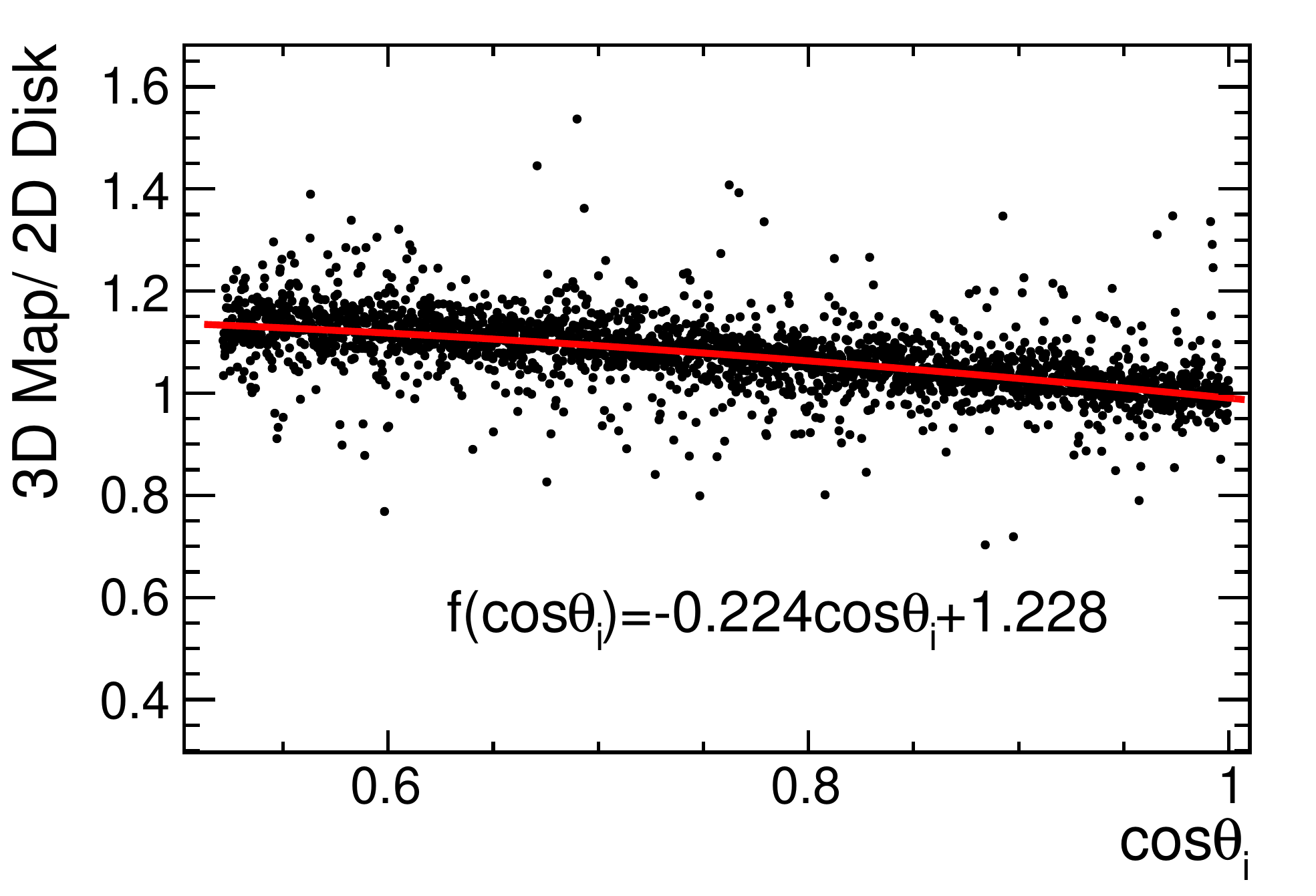}
    \caption{The relationship distribution of the 3D map and the 2D disk model solid angle versus $\cos{\theta_i}$. The Y-axis represents the value of the 3D divided by the 2D solid angle.}
    \label{Fig: SolidAngleFit}
\end{figure}

Figure~\ref{Fig: SolidAngleFit} shows the ratio between the 3D and 2D results versus $\cos{\theta_i}$. 
A linear function $f(\cos{\theta_i})$ is fitted to the figure and the result is applied as a correction to the 2D disk assumption
\begin{equation}
\Omega_i= f(\cos{\theta_i})\frac{\pi r_{\mathrm{PMT}}^2}{4\pi R_i^2}\cos{\theta_i}.
\label{Eq: SolidAngle}
\end{equation}

\subsubsection{PMT Detection Efficiency}\label{Sec: DE}
The probability that an optical photon is converted into PE when it reaches the PMT photocathode depending on the quantum efficiency and the collection efficiency, which are referred to jointly as the detection efficiency, DE. In this work, the peak of the quantum efficiency spectrum is 34\% at \SI{390}{nm}, and the collection efficiency is 100\%~\citep{HighQEByHama}. Therefore, the DE spectrum is the same as the quantum efficiency spectrum. 

Different wavelength photons correspond to different DE, but the readout system cannot distinguish the wavelength of the photons. Therefore, we try to avoid this problem with an effective DE,
which is the average DE calculated by convoluting the wavelength spectrum for the photons arriving at the PMT and the latter’s wavelength dependent quantum efficiency.

With full MC simulation, we observe that the light spectrum from scintillation or Cherenkov is almost unchanged beyond \SI{500}{mm} in SlowLS or mineral oil, because  
the short-wavelength photons (<\SI{360}{nm}) have been quickly absorbed. Only the long-wavelength photons have been retained.
The scintillation and Cherenkov photon spectra of the arriving photons are obtained with the full MC simulation for the effective DE calculation. As a result, the effective DE of scintillation photons, $\epsilon^{\mathrm{S}}_{\mathrm{eff}}$ is 29.9\%, and the effective DE of Cherenkov photons, $\epsilon^{\mathrm{C}}_{\mathrm{eff}}$ is 16.0\%. 
The effective DE of Cherenkov photons is smaller than that of scintillation photons because a wider Cherenkov light spectrum is chosen as seen in figure~\ref{CSPhotonSpec} and the DE for short wavelengths Cherenkov light is low.

\subsection{Direct and Indirect Photons}

Although the refractive indices of the medium in the CSD are similar, the PMT glass and alkali metal coating create a mirror-like surface where photons are strongly reflected. In addition, the photons undergo Rayleigh scattering in the medium, which changes the direction of the photons. These two effects will make the predicted number of PEs and the TOF of PEs inaccurate. Therefore, we must consider the effect of reflection and scattering on the reconstruction. 

In this subsection, the proportion of the direct and indirect photons in different positions of the detector, the relationship between the position of the direct and indirect photons, and the time profile of the direct and indirect photons are described.

\subsubsection{Proportion of Direct and Indirect Photons} \label{Sec: DirectandNon}
The definition of direct photons is the detected photons that enter the PMT photocathode directly (ignoring the refraction effects), and indirect photons are the detected photons influenced by reflection and scattering. The reflection and scattering effects are independent of the electron energy, so the variation in the energy of the electrons is not taken into account. Only the effect of position is taken into account, since reflection and scattering are only related to the wavelength of the photon and the photon flight distance. 

\begin{figure}[!htbp]
\centering
\subfigure{\includegraphics[scale=0.37]{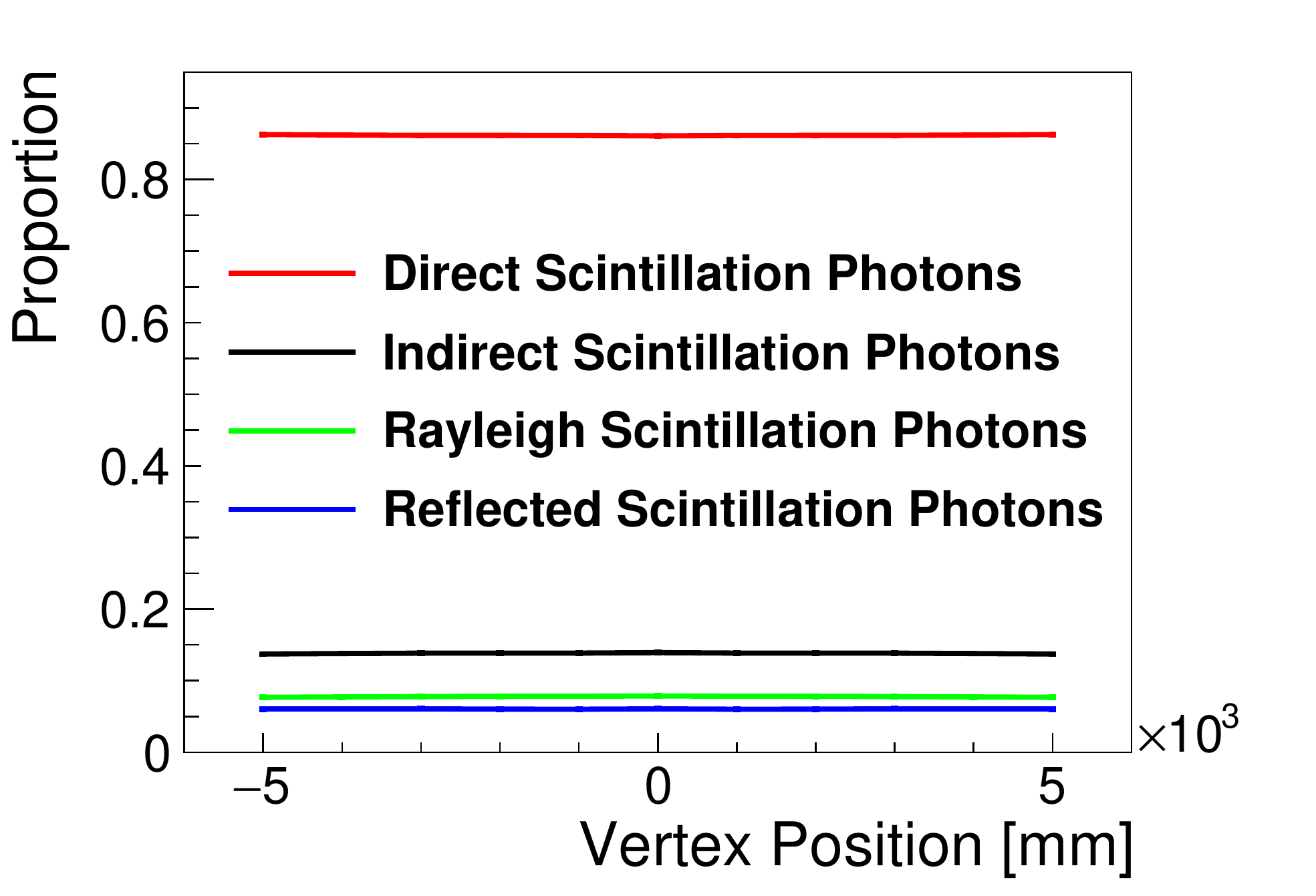}}
\subfigure{\includegraphics[scale=0.37]{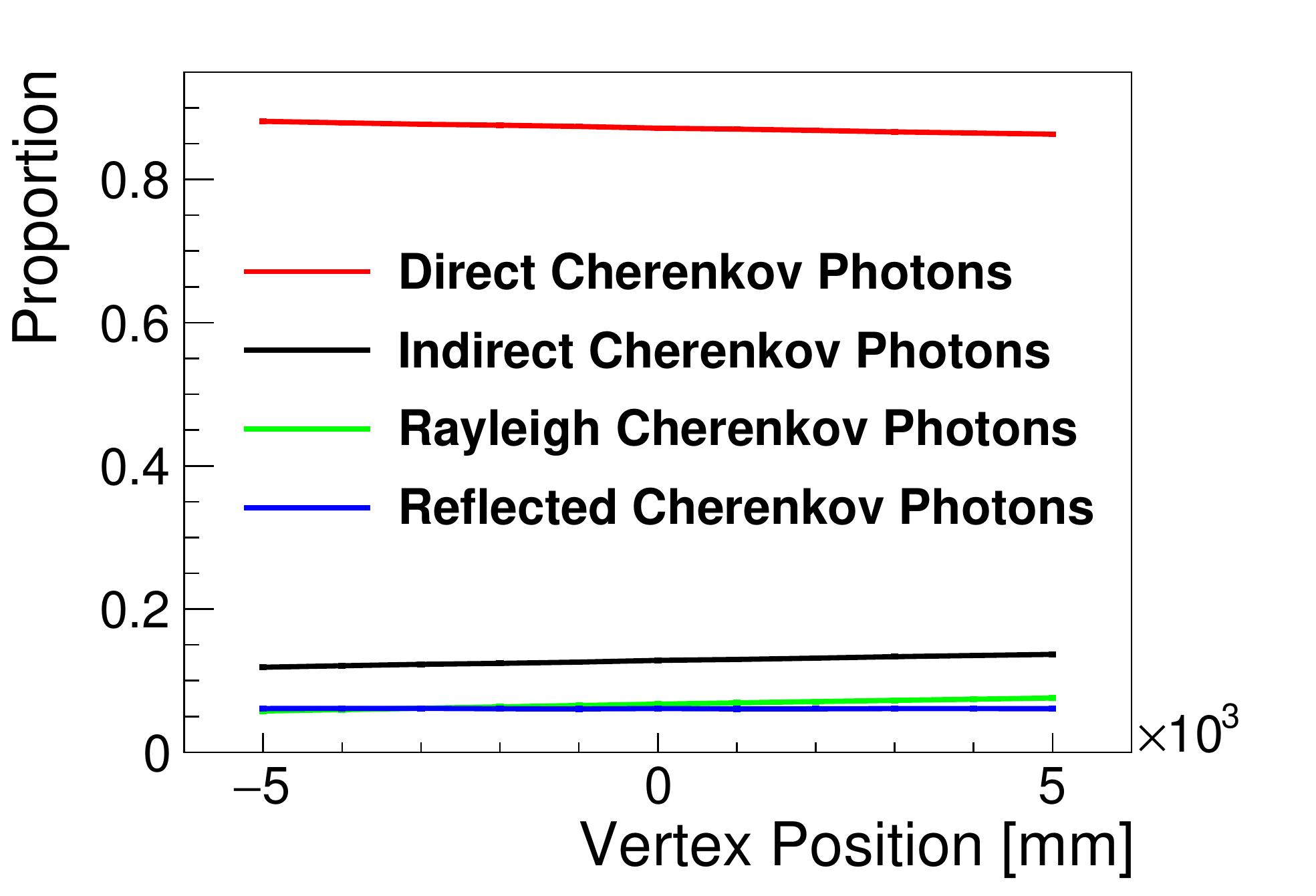}}
\caption{The distributions of the proportion of direct and indirect photons (including reflected and Rayleigh scattering photons) versus the vertex position for scintillation (left) and Cherenkov light (right) are plotted.}
\label{fig:oaspl_a}
\end{figure}

Therefore, a proportion of direct to indirect photons is obtained at different detector positions by the full MC simulation. A total of 10,000 electrons are simulated from \SI{-5}{m} to \SI{5}{m} on the detector X-axis, respectively, and their normalized direction ($p_x$, $p_y$, $p_z$) is (-1,~0,~0) in O-xyz. The number of direct and indirect photons for scintillation or Cherenkov light is obtained by the full MC simulation. As a result, the proportion of direct and indirect photons (including reflected and Rayleigh scattering photons) are shown in figure~\ref{fig:oaspl_a}.

The proportion of direct and indirect scintillation photons remain essentially the same at different positions in the SlowLS, with 85\% and 15\% of these two types of photons, respectively.

The maximum difference in the proportion of indirect Cherenkov photons at different positions is nearly 2\%. Because the chance of Rayleigh scattering is higher when the optical path of Cherenkov light is longer, the proportion of indirect Cherenkov photons increases. Nevertheless, to simplify the calculation, the proportion of indirect Cherenkov photons is considered to be 15\%, while the proportion of direct Cherenkov photons is approximately 85\%.
\subsubsection{Virtual Vertex of Indirect Photons}\label{Sec: VirtualVetex}

As can be seen in the previous subsection, almost 15\% of the photons are indirect photons, so their effect on the reconstruction cannot be ignored.
We attempt to develop a simplified function to estimate the number of indirect photons for each PMT. Because of reflection and scattering, the spatial distribution of the indirect photons is more fuzzy than that of direct photons, while the indirect photons preserve some of the qualities of the direct photons. Therefore, we explore the association between the direct and indirect photons by the full MC simulation and a vertex position reconstruction.

\begin{figure}[!htbp]
\centering
\subfigure{\includegraphics[scale=0.37]{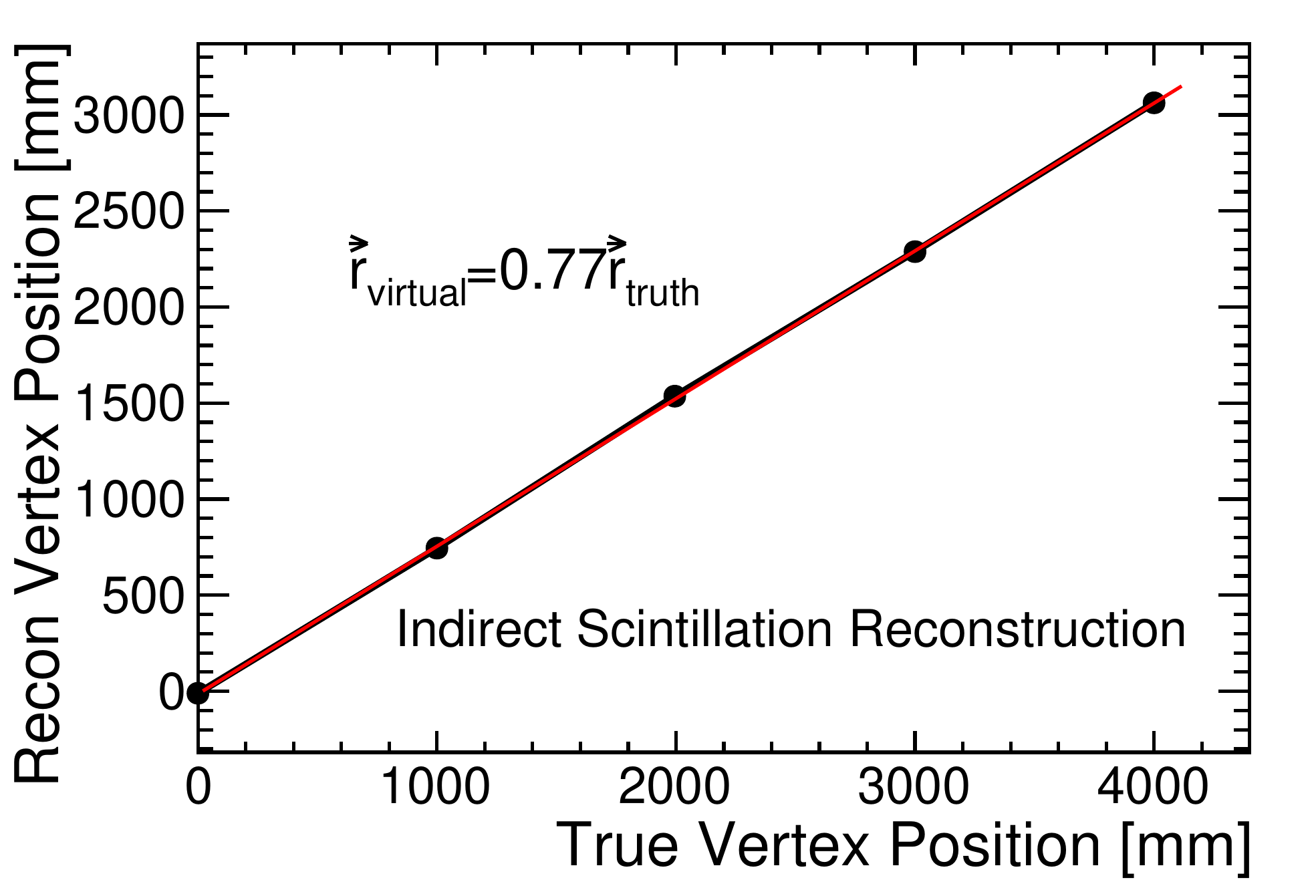}}
\subfigure{\includegraphics[scale=0.37]{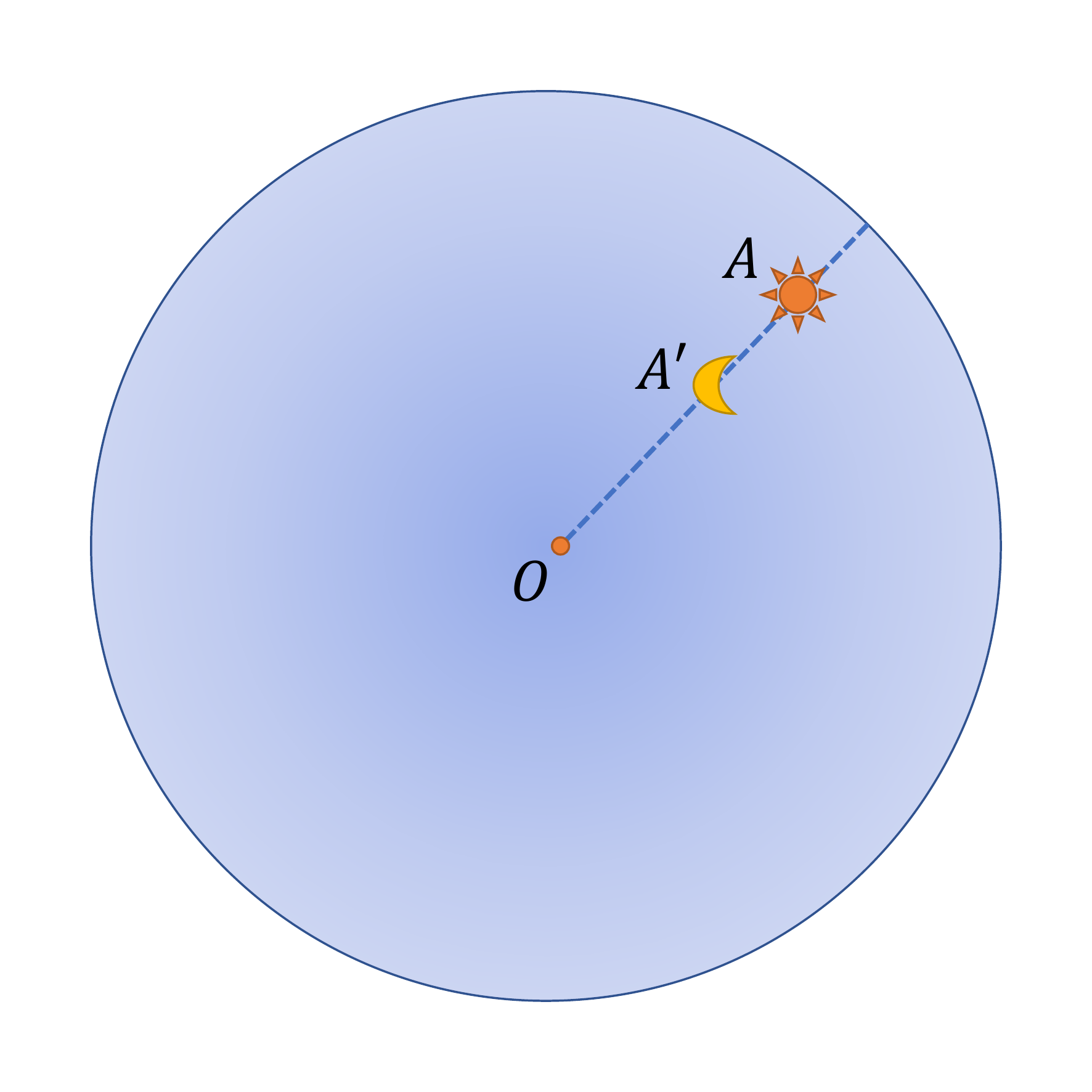}}
\caption{Left: Comparison of the true vertex position, $\vec{r}_{{\mathrm{truth}}}$, and the position as reconstructed using the indirect scintillation photons, $\vec{r}_{{\mathrm{virtual}}}$. There is an approximately linear relationship between the two vertex positions. Right: Schematic diagram of the virtual and real light sources. The center of the detector is $O$, and points $A$ and $A^{'}$ are the real light source (direct photons) and virtual light source (indirect photons), respectively.}
\label{Fig: NonSPERecon}
\end{figure}

A total of 100,000 electrons are simulated with a kinetic energy of \SI{5}{MeV} and fixed them at the X-axis coordinates of 0, \SI{1}{m}, \SI{2}{m}, \SI{3}{m}, and \SI{4}{m}. 
The indirect scintillation photons are selected from the full MC simulation, and subsequently the vertex positions of these indirect photons are reconstructed using the charge likelihood function (latter section), i.e., virtual vertex position, $\vec{r}_{{\mathrm{virtual}}}$.
We compare the $\vec{r}_{{\mathrm{virtual}}}$ with the true vertex position, $\vec{r}_{{\mathrm{truth}}}$, and the left panel of figure~\ref{Fig: NonSPERecon} shows the results.

As we can see, the virtual vertex position can be assumed to be a point, $A'$, closer to the center than the true vertex position, $A$, as shown on the right side of figure~\ref{Fig: NonSPERecon}, because the distribution of indirect scintillation photons is more uniform ($\vec{r}_{{\mathrm{virtual}}}=0.77\vec{r}_{{\mathrm{truth}}}$). Moreover, the same approach is applied to the indirect Cherenkov photons. %($\vec{r}_{{\mathrm{virtual}}}=0.85\vec{r}_{{\mathrm{truth}}}$)

\subsubsection{Time Profile of Direct and Indirect Photons}\label{Sec: TimeRes}
By using the effective group velocities in the different medium, the TOF of scintillation, $\mathrm{TOF}_{i,\mathrm{m}}^{\mathrm{S,pred}}$ and Cherenkov photons $\mathrm{TOF}_{i,\mathrm{m}}^{\mathrm{C,pred}}$ are predicted as follows,

\begin{equation}
\mathrm{TOF}_{i,\mathrm{m}}^{\mathrm{S,pred}}=\frac{R_{i,\mathrm{m}}(x,y,z)}{v_{\mathrm{m}}^{\mathrm{S,eff}}},
\label{Eq: SPETOF}
\end{equation}
\begin{equation}
\mathrm{TOF}_{i,\mathrm{m}}^{\mathrm{C,pred}}=\frac{R_{i,\mathrm{m}}(x,y,z)}{v_{\mathrm{m}}^{\mathrm{C,eff}}},
\label{Eq: CPETOF}
\end{equation}

\noindent where $R_{i,\mathrm{m}}(x,y,z)$ is the distance of the photon in medium $\mathrm{m}$ as it flies toward the $i$th PMT for vertex position $(x,y,z)$. $v_{\mathrm{m}}^{\mathrm{S,eff}}$ and $v_{\mathrm{m}}^{\mathrm{C,eff}}$ are the effective group velocities of scintillation and Cherenkov photons (section~\ref{Sec: GroupSpeed}). 

For the PE time $t_{ij}$ of the $j$th PE of the $i$th PMT, the time residual of scintillation PEs, $t_{ij}^{\mathrm{S},\mathrm{res}}$, or Cherenkov PEs, $t_{ij}^{\mathrm{C},\mathrm{res}}$, can be written as

\begin{equation}
t_{ij}^{\mathrm{S},\mathrm{res}}=t_{ij}-t_{\mathrm{event}}-\sum_{\mathrm{m}}{\mathrm{TOF}_{i,\mathrm{m}}^{\mathrm{S,pred}}}-t_{\mathrm{shift}},
\label{Eq: STimeres}
\end{equation}
\begin{equation}
t_{ij}^{{\mathrm{C}},\mathrm{res}}=t_{ij}-t_{\mathrm{event}}-\sum_{\mathrm{m}}{\mathrm{TOF}_{i,\mathrm{m}}^{\mathrm{C,pred}}}-t_{\mathrm{shift}},
\label{Eq: CTimeres}
\end{equation}

\noindent where $t_{\mathrm{event}}$ is the event time being fit, and $t_{\mathrm{shift}}$ consists of the electronic time and TT, which can be obtained by a calibration. 

However, the predicted TOF is different from the true TOF because the true TOF will be larger than the predicted TOF under the influence of reflection and scattering, so the time residual needs to consider the influence of indirect photons. 

We obtain the time residual p.d.f. of direct and indirect photons by the full MC simulation. The scintillation time profile of $\tau_r=\SI{7.7}{ns}$ and $ \tau_d=\SI{37}{ns}$ is used as an example. The \SI{5}{MeV} electrons are simulated at various positions \SI{-4}{m} to \SI{4}{m} in a fixed direction (1,~0,~0), and the time residual p.d.f.~is obtained by using eq.~(\ref{Eq: STimeres}) or eq.~(\ref{Eq: CTimeres}). Moreover, the vertex position $(x,y,z)$ used in the calculation of $R_{i,\mathrm{m}}(x,y,z)$ is the true vertex position.

\begin{figure}[!htbp]
\centering
\subfigure{\includegraphics[scale=0.37]{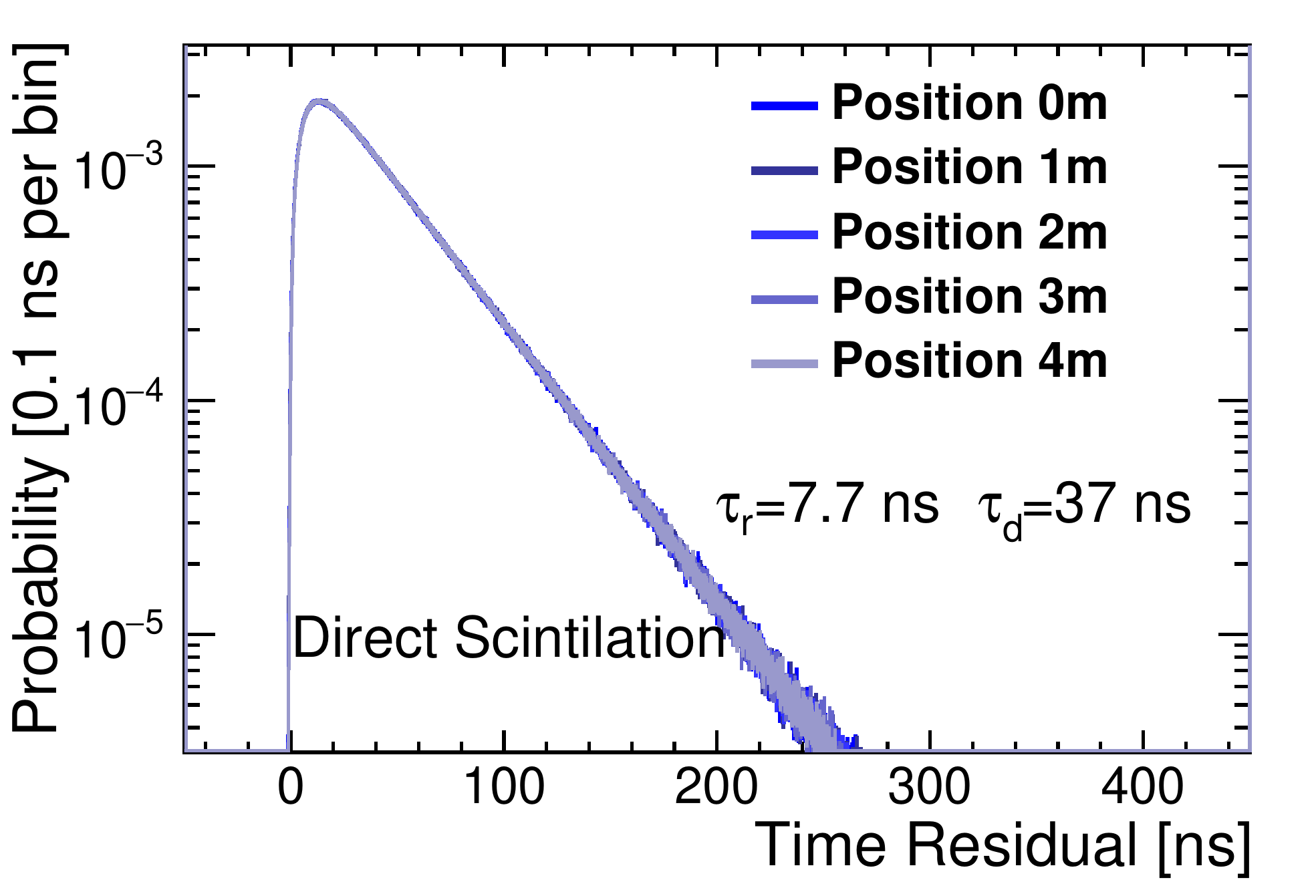}}
\subfigure{\includegraphics[scale=0.37]{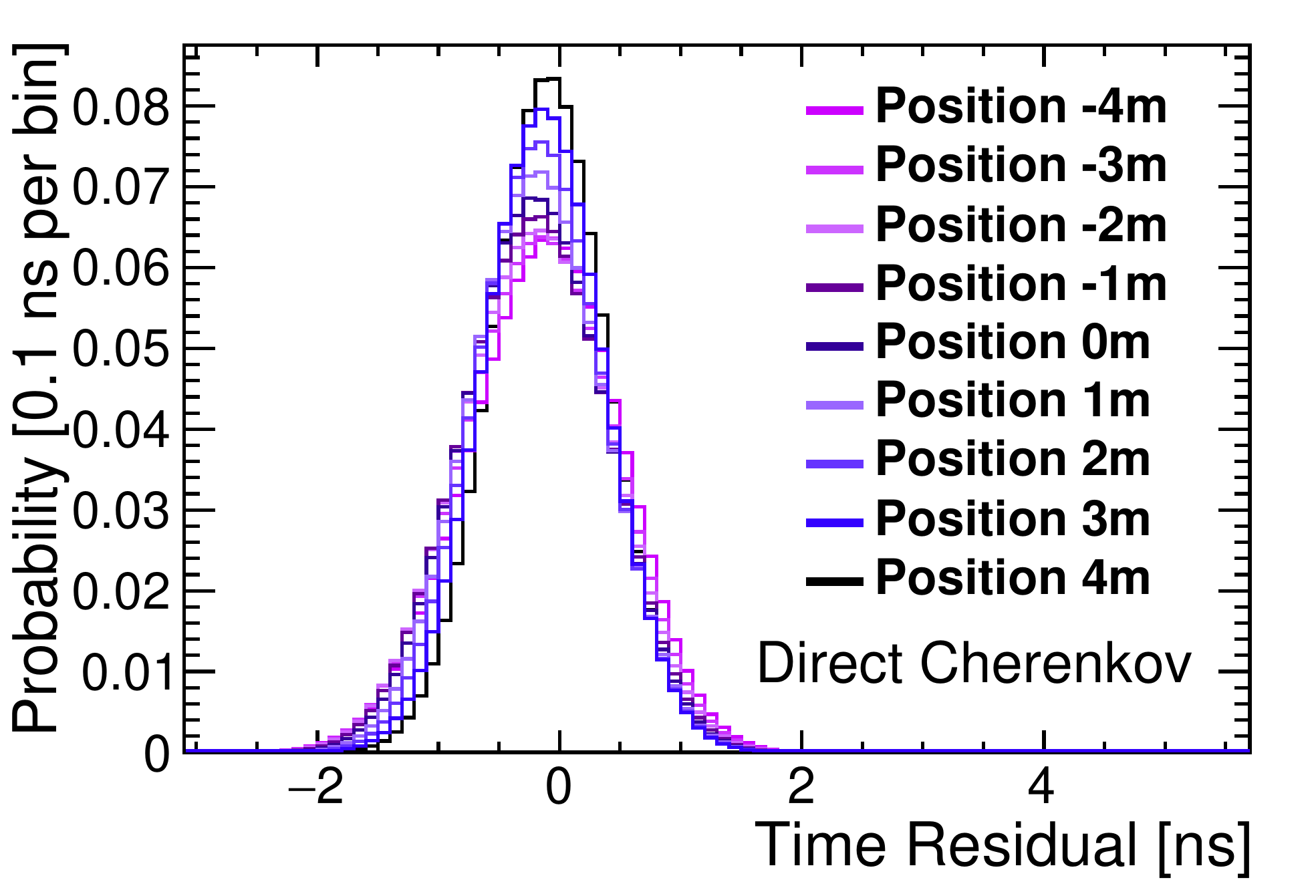}}
\subfigure{\includegraphics[scale=0.37]{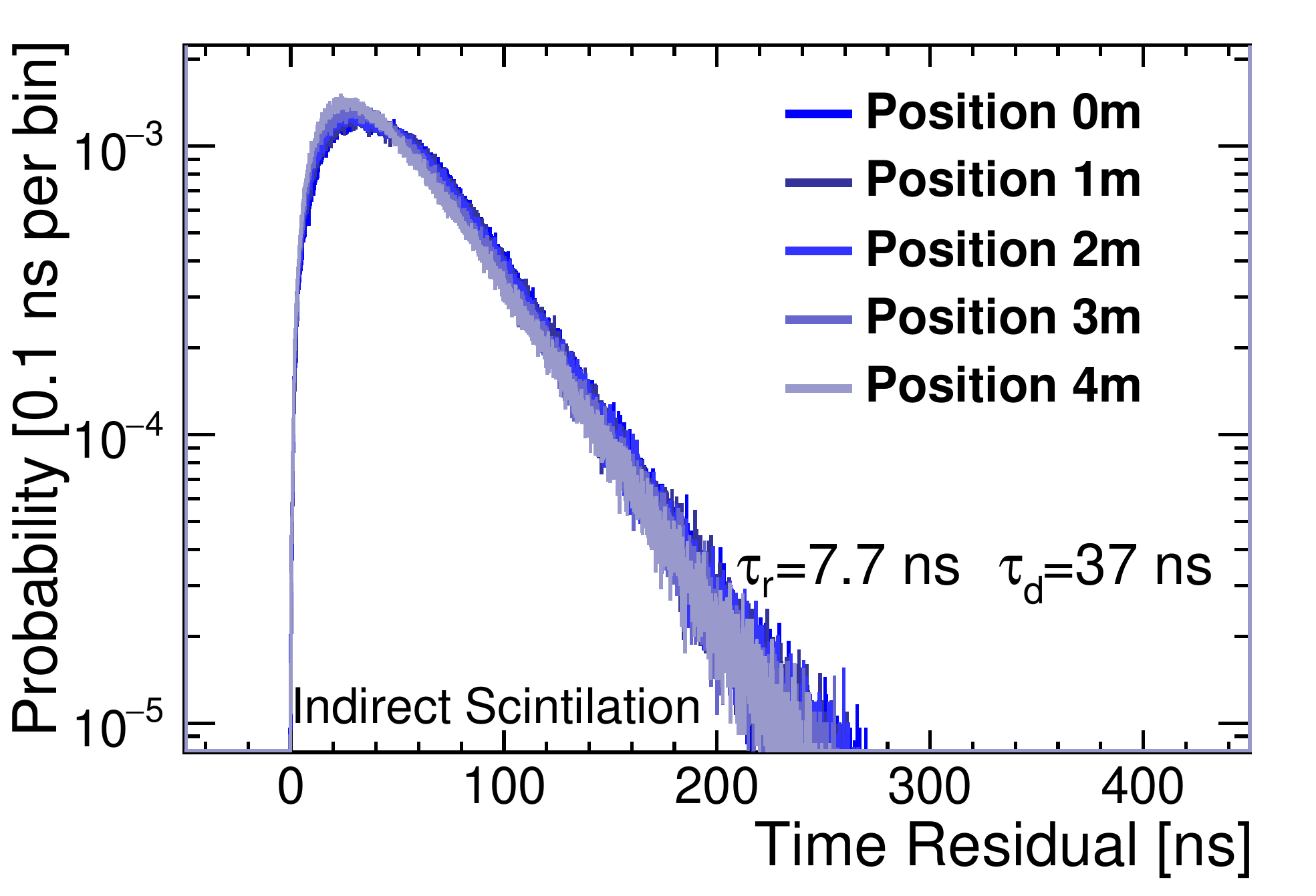}}
\subfigure{\includegraphics[scale=0.37]{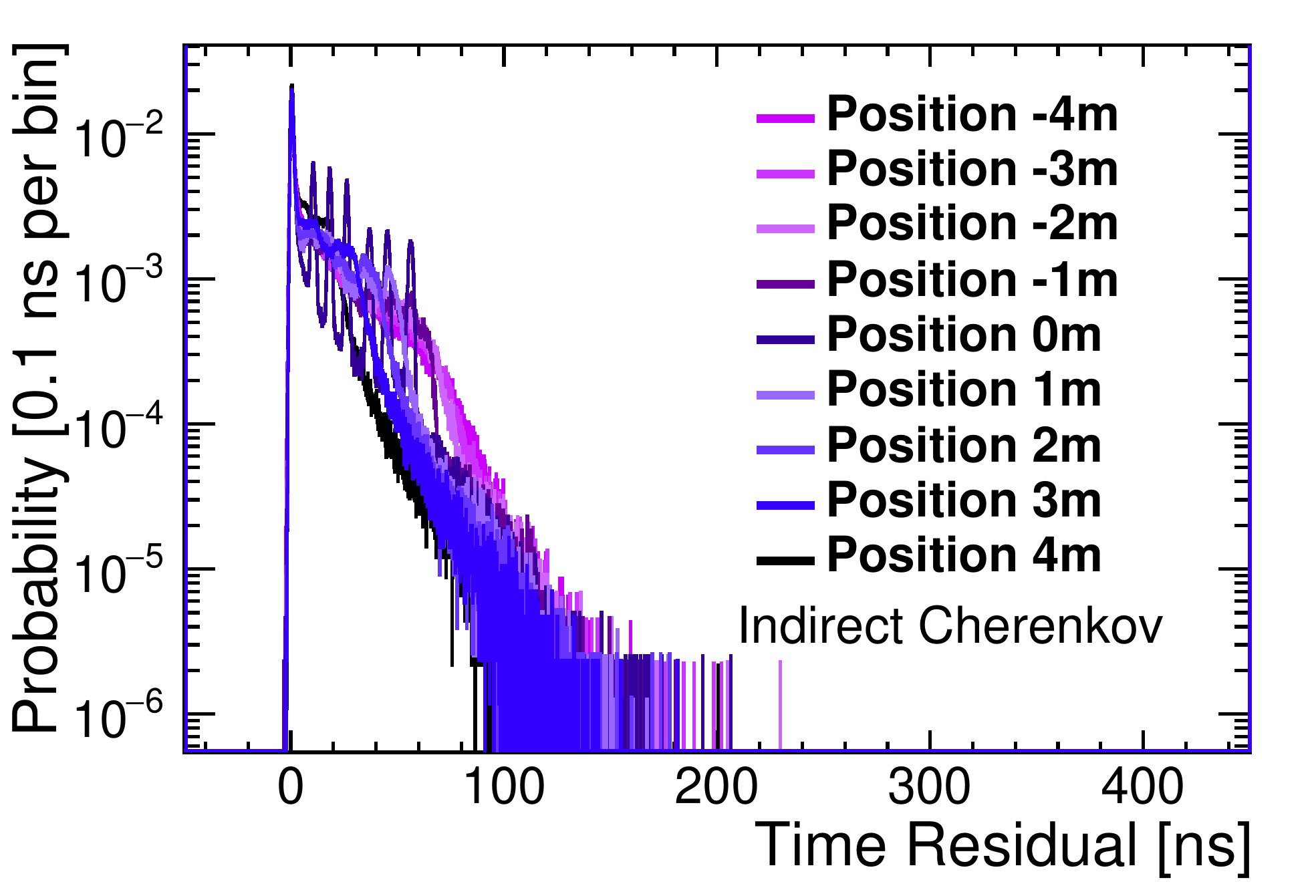}}
\caption{Four charts are used to display the time residual p.d.f.~for direct photons and indirect photons (including scintillation and Cherenkov photons). Top left: The direct scintillation photon time residual p.d.f.~with a rise time of \SI{7.7}{ns} and decay time of \SI{37}{ns} at (0,~0,~0), (\SI{1}{m},~0,~0) ... (\SI{4}{m},~0,~0) are plotted. Top right: The direct Cherenkov photon time residual p.d.f.~at (\SI{-4}{m},~0,~0), (\SI{-3}{m},~0,~0) ... (\SI{4}{m},~0,~0) are plotted. Bottom left: The indirect scintillation photon time residual p.d.f.~at (0,~0,~0), (\SI{1}{m},~0,~0) ... (\SI{4}{m},~0,~0) with a \SI{7.7}{ns} rise time and \SI{37}{ns} decay time are plotted. Bottom right: The indirect Cherenkov photon time residual p.d.f.~at (\SI{-4}{m},~0,~0), (\SI{-3}{m},~0,~0) ... (\SI{4}{m},~0,~0) are plotted.} 
\label{Time Profile}
\end{figure}

The positions of scintillation photons are 0 to \SI{4}{m} because its emission direction is isotropic. The four time residual p.d.f.~of direct and indirect photons (including scintillation and Cherenkov photons) are shown in figure~\ref{Time Profile}.

It can be seen that the time residual p.d.f.~of the four types of light vary greatly. Details of the discussion are as follows.

\paragraph{Direct Scintillation} For the direct scintillation photon, the time residual p.d.f.~is basically the same as the scintillation time profile at different positions, as shown in the top left of figure~\ref{Time Profile}.

\paragraph{Direct Cherenkov} For the direct Cherenkov photon, the time resolution of the time residual p.d.f.~is related to the flight distance of Cherenkov light due to the dispersion effect, and the longer the flight distance is, the larger the time resolution, as shown in the top right of figure~\ref{Time Profile}.

\paragraph{Indirect Scintillation} For the indirect scintillation photons, the time residual p.d.f.~is wider than that of direct scintillation photons, which is caused by reflection and scattering. Additionally, the variation of the time residual p.d.f.~at different positions is small and can be neglected, as shown in the bottom left of figure~\ref{Time Profile}.

\paragraph{Indirect Cherenkov} For the indirect Cherenkov photon, the time residual p.d.f.~is only related to the structure of the detector, as shown in the bottom right of figure~\ref{Time Profile}. The reflection peaks of the indirect Cherenkov photon time residual p.d.f.~at different positions are different. To simplify the calculation, we assume that the time residual p.d.f.~of the indirect Cherenkov photons does not vary with position, since the effect of indirect Cherenkov photons on the directional reconstruction is small.

In addition, the fitting of the time residual p.d.f.~is discussed in the next section.

\section{Fitting Model Construction}\label{Sec: Model}

In this section, a likelihood function is constructed according to the signal process discussed in the last section. The prediction of the charge and time by the likelihood function is compared with the full MC simulation data for correctness verification.

\subsection{Likelihood Function of Reconstruction}

A likelihood function is created similar to  SNO~\citep{SNOAngle,Direction2013SNO3} and applied to the event's reconstruction. It depends on $E$, $x$, $y$, $z$, $t_{\mathrm{event}}$, $\vec{d}_{\mathrm{Fit}}$, which are described in table~\ref{Table: PE}. The likelihood function is mainly composed of charge and time terms, and its specific form is as follows:

\begingroup
\setlength{\tabcolsep}{10pt} % Default value: 6pt
\renewcommand{\arraystretch}{1.6} % Default value: 1
\begin{table}[!htbp]
\centering
\begin{tabular}{cp{11cm}<{\centering}}
\hline
Parameters & Explanation  \\ \hline

$i$  &                The $i$th PMT \\

$ij$  &                The $j$th PE of the $i$th PMT \\

$E$  &                Event visible energy to be fit \\

$x,y,z$  &           Vertex position to be fit \\
$t_{\mathrm{event}}$  &           Event time to be fit \\

$\vec{d}_{\mathrm{Fit}}$  &           Event direction to be fit\\

$\mathrm{m}$  &           medium include the SlowLS, acrylic and mineral oil  \\

$x^{\mathrm{S}}_{\mathrm{vir}},y^{\mathrm{S}}_{\mathrm{vir}},z^{\mathrm{S}}_{\mathrm{vir}}$  &           Virtual vertex position of scintillation photons [section~\ref{Sec: VirtualVetex}] \\

$x^{\mathrm{C}}_{\mathrm{vir}},y^{\mathrm{C}}_{\mathrm{vir}},z^{\mathrm{C}}_{\mathrm{vir}}$  &           Virtual vertex position of Cherenkov photons [section~\ref{Sec: VirtualVetex}] \\

$f_{\mathrm{D}}$         &       Direct photon fraction (in this work is 0.85)   [section~\ref{Sec: DirectandNon}]  \\

$f_{k_{\mathrm{B}}}$         &       Birks’ law effect factor [section~\ref{Sec: YS}]  \\

$Y_{\mathrm{s}} $         &        Scintillation light yield [eq.~(\ref{Eq: YS})]  \\

$Y_{\mathrm{c}} $         &       Cherenkov light yield  [eq.~(\ref{Eq: YC})]   \\
 $g(\cos\alpha_i)$     &      Cherenkov light angular distribution [section~\ref{Sec: YC}] \\

$T_{i,\mathrm{m}}^{\mathrm{S}}$         &   Phototransmission for scintillation photons [eq.~(\ref{Eq: ABSCPE})] \\

$T_{i,\mathrm{m}}^{\mathrm{C}}$         &      Phototransmission for Cherenkov photons [eq.~(\ref{Eq: ABSCPE})] \\

$\Omega_i$      &      Solid angle [eq.~(\ref{Eq: SolidAngle})]   \\

$\epsilon_{i}^{\mathrm{S,PMT}}$         &         Effective DE for  scintillation photons [section~\ref{Sec: DE}]  \\

$\epsilon_{i}^{\mathrm{C,PMT}}$         &        Effective DE for Cherenkov photons [section~\ref{Sec: DE}] \\

$t_{ij}^{\mathrm{S},\mathrm{res}}$ &          Scintillation time residual [eq.~(\ref{Eq: STimeres})]\\

$t_{ij}^{\mathrm{C},\mathrm{res}}$ &          Cherenkov light time residual [eq.~(\ref{Eq: CTimeres})]\\

\hline
\end{tabular}
\caption{The description of the parameters.}
\label{Table: PE}
\end{table}
\endgroup

\begin{equation}
\mathcal{L}(n_{i}^{\mathrm{Obs}},t_{ij}|E,x,y,z,t_{\mathrm{event}},\vec{d}_{\mathrm{Fit}}) = \prod_i^{N_{\mathrm{PMT}}}P_i^{C} {\prod_j^{n_i^{\mathrm{Obs}}} {P_{ij}^{T}}},
\label{Eq: ReconLikelihood}
\end{equation}

\noindent where $n_{i}^{\mathrm{Obs}}$ is the number of PEs of the $i$th PMT, $t_{ij}$ is the $j$th PE time of the $i$th PMT, $N_{\mathrm{PMT}}$ is the total number of PMTs, $P_{i}^{C}$ is the probability of charge of the $i$th PMT, and $P_{ij}^{T}$ is the probability of the $j$th PE time of the $i$th PMT.

$P_{i}^{\mathrm{C}}$ is presented as follows based on previous research into the signaling process:
\begin{equation}
P_{i}^{C} = f_{\mathrm{Poisson}}(n_i^{\mathrm{Obs}};n_i^{\mathrm{SD}}+n_i^{\mathrm{SI}}+n_i^{\mathrm{CD}}+n_i^{\mathrm{CI}}+n_i^{\mathrm{DN}}),
\label{Eq: ChargeLikelihood}
\end{equation}

\noindent where $f_{\mathrm{Possion}}$ is a Poisson distribution. The number of direct and indirect scintillation PEs of the $i$th PMT is represented by $n_i^{\mathrm{SD}}$ and $n_i^{\mathrm{SI}}$, respectively. The number of direct and indirect Cherenkov PEs of the $i$th PMT are $n_i^{\mathrm{CD}}$ and $n_i^{\mathrm{CI}}$, respectively. The number of signals caused by dark noise for the $i$th PMT is $n_i^{\mathrm{DN}}$. 

The definition of $P_{ij}^{\mathrm{T}}$ is,
\begin{equation}
P_{ij}^{T} = \frac{n_{i}^{\mathrm{SD}}}{n_i^{\mathrm{Exp}}}f^{\mathrm{SD}}({t_{ij}^{\mathrm{S},\mathrm{res}}})+\frac{n_i^{\mathrm{SI}}}{n_i^{\mathrm{Exp}}}f^{\mathrm{SI}}({t_{ij}^{\mathrm{S},\mathrm{res}}})+\frac{n_i^{\mathrm{CD}}}{n_i^{\mathrm{Exp}}}f^{\mathrm{CD}}({t_{ij}^{\mathrm{C},\mathrm{res}}})+\frac{n_i^{\mathrm{CI}}}{n_i^{\mathrm{Exp}}}f^{\mathrm{CI}}({t_{ij}^{\mathrm{C},\mathrm{res}}})+\frac{n_i^{\mathrm{DN}}}{n_i^{\mathrm{Exp}}}f^{\mathrm{DN}},
\end{equation}

\noindent where ${n_i^{\mathrm{Exp}}}=n_i^{\mathrm{SD}}+n_i^{\mathrm{SI}}+n_i^{\mathrm{CD}}+n_i^{\mathrm{CI}}+n_i^{\mathrm{DN}}$. $f^{\mathrm{SD}}$ and $f^{\mathrm{SI}}$ present the time residual p.d.f.~for the direct and indirect scintillation photons, respectively. The time residual p.d.f.~for the direct and indirect Cherenkov photons, respectively, is $f^{\mathrm{CD}}$ and $f^{\mathrm{CI}}$. The time p.d.f.~of the dark noise is $f^{\mathrm{DN}}$.

\subsection{Fitting Model of the Charge Profile}\label{Sec: ChargeProfile}
Based on the simplification of the signal process, the expressions for calculating the number of PEs for the four types of light are developed in the following form.

\paragraph{Direct Scintillation }

According to section~\ref{Sec: SignalProcess}, the following factors must be taken into account when calculating the direct scintillation PE:
\begin{itemize}
\item The scintillation light yield.
\item The quenching effect resulting from Birks' law.
\item The proportion of the direct photons.
\item The attenuation of scintillation photons in the material.
\item The solid angle accepted by the PMT.
\item The effective DE of scintillation photons.
\end{itemize}

Consequently, the number of direct scintillation PEs, $n_i^{\mathrm{SD}}$, is provided by

\begin{equation}
n_i^{\mathrm{SD}} = f_{\mathrm{D}}f_{k_{\mathrm{B}}} Y_{\mathrm{s}}(E)  \prod_{\mathrm{m}}^{}{T_{i,\mathrm{m}}^{\mathrm{S}}(x,y,z)} \Omega_i(x,y,z)\epsilon^{\mathrm{S,PMT}}.
\label{Eq: N_sci}
\end{equation}
 
Table~\ref{Table: PE} displays all parameters that are used in the calculation.

\paragraph{Direct Cherenkov}
The following factors need to be considered when calculating the direct Cherenkov PE, according to section~\ref{Sec: SignalProcess}.

\begin{itemize}
\item Cherenkov light yield (300-\SI{700}{nm}).
\item The proportion of the direct photons.
\item The attenuation of Cherenkov photons in the material.
\item The solid angle accepted by the PMT.
\item The angular distribution of Cherenkov light.
\item The effective DE of Cherenkov photons.
\end{itemize}

The number of direct Cherenkov PEs, $n_i^{\mathrm{CD}}$, is therefore defined as:

\begin{equation}
n_i^{\mathrm{CD}} =f_{\mathrm{D}} Y_{\mathrm{c}}(E)  \prod_{\mathrm{m}}^{}{T_{i,\mathrm{m}}^{\mathrm{C}}(x,y,z)} \Omega_i(x,y,z)\epsilon^{\mathrm{C,PMT}} g(\cos\alpha_i(x,y,z,\vec{d}_{\mathrm{Fit}})).
\label{Eq: N_chev}
\end{equation}

All of the input variables utilized in the calculation are shown in table~\ref{Table: PE}.

\paragraph{Indirect Scintillation }
The indirect scintillation calculations are fairly similar to those for the direct photons (eq.~\ref{Eq: N_sci}) but with the virtual vertex position $(x^{\mathrm{S}}_{\mathrm{vir}},y^{\mathrm{S}}_{\mathrm{vir}},z^{\mathrm{S}}_{\mathrm{vir}})$. Therefore, it is possible to calculate the number of indirect scintillation PEs, $n_i^{\mathrm{SI}}$, as:

\begin{equation}
n_i^{\mathrm{SI}} = (1-f_{\mathrm{D}})f_{k_{\mathrm{B}}} Y_{\mathrm{s}}(E)  \prod_{\mathrm{m}}^{}{T_{i,\mathrm{m}}^{\mathrm{S}}(x^{\mathrm{S}}_{\mathrm{vir}},y^{\mathrm{S}}_{\mathrm{vir}},z^{\mathrm{S}}_{\mathrm{vir}})}  \Omega_i(x^{\mathrm{S}}_{\mathrm{vir}},y^{\mathrm{S}}_{\mathrm{vir}},z^{\mathrm{S}}_{\mathrm{vir}})\epsilon^{\mathrm{S,PMT}}.
\end{equation}

The parameters that are utilized in the calculation are shown in table~\ref{Table: PE}.

\paragraph{Indirect Cherenkov}
Cherenkov light that is not directly can be calculated in a manner that is quite similar to that of directly Cherenkov light (eq.~\ref{Eq: N_chev}) but with the virtual vertex position $(x^{\mathrm{C}}_{\mathrm{vir}},y^{\mathrm{C}}_{\mathrm{vir}},z^{\mathrm{C}}_{\mathrm{vir}})$.
As a result, it is possible to display the number of indirect Cherenkov PEs, $n_i^{\mathrm{CI}}$, as:

\begin{equation}
n_i^{\mathrm{CI}} =(1-f_{\mathrm{D}}) Y_{\mathrm{c}}(E) \prod_{\mathrm{m}}^{}{T_{i,\mathrm{m}}^{\mathrm{C}}(x^{\mathrm{C}}_{\mathrm{vir}},y^{\mathrm{C}}_{\mathrm{vir}},z^{\mathrm{C}}_{\mathrm{vir}})}  \Omega_i(x^{\mathrm{C}}_{\mathrm{vir}},y^{\mathrm{C}}_{\mathrm{vir}},z^{\mathrm{C}}_{\mathrm{vir}})\epsilon^{\mathrm{C,PMT}} g(\cos\alpha_i(x^{\mathrm{C}}_{\mathrm{vir}},y^{\mathrm{C}}_{\mathrm{vir}},z^{\mathrm{C}}_{\mathrm{vir}},\vec{d}_{\mathrm{Fit}})).
\end{equation}

All of the inputs used in the calculation are shown in table~\ref{Table: PE}.

\paragraph{Dark Noise }
Only the dark noise rate, $\mathrm{F}_{\mathrm{DN}}$, and time window, $T_{\mathrm{Window}}$, of the PMT are relevant to the dark noise-produced PE. In this work, $T_{\mathrm{Window}}$ is \SI{500}{ns}, and $\mathrm{F}_{\mathrm{DN}}$ is \SI{2000}{Hz}. Therefore, the number of the signals caused by the dark noise, $n_i^{\mathrm{DN}}$, is calculated as follows:

\begin{equation}
n_i^{\mathrm{DN}} = \mathrm{F_{\mathrm{DN}}}\cdot \mathrm{T_{\mathrm{Window}}}.
\end{equation}

\subsection{Fitting Model of the Time Profile}

The time residual p.d.f.~of various types of light are discussed in section~\ref{Sec: TimeRes}, and the time residual p.d.f.~are fit separately here.

\paragraph{Direct Scintillation }

The definition of the time residual p.d.f.~of the direct scintillation photon, $f^{\mathrm{SD}}(t_{ij}^{\mathrm{S},\mathrm{res}})$, is

\begin{equation}
f^{\mathrm{SD}}(t_{ij}^{\mathrm{S},\mathrm{res}})=\frac{\tau_{r}^{\mathrm{SD}}+\tau_{d}^{\mathrm{SD}}}{{\tau_{d}^{\mathrm{SD}}}^2}(1-e^{-t_{ij}^{\mathrm{S},\mathrm{res}}/\tau_{r}^{\mathrm{SD}}})\cdot e^{-t_{ij}^{\mathrm{S},\mathrm{res}}/\tau_{d}^{\mathrm{SD}}}\otimes \mathrm{Gauss}(0,\sigma^{\mathrm{SD}}_{t}),
\label{Eq: SPETime}
\end{equation}

\noindent where ${\tau_{r}^{\mathrm{SD}}}$ and $\tau_{d}^{\mathrm{SD}}$ are the are the rise time and decay time of the direct scintillation PEs, respectively, and $\sigma^{\mathrm{SD}}_{t}$ is the time resolution for the direct scintillation photon time residual p.d.f.. By fitting the upper left panel of figure~\ref{Time Profile}, ${\tau_{r}^{\mathrm{SD}}}$ and $\tau_{d}^{\mathrm{SD}}$ are \SI{7.7}{ns} and \SI{37}{ns}, respectively. $\sigma^{\mathrm{SD}}_{t}$ in this study is \SI{0.0398}{ns}.

\paragraph{Direct Cherenkov }

\begin{figure}[!htbp]
\centering
\includegraphics[scale=0.4]{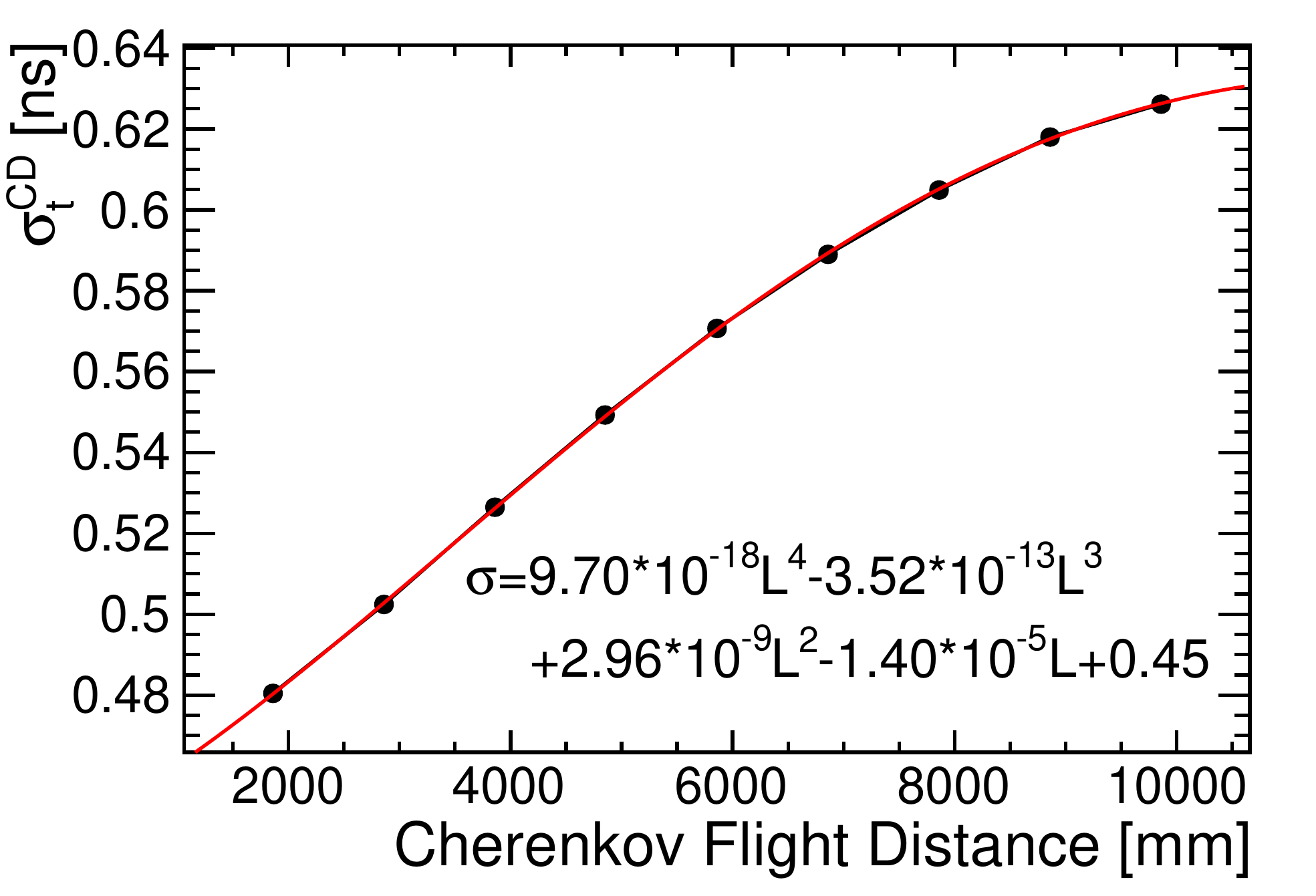}
\caption{Cherenkov light flight distance versus the time resolution, $\sigma^{\mathrm{CD}}_{t}$, is plotted. }
\label{Fig: CPE-TTS}
\end{figure}

The time of Cherenkov light produced by several-MeV electrons is usually within \SI{1}{ns}, while the generation of scintillation light takes at least a few nanoseconds. Therefore, the time distribution of Cherenkov light is treated as a $\delta$ function. A Gaussian function is used in the fitting of the time residual p.d.f.~of the Cherenkov light.
The definition of the time residual p.d.f.~of the direct Cherenkov photons, $f^{\mathrm{CD}}(t_{ij}^{\mathrm{C},\mathrm{res}})$, is
\begin{equation}
f^{\mathrm{CD}}(t_{ij}^{\mathrm{C},\mathrm{res}})=\delta(t_{ij}^{\mathrm{C},\mathrm{res}})\otimes \mathrm{Gauss}(0,\sigma^{\mathrm{CD}}_{t}).
\label{Eq: CPETime}
\end{equation}
\noindent where $\sigma^{\mathrm{CD}}_{t}$ is the time resolution for the time residual p.d.f..
As shown in the upper right of figure~\ref{Time Profile}, the resolution of the time residual p.d.f.~ becomes larger as the Cherenkov light flight distance becomes longer, due to the dispersion effects. Then, we fit the relationship between the time resolution and flight distance using a fourth-order polynomial, as shown in figure~\ref{Fig: CPE-TTS}. As a result, the time resolution $\sigma^{\mathrm{CD}}_{t}$ is used as the fourth-order polynomial in eq.~(\ref{Eq: CPETime}).

\paragraph{Indirect Scintillation }

The indirect scintillation photon time residual p.d.f., $f^{\mathrm{SI}}(t_{ij}^{\mathrm{S},\mathrm{res}})$, is described as

\begin{equation}
f^{\mathrm{SI}}(t_{ij}^{\mathrm{S},\mathrm{res}})=\frac{\tau_{r}^{\mathrm{SI}}+\tau_{d}^{\mathrm{SI}}}{{\tau_{d}^{\mathrm{SI}}}^2}(1-e^{-t_{ij}^{\mathrm{S},\mathrm{res}}/\tau_{r}^{\mathrm{SI}}})\cdot e^{-t_{ij}^{\mathrm{S},\mathrm{res}}/\tau_{d}^{\mathrm{SI}}}\otimes \mathrm{Gauss}(0,\sigma^{\mathrm{SI}}_{t}),
\label{Eq: NonSPETime}
\end{equation}

\noindent where ${\tau_{r}^{\mathrm{SI}}}$ and $\tau_{d}^{\mathrm{SI}}$ are the rise time and decay time of the indirect scintillation PEs, respectively, and $\sigma^{\mathrm{SI}}_{t}$ is the time resolution for the indirect scintillation PEs time residual p.d.f.. 

\begin{figure}[!htbp]
\centering
\subfigure{\includegraphics[scale=0.37]{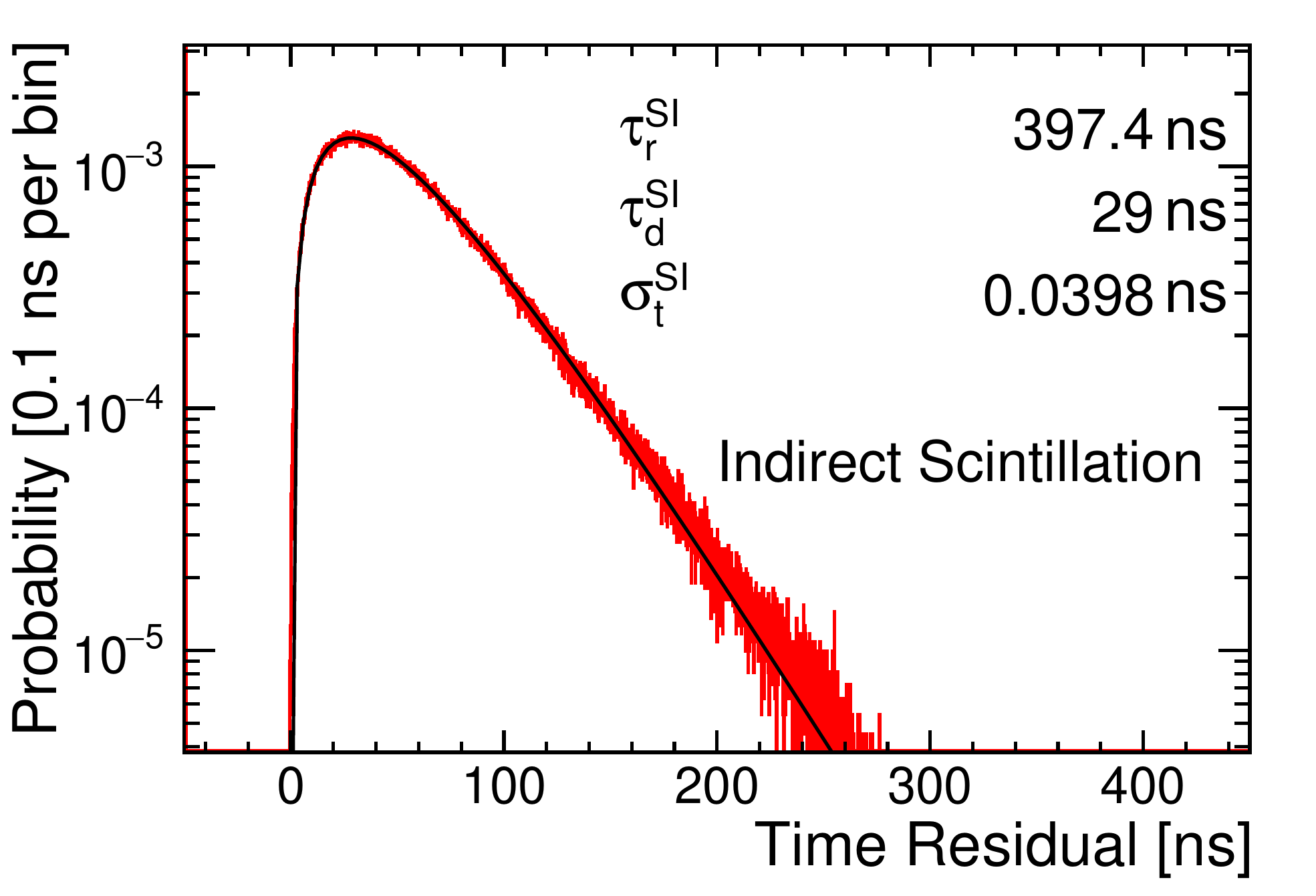}}
\subfigure{\includegraphics[scale=0.37]{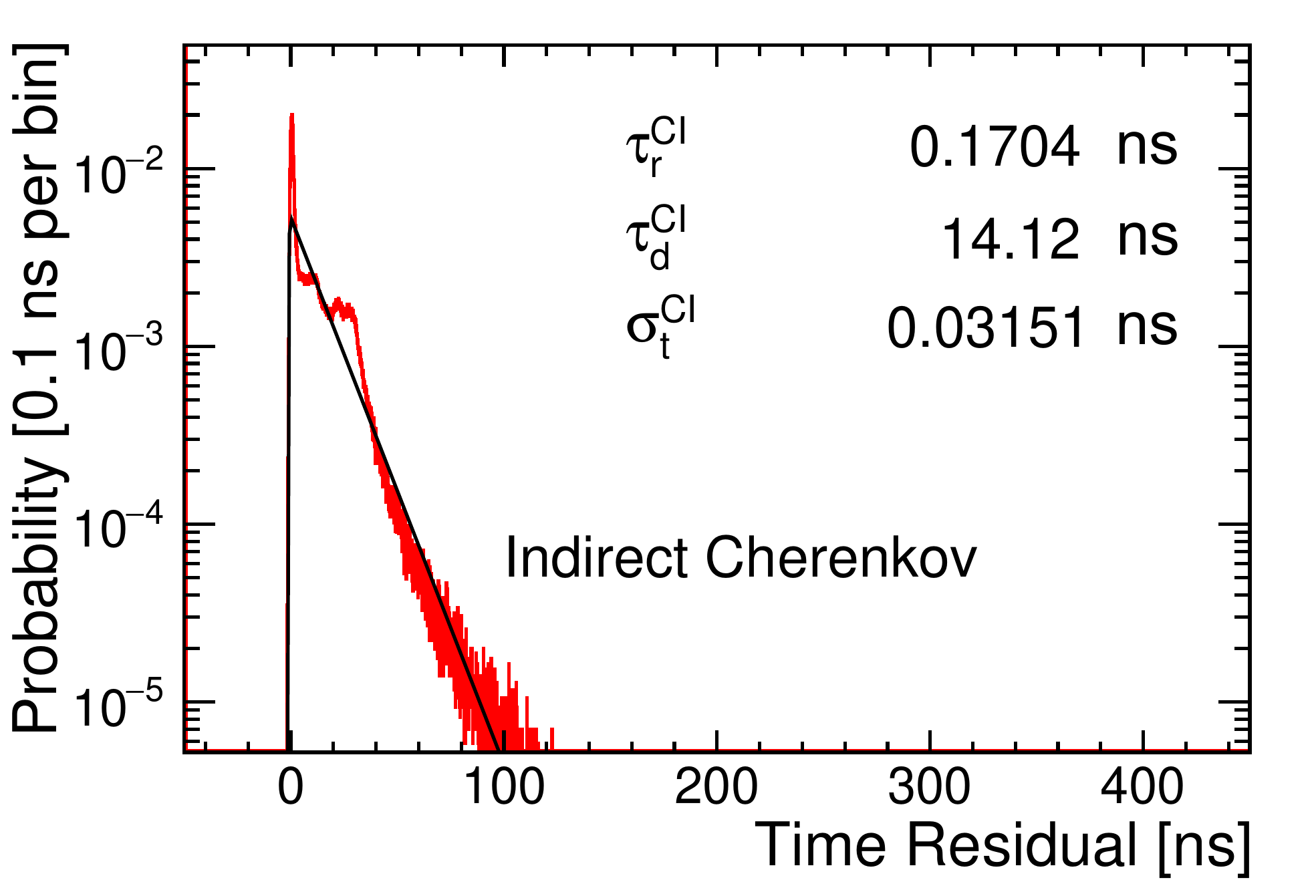}}
\caption{Fitting of indirect scintillation (left) and Cherenkov (right) time residuals p.d.f.~and the fit results are shown.}
\label{Fig: NonPETimeFit}
\end{figure}

The time residual p.d.f.~of different positions is shown on the bottom left of figure~\ref{Time Profile} to be very similar.
Therefore, we select the time residual p.d.f.~of (\SI{3}{m},~0,~0) for fitting using eq.~(\ref{Eq: NonSPETime}), and the fitting results are displayed in the left panel of figure~\ref{Fig: NonPETimeFit}.

\paragraph{Indirect Cherenkov }

The time residual p.d.f.~of the indirect Cherenkov photon, $f^{\mathrm{CI}}(t_{ij}^{\mathrm{C},\mathrm{res}})$, is described as
\begin{equation}
f^{\mathrm{CI}}(t_{ij}^{\mathrm{C},\mathrm{res}})=\frac{\tau_{r}^{\mathrm{CI}}+\tau_{d}^{\mathrm{CI}}}{{\tau_{d}^{\mathrm{CI}}}^2}(1-e^{-t_{ij}^{\mathrm{C},\mathrm{res}}/\tau_{r}^{\mathrm{CI}}})\cdot e^{-t_{ij}^{\mathrm{C},\mathrm{res}}/\tau_{d}^{\mathrm{CI}}}\otimes \mathrm{Gauss}(0,\sigma^{\mathrm{CI}}_{t}),
\label{Eq: NonCPETime}
\end{equation}

\noindent where ${\tau_{r}^{\mathrm{CI}}}$ and $\tau_{d}^{\mathrm{CI}}$ are the rise time and decay time of the indirect Cherenkov PEs, respectively, and $\sigma^{\mathrm{CI}}_{t}$ is the time resolution for the indirect Cherenkov PEs time residual p.d.f.. Eq.~(\ref{Eq: NonCPETime}) is used to fit the p.d.f.~at (\SI{3}{m},~0,~0), and the right panel of figure~\ref{Fig: NonPETimeFit} displays the results. The origin of structure in this figure mainly originates from reflections on the PMT. For example, the first peak in the right panel of figure~\ref{Fig: NonPETimeFit} is primarily due to the single Cherenkov light reflection from the PMT near the first hit, while the rest are due to multiple reflections.

\paragraph{Dark Noise }

The dark noise time profile is independent of the time residual and only relates to the dark noise rate and time window, $T_{\mathrm{Window}}$. The time distribution of dark noise is uniform within the time window, so it is simple to calculate the dark noise time p.d.f., $f^{\mathrm{DN}}$, using the following formula:
\begin{equation}
f^{\mathrm{DN}} = 1/T_{\mathrm{Window}}.
\end{equation}

\subsection{Validation of the fit Model with Full MC Simulation}
The calculation of the TOF for PEs plays a major role in the comparison of time profile. The estimation of the effective group velocity and vertex position affects the TOF calculation. The effective group velocity is obtained by the full MC simulation (section~\ref{Sec: GroupSpeed}). The time residual distribution calculated using the true vertex positions is also obtained by the full MC simulation. The model is consistent with the full MC simulation.

The results of the model computation in the charge profile are checked with those of the full simulation to ensure its accuracy. We select two positions for the comparison. One is near the center of the detector, (\SI{1}{m},~\SI{1}{m},~0), and the other is near the detector boundary, (\SI{3}{m},~\SI{2}{m},~\SI{2}{m}). 
500,000 electrons with \SI{5}{MeV} kinetic energy are simulated in the direction (1,~0,~0) in a high scintillation light yield \SI{5000}{photons/MeV} , i.e.~$\mathrm{LS_{\mathrm{376~PE}}^{\mathrm{7.7~ns}}}$, sample.

\begin{figure}[!htbp]
\centering
\subfigure{\includegraphics[scale=0.24]{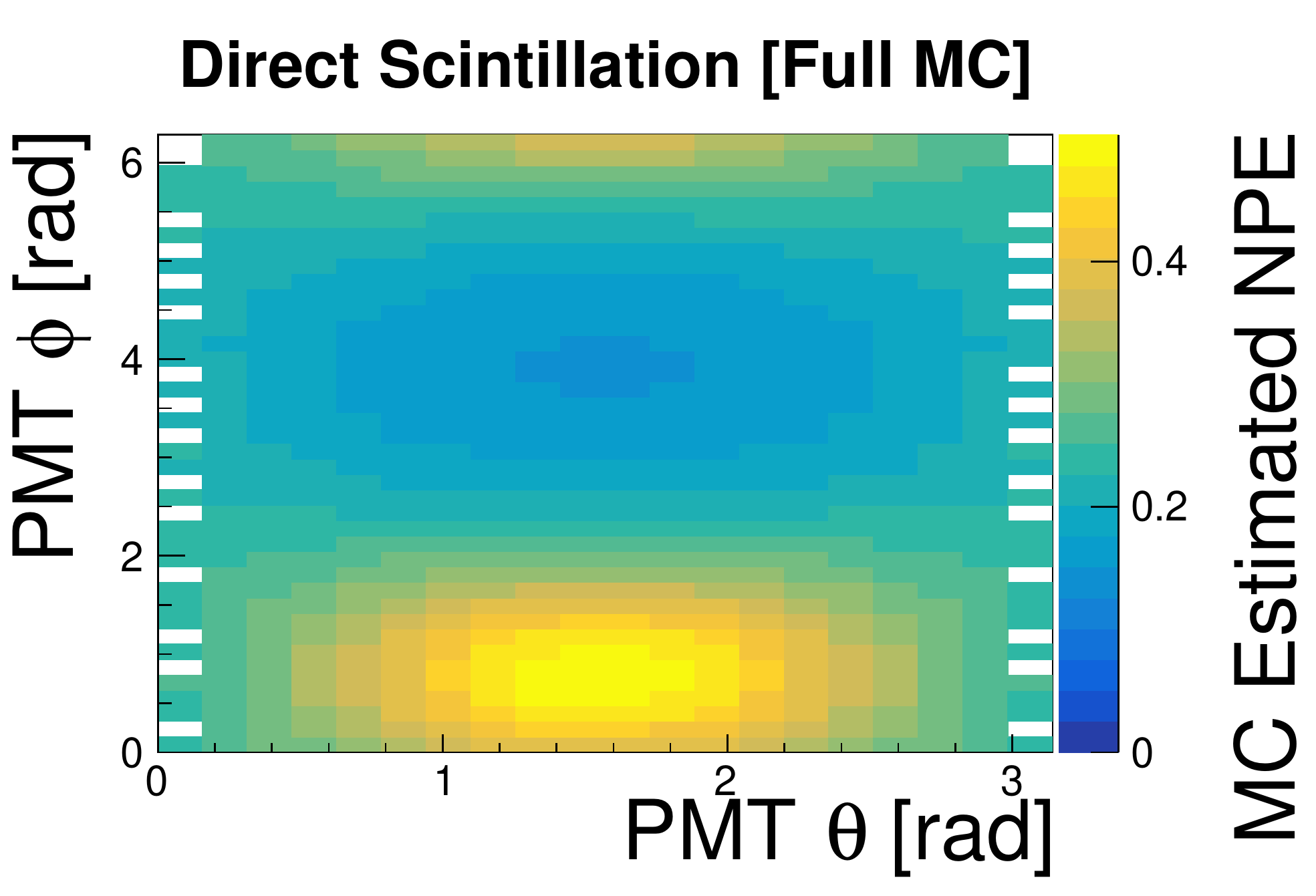}}
\subfigure{\includegraphics[scale=0.24]{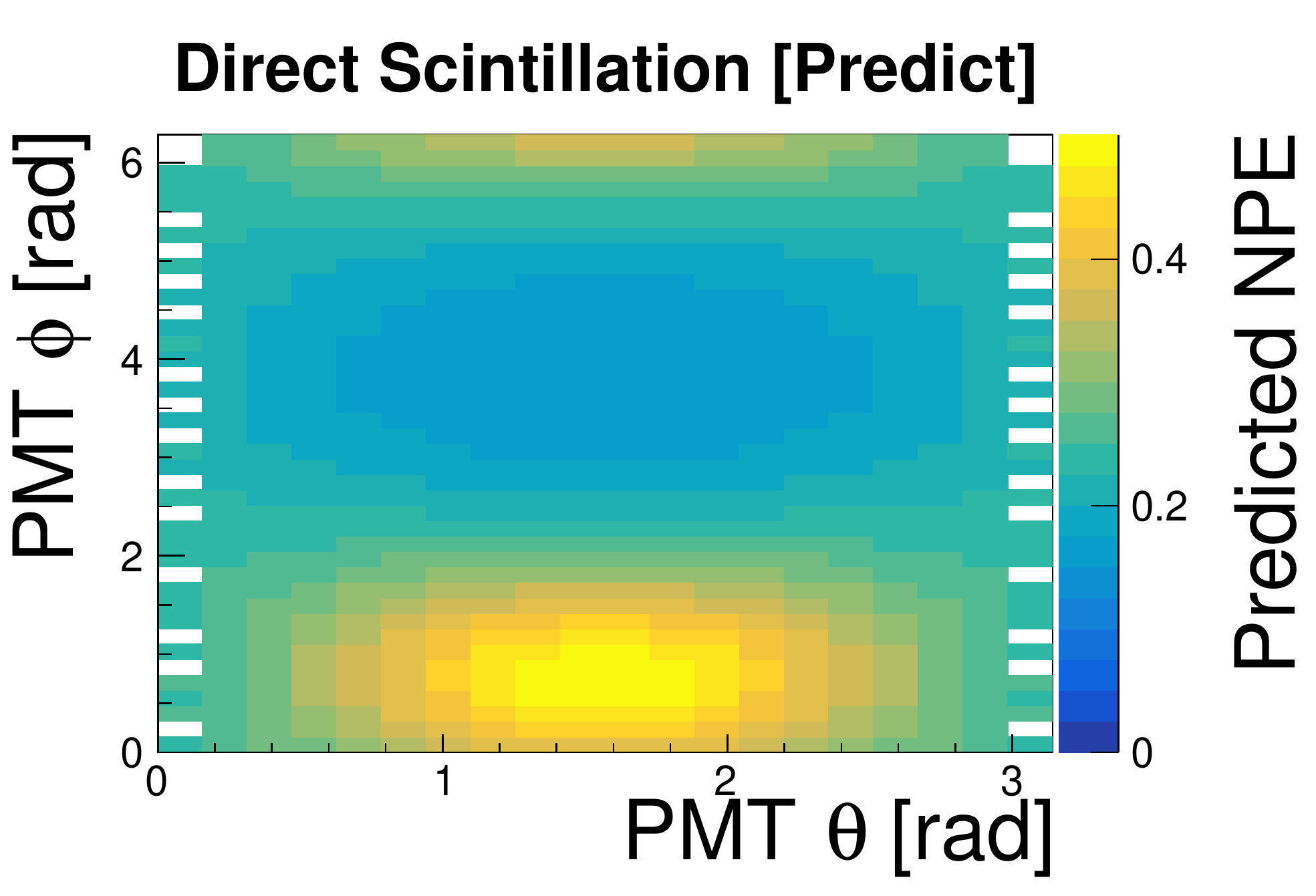}}
\subfigure{\includegraphics[scale=0.24]{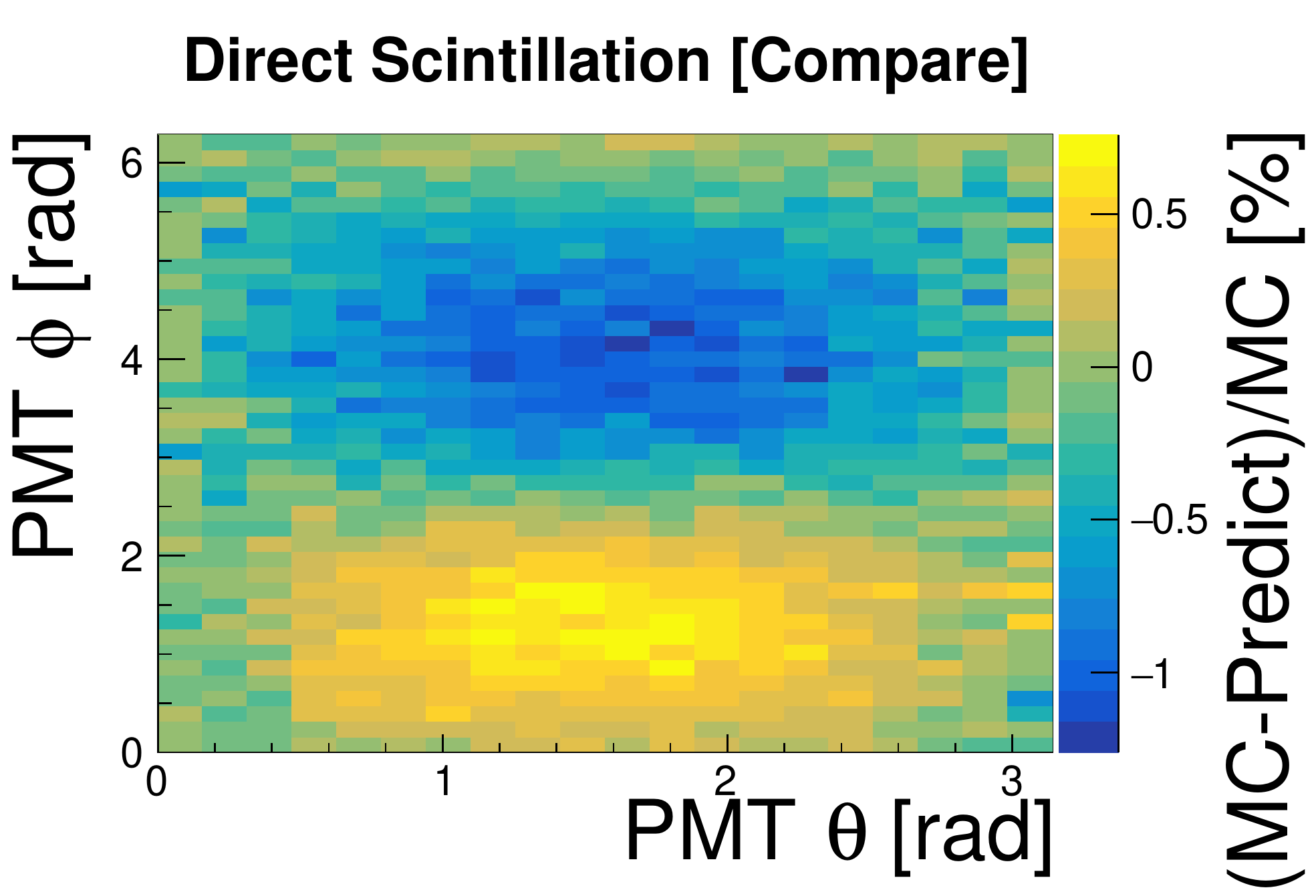}}
\subfigure{\includegraphics[scale=0.24]{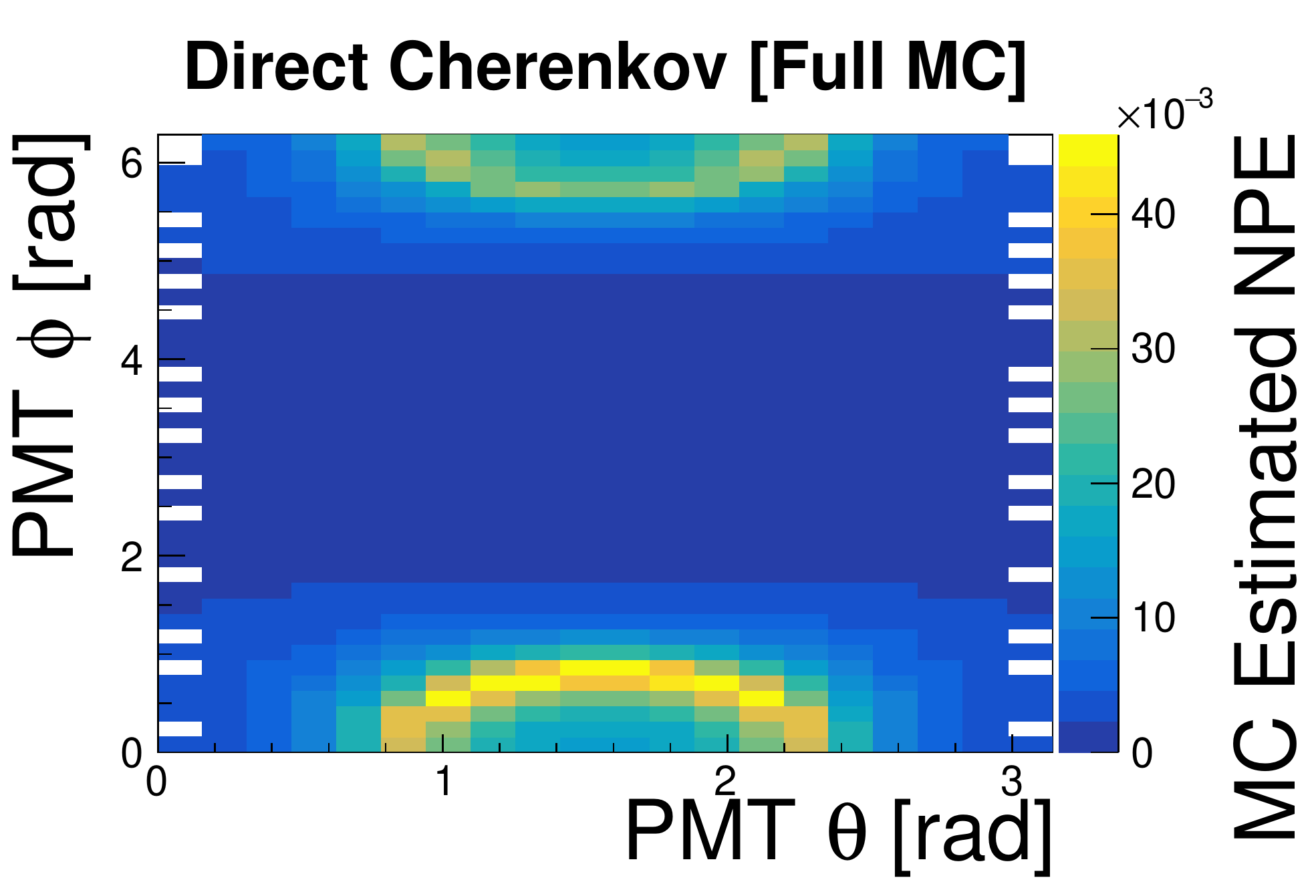}}
\subfigure{\includegraphics[scale=0.24]{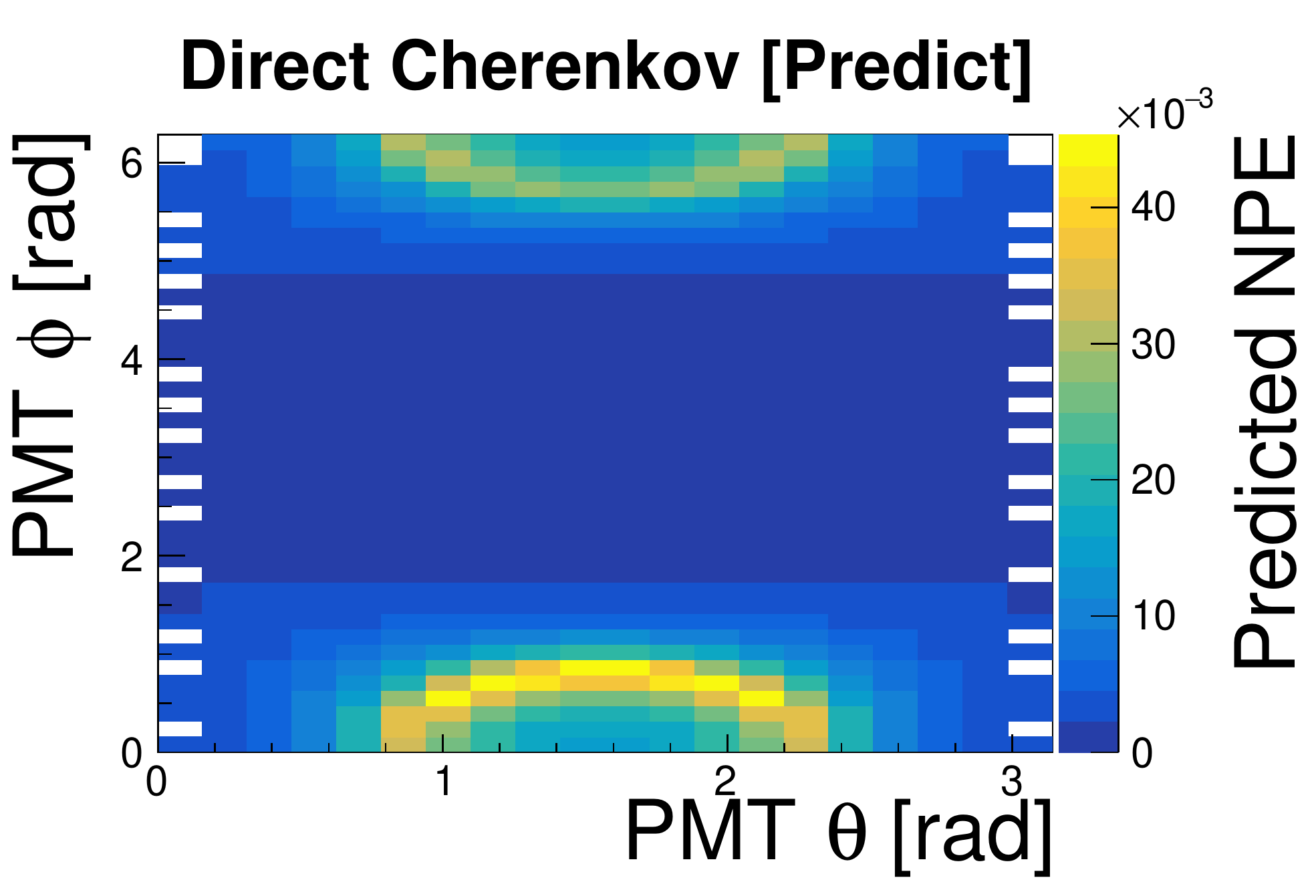}}
\subfigure{\includegraphics[scale=0.24]{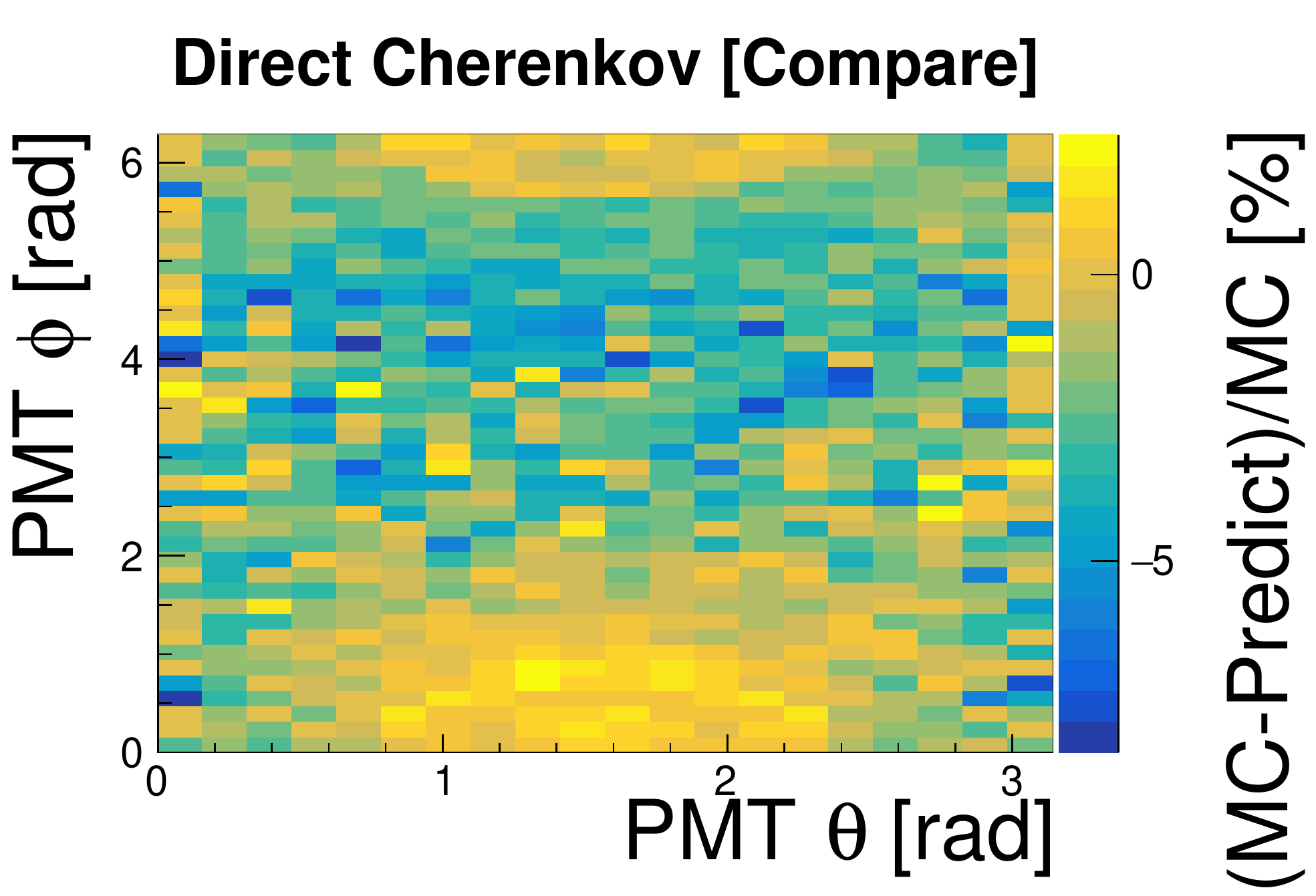}}
\subfigure{\includegraphics[scale=0.24]{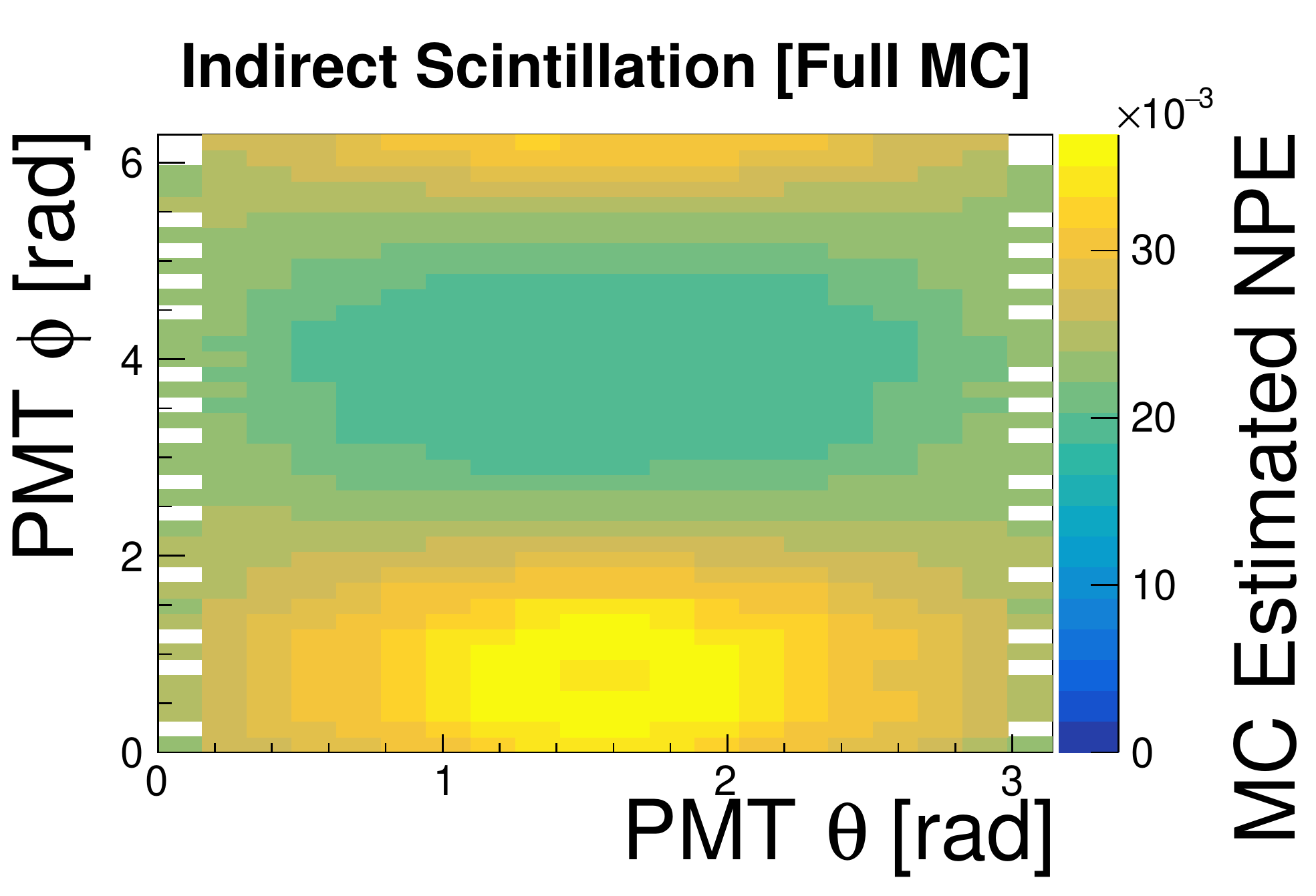}}
\subfigure{\includegraphics[scale=0.24]{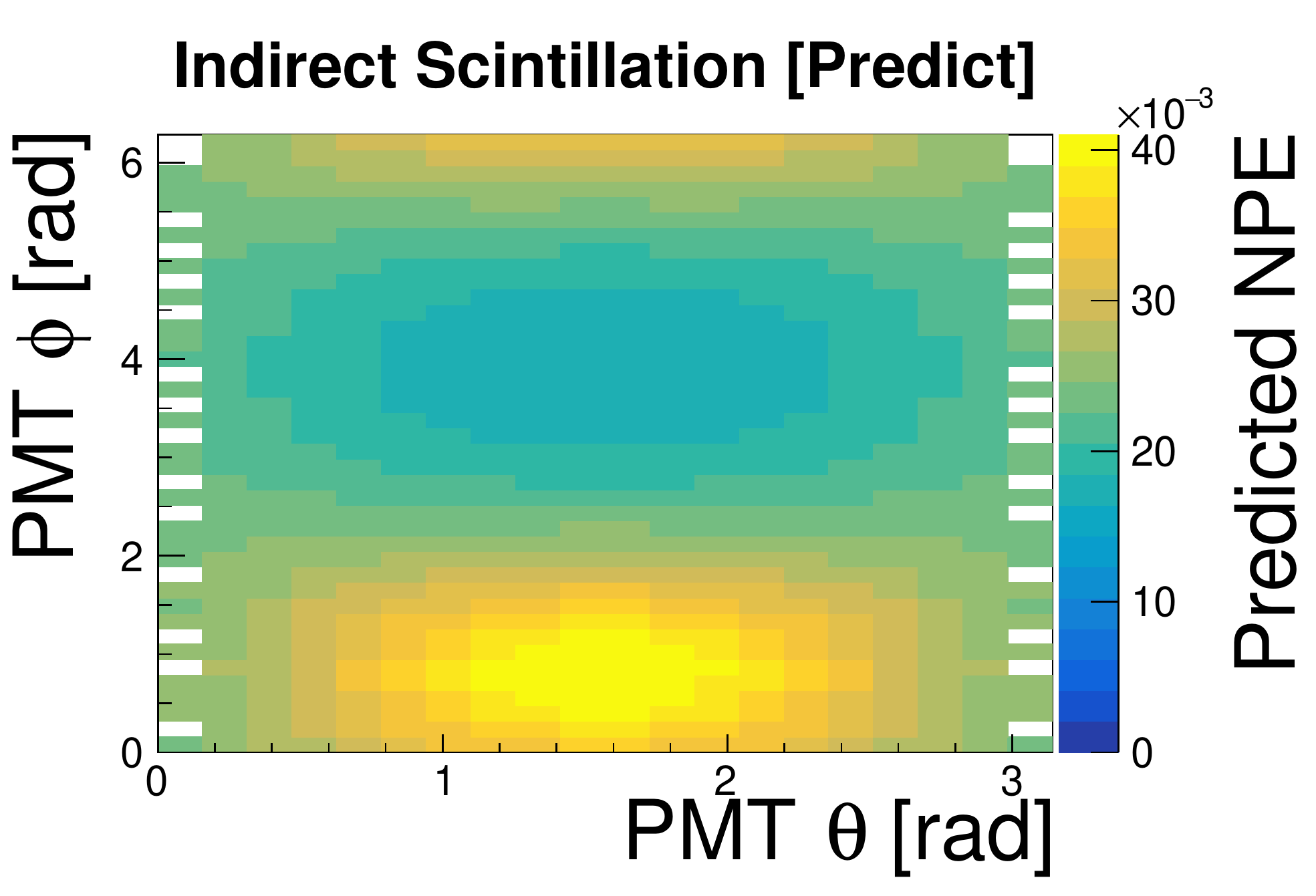}}
\subfigure{\includegraphics[scale=0.24]{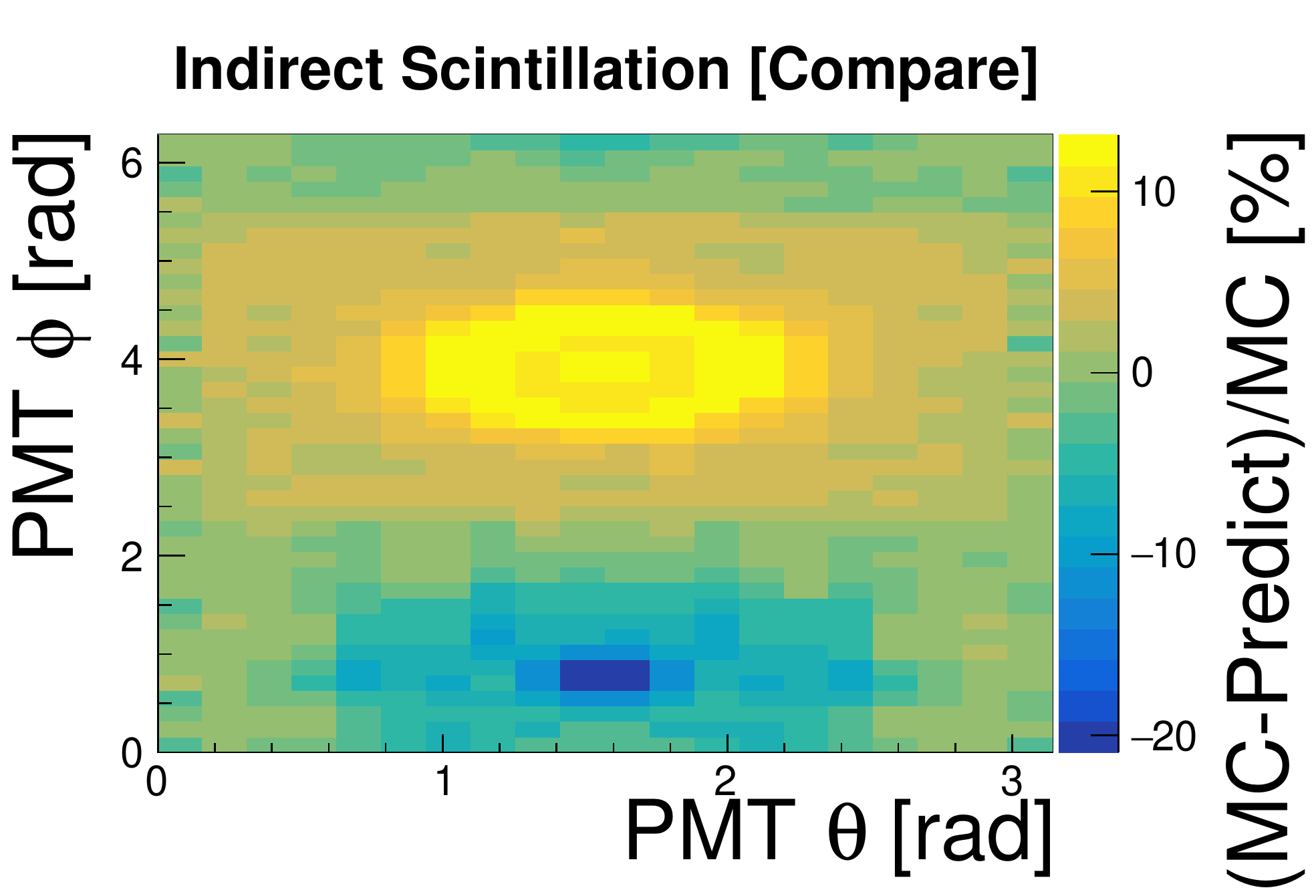}}
\subfigure{\includegraphics[scale=0.24]{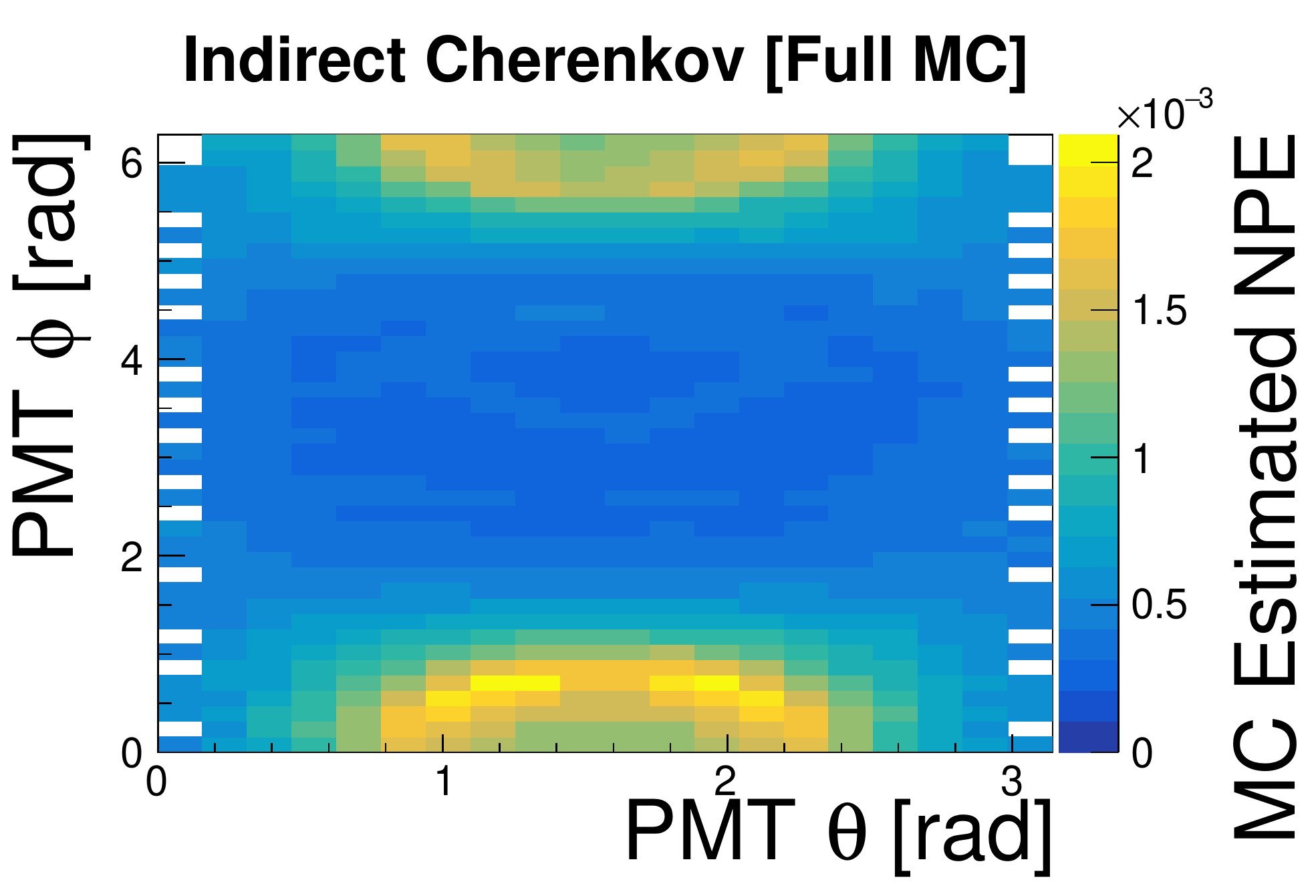}}
\subfigure{\includegraphics[scale=0.24]{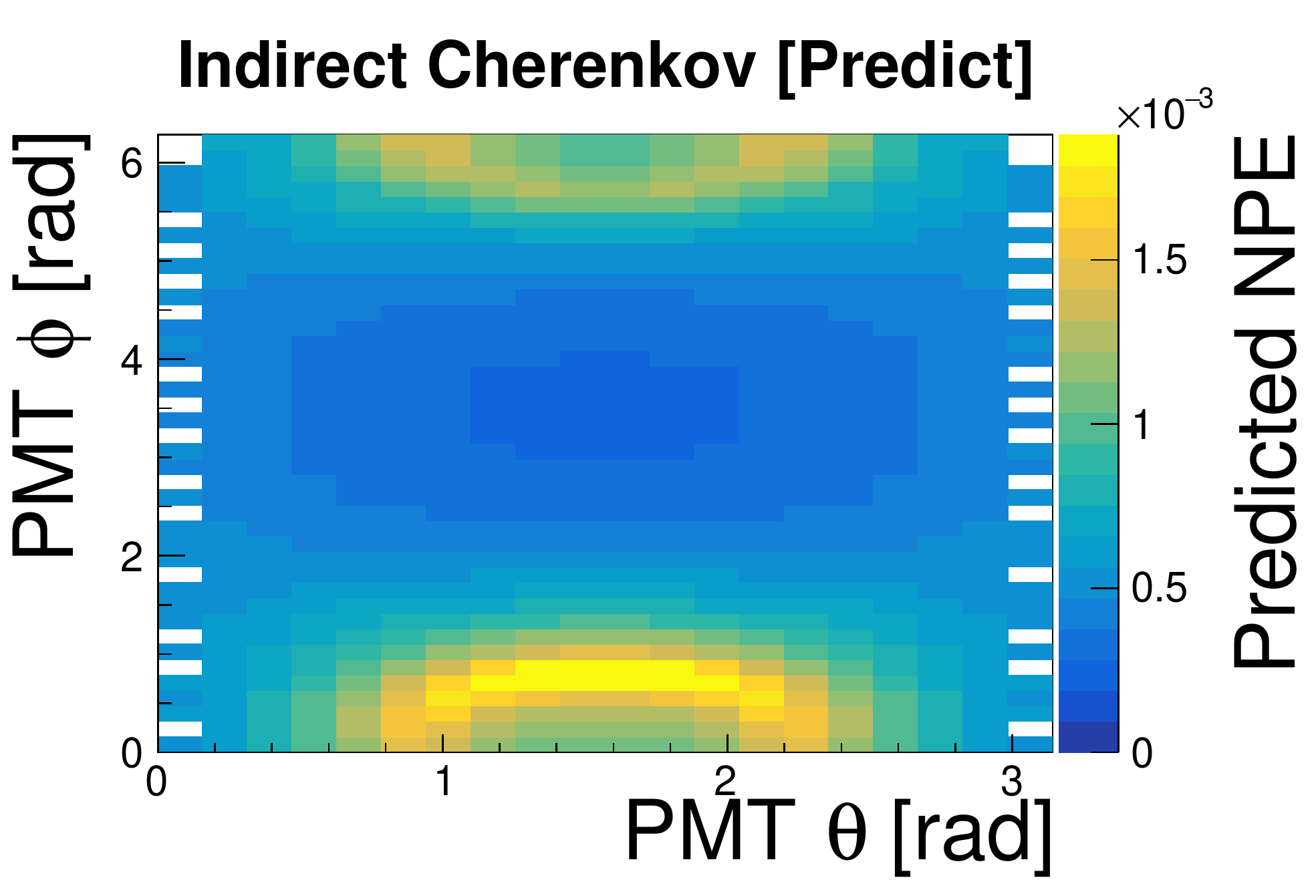}}
\subfigure{\includegraphics[scale=0.24]{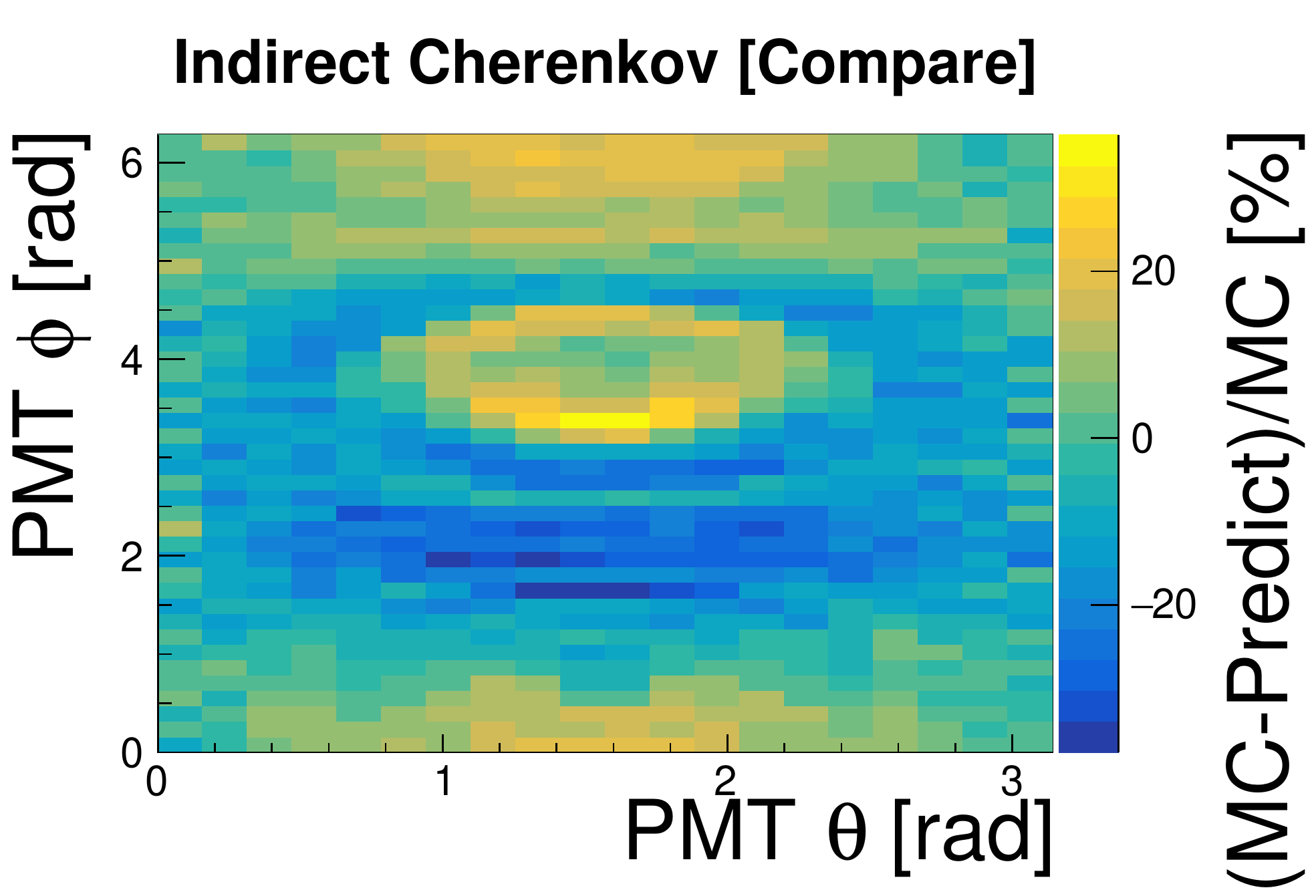}}
\caption{A total of 500,000 electrons are simulated in the direction of (1,~0,~0) at (\SI{1}{m},~\SI{1}{m},~\SI{0}) with a kinetic energy of \SI{5}{MeV}. The number of the full MC-estimated PEs and the model-predicted PEs are compared. The comparison of the direct scintillation PEs is the first row. The comparison of the direct Cherenkov PEs is the second row. The comparison of the indirect scintillation row is the third row. The comparison of the indirect Cherenkov light is the fourth row.}
\label{fig: SD ComPared2}
\end{figure}

\begin{figure}[!htbp]
\centering
\subfigure{\includegraphics[scale=0.24]{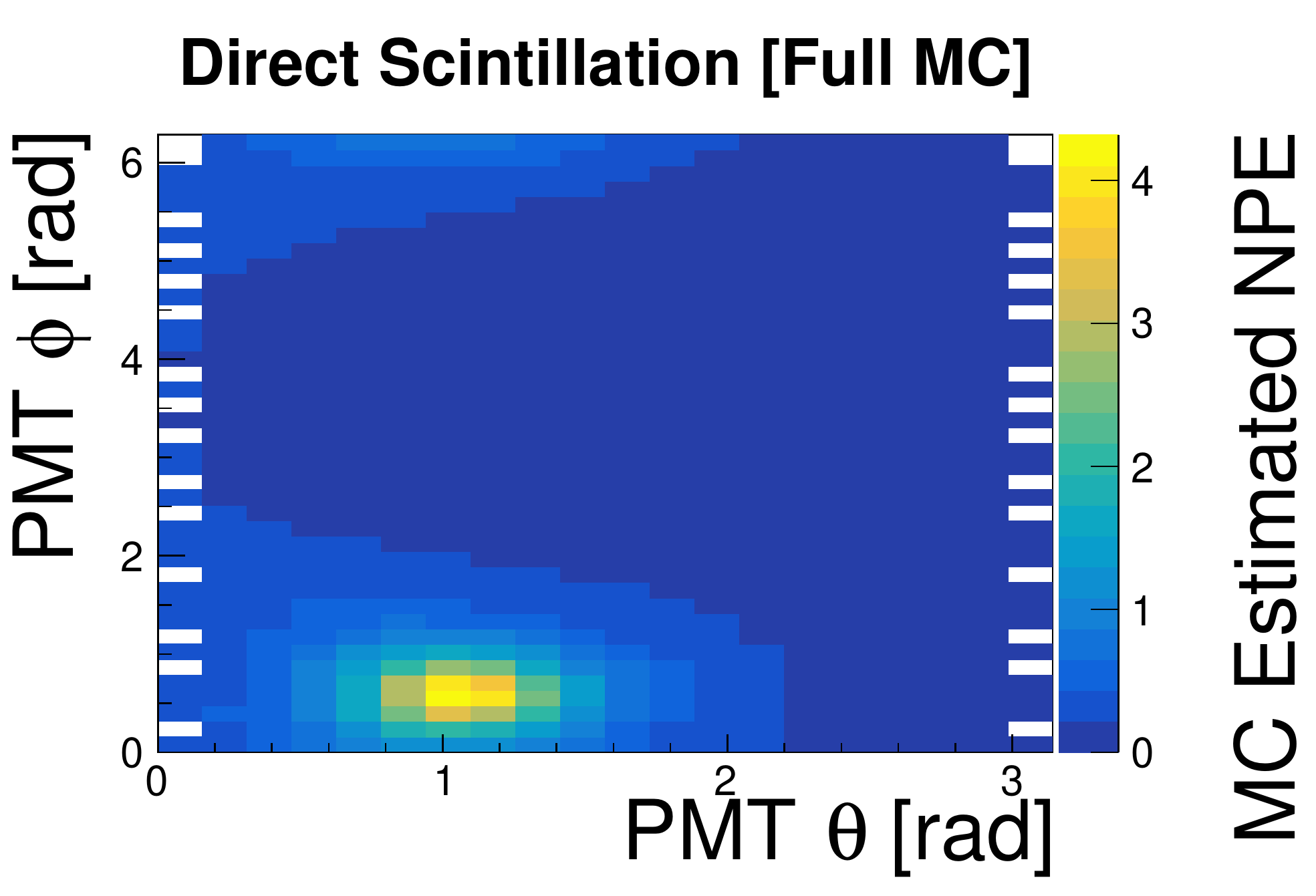}}
\subfigure{\includegraphics[scale=0.24]{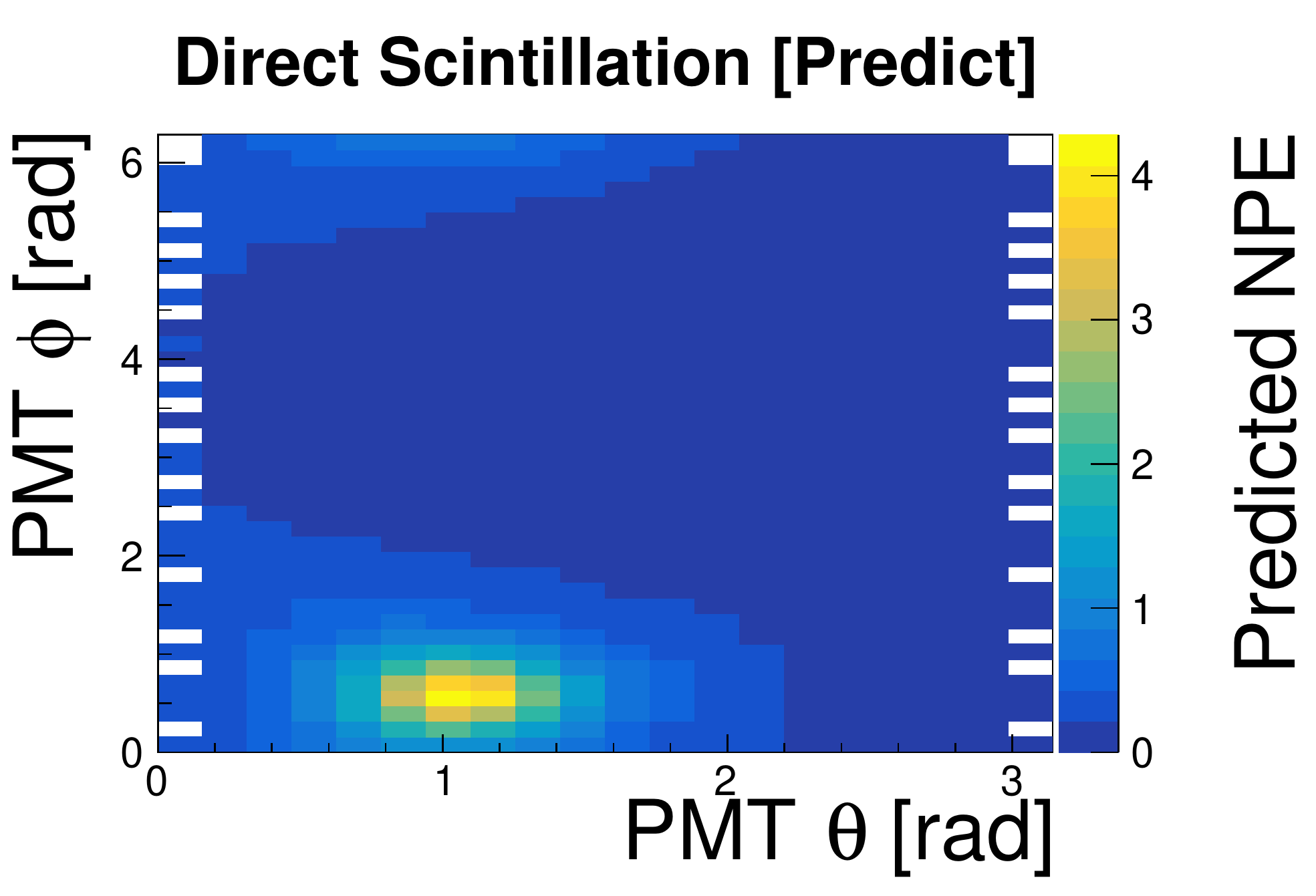}}
\subfigure{\includegraphics[scale=0.24]{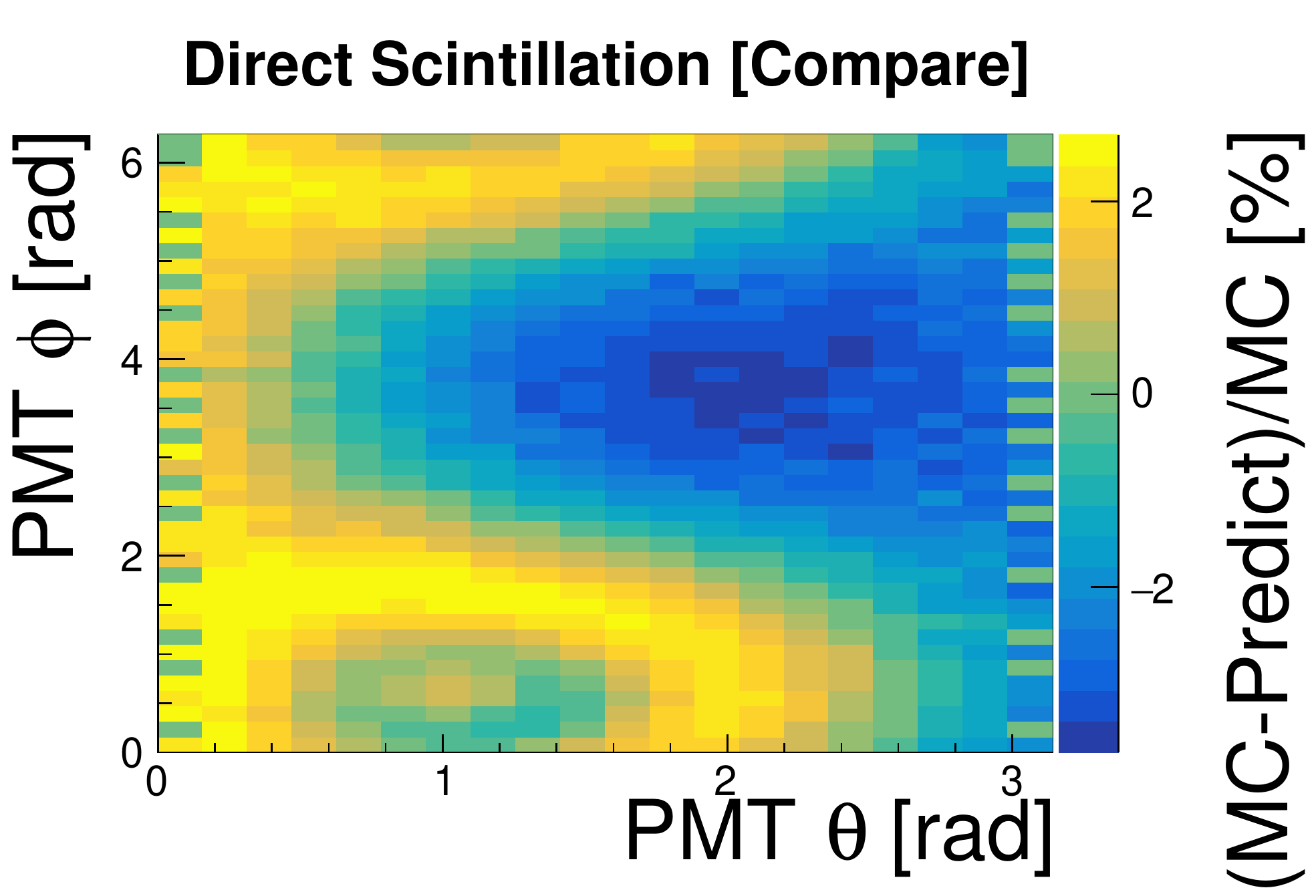}}
\subfigure{\includegraphics[scale=0.24]{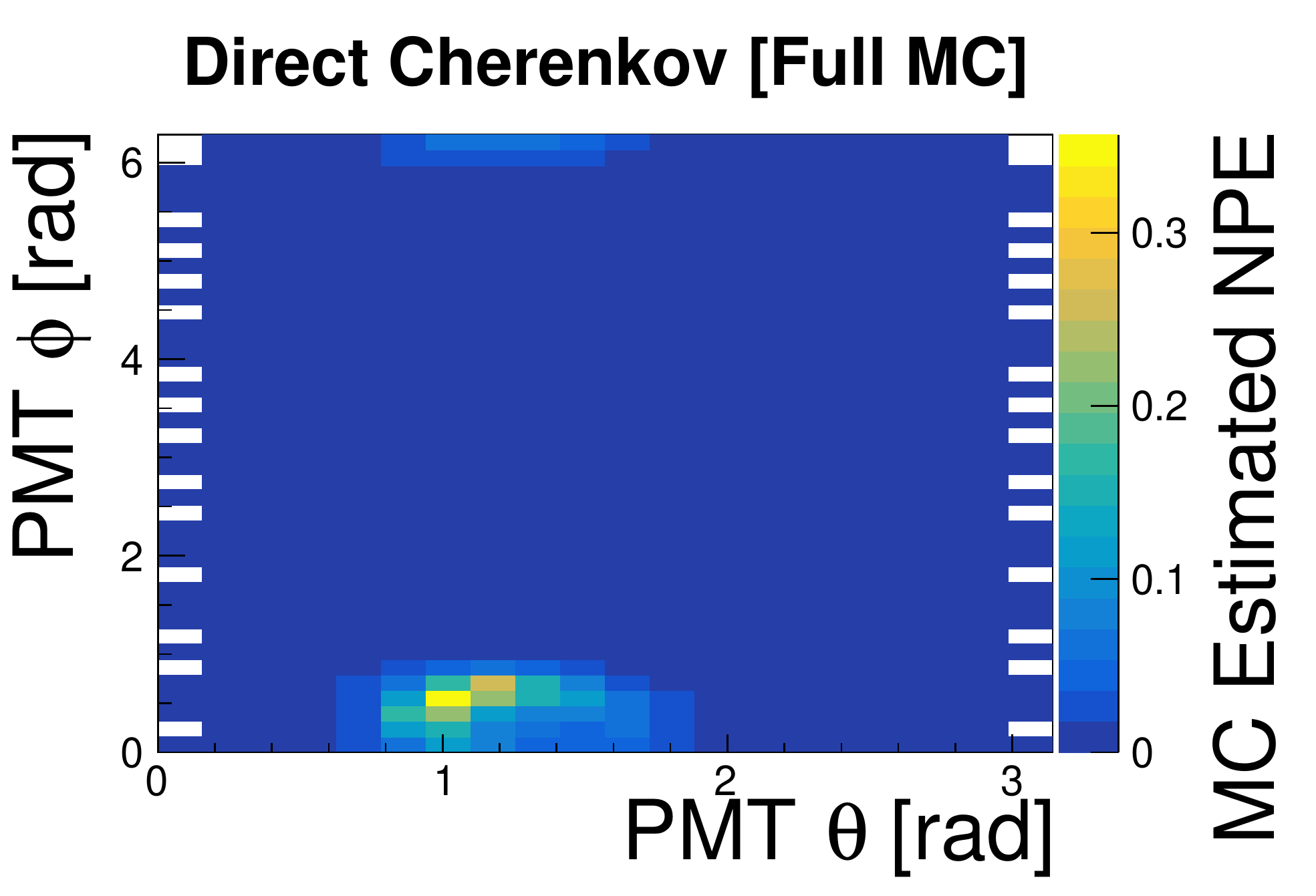}}
\subfigure{\includegraphics[scale=0.24]{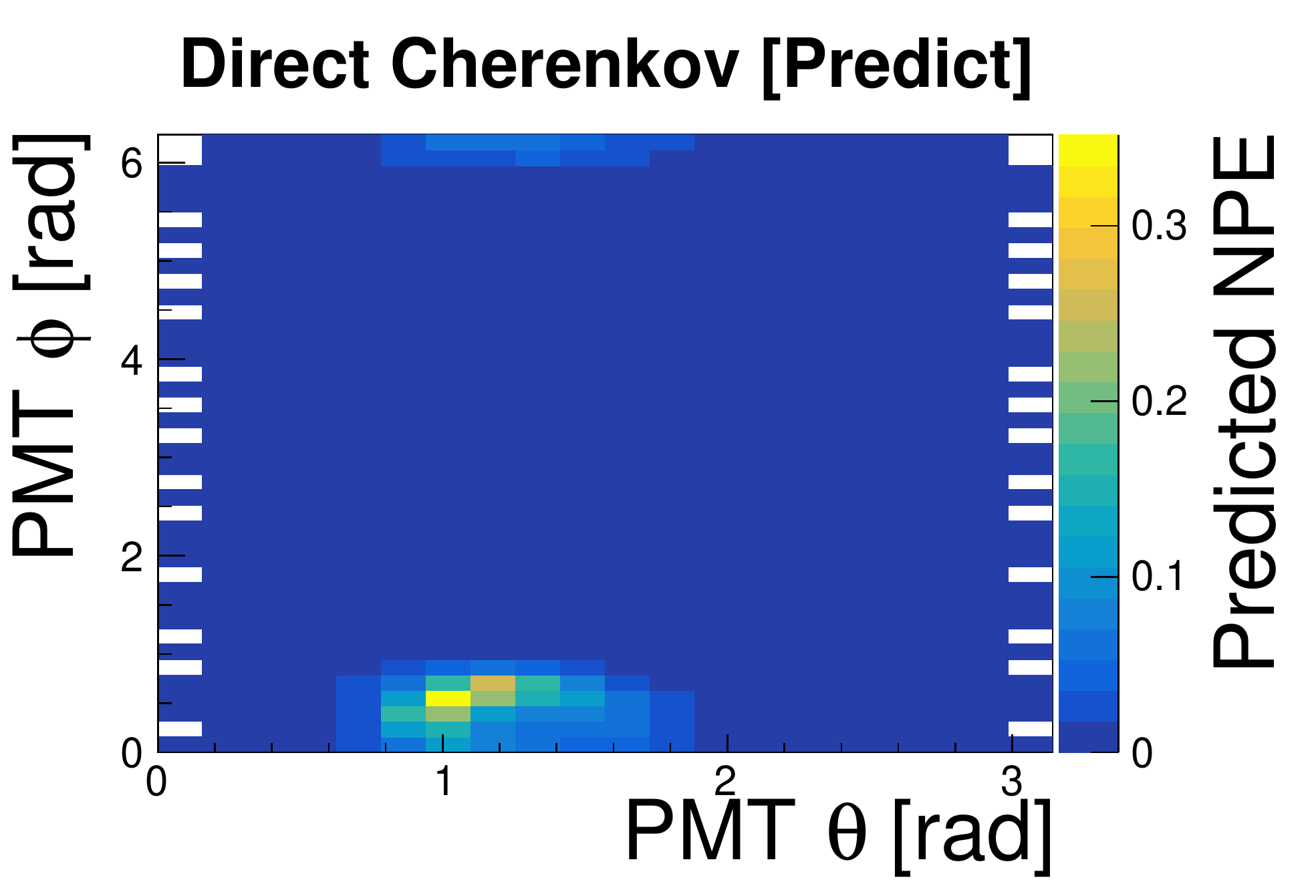}}
\subfigure{\includegraphics[scale=0.24]{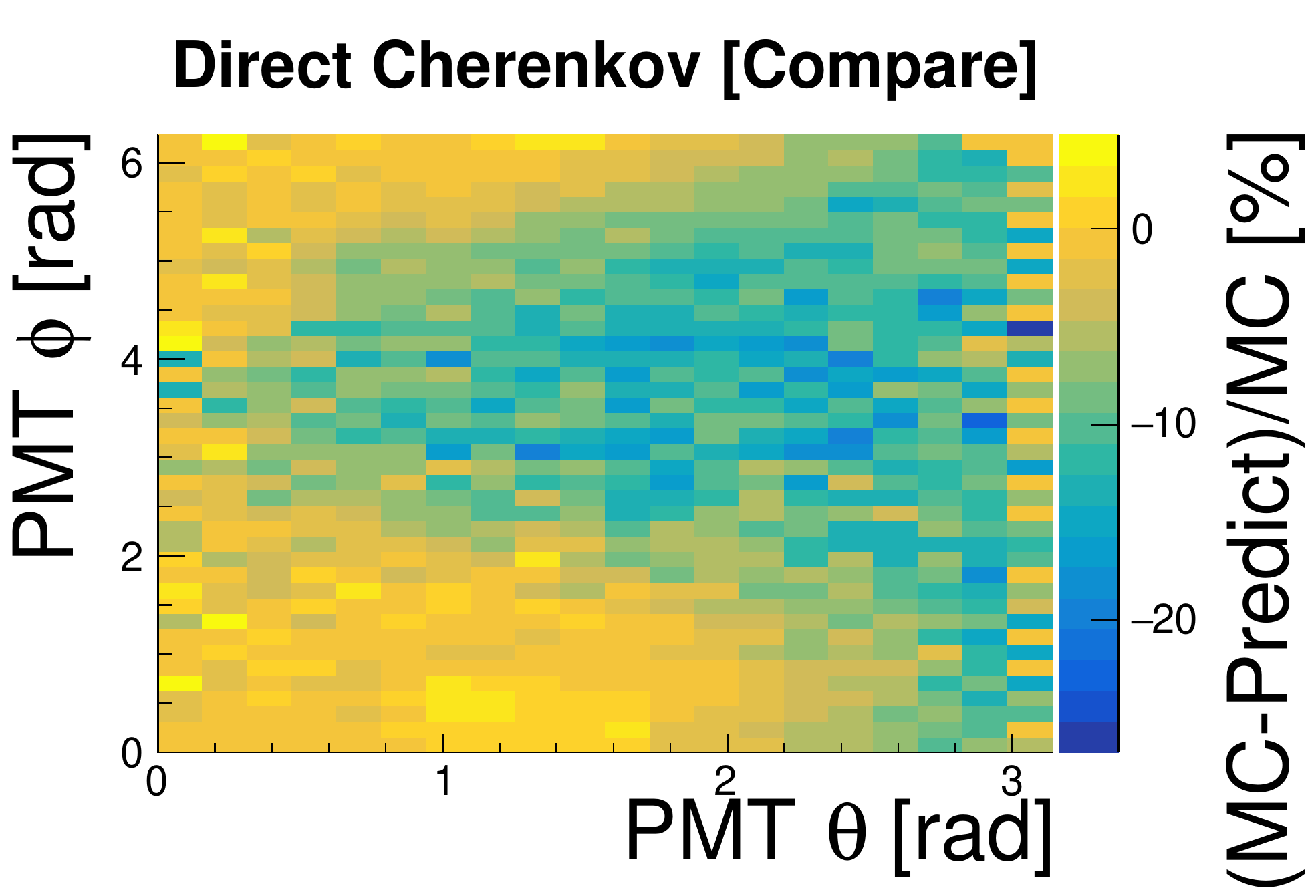}}
\subfigure{\includegraphics[scale=0.24]{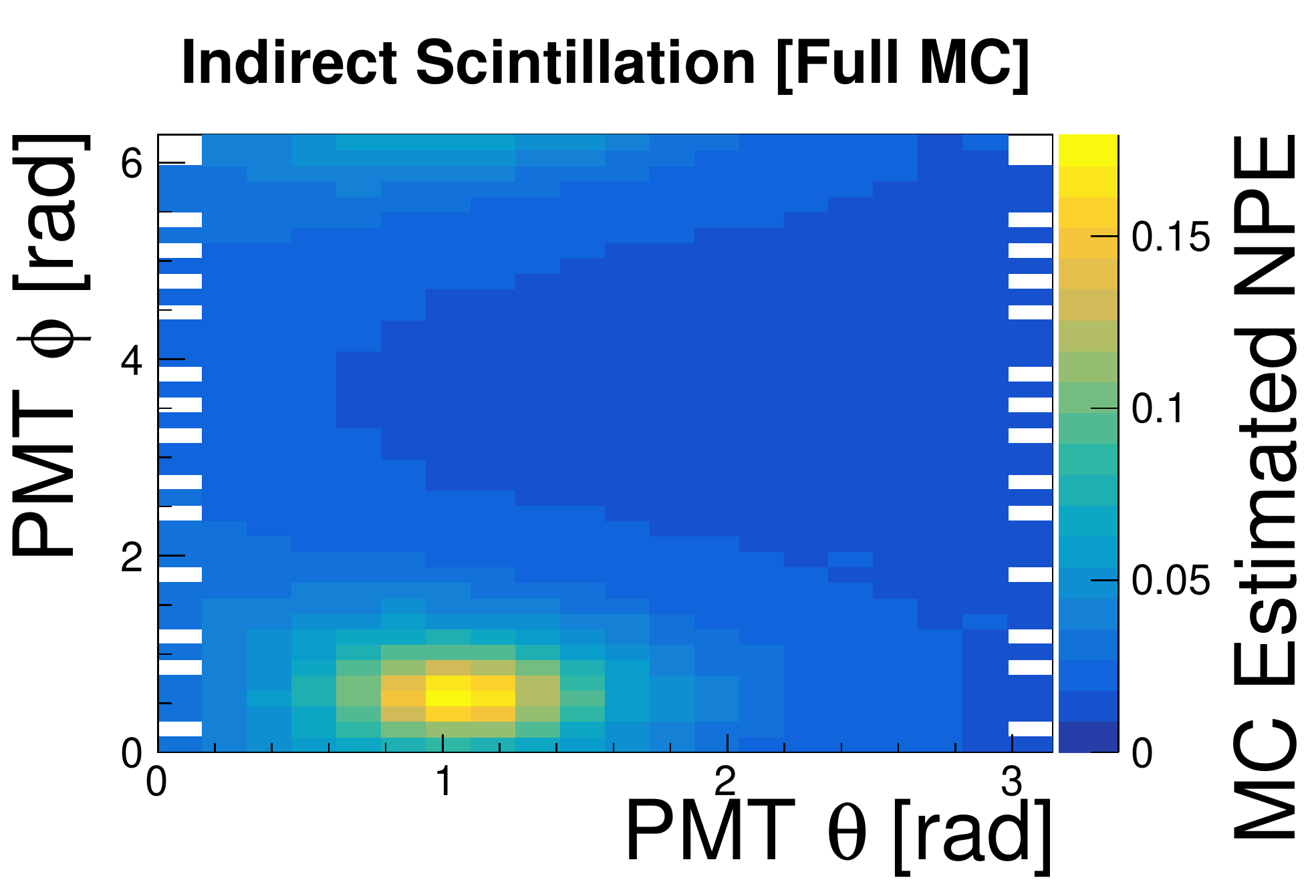}}
\subfigure{\includegraphics[scale=0.24]{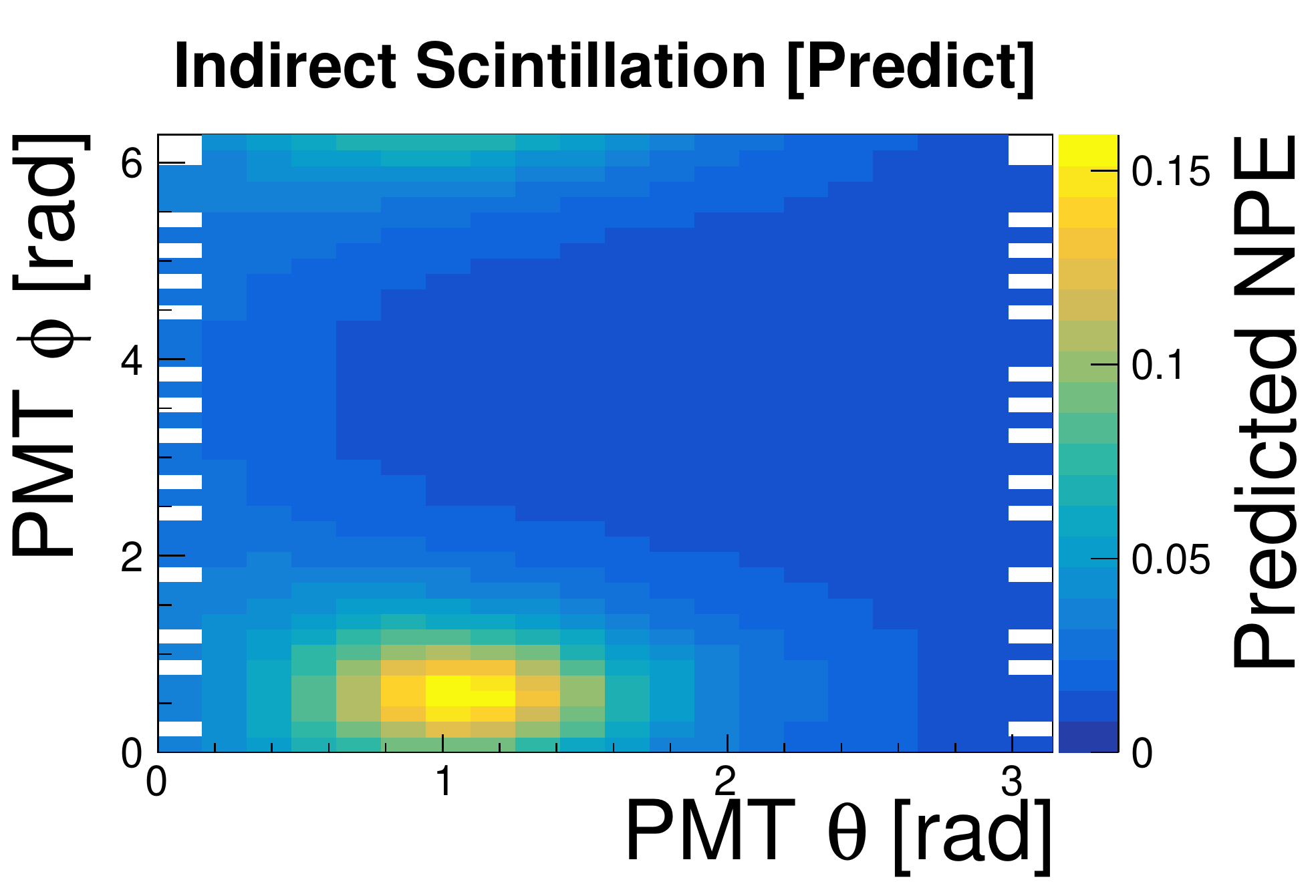}}
\subfigure{\includegraphics[scale=0.24]{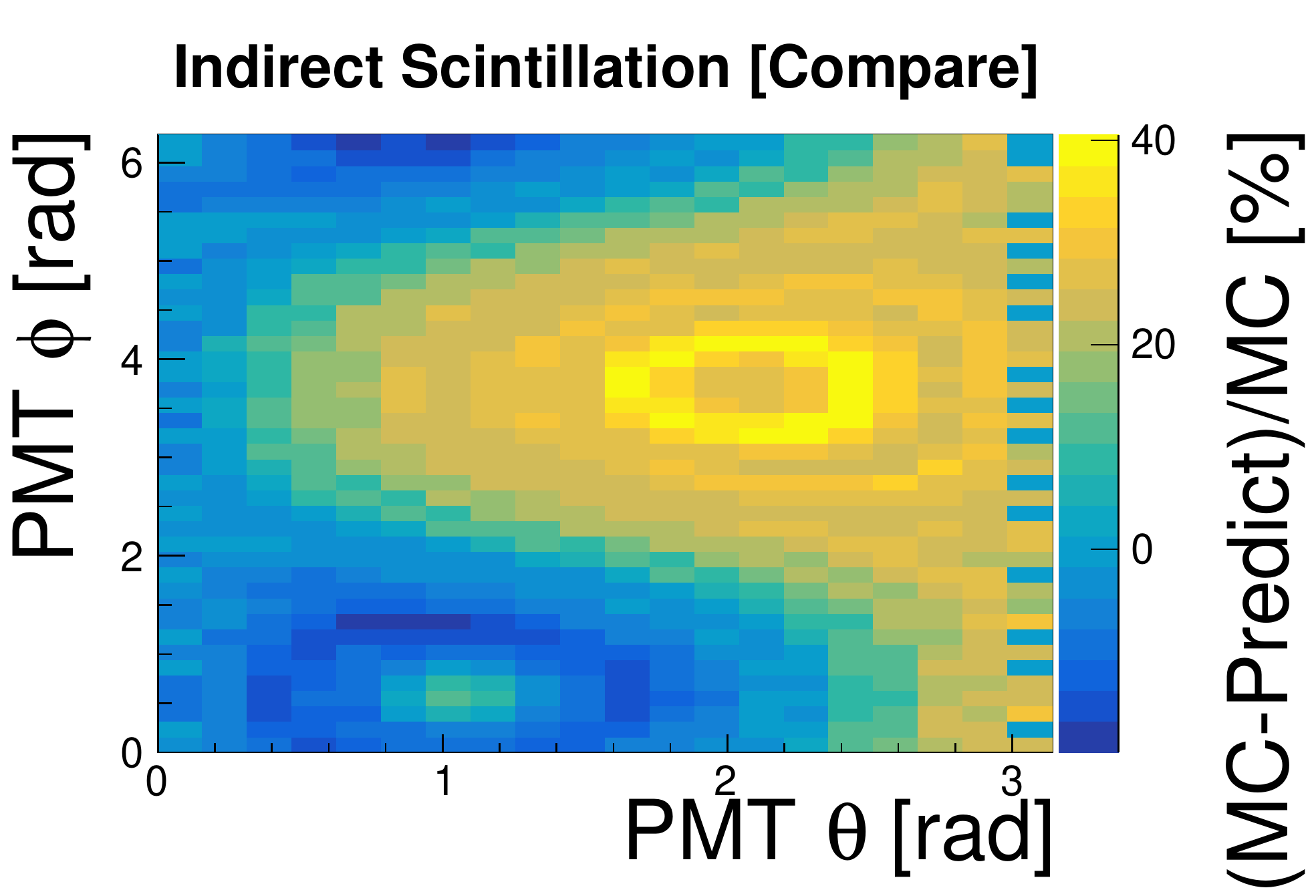}}
\subfigure{\includegraphics[scale=0.24]{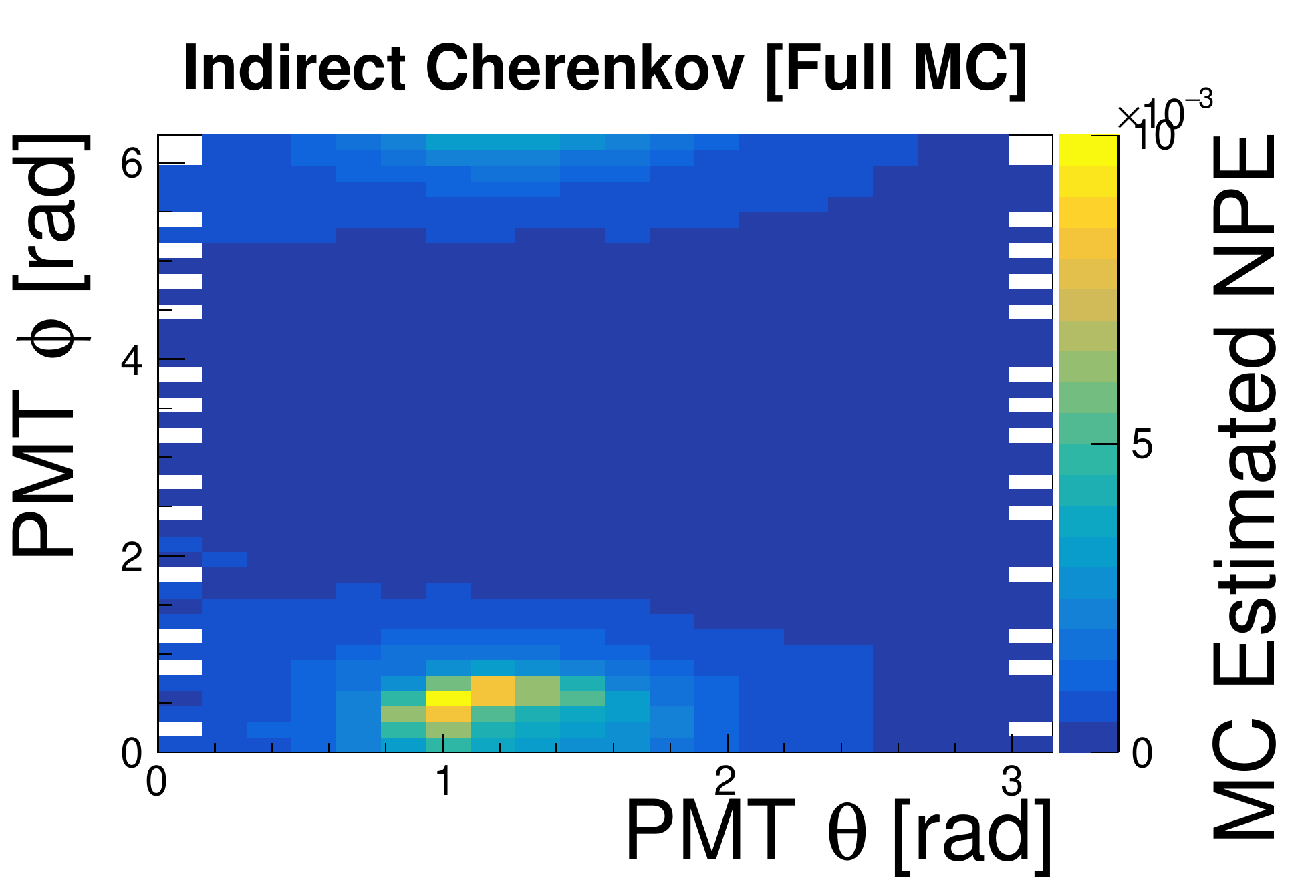}}
\subfigure{\includegraphics[scale=0.24]{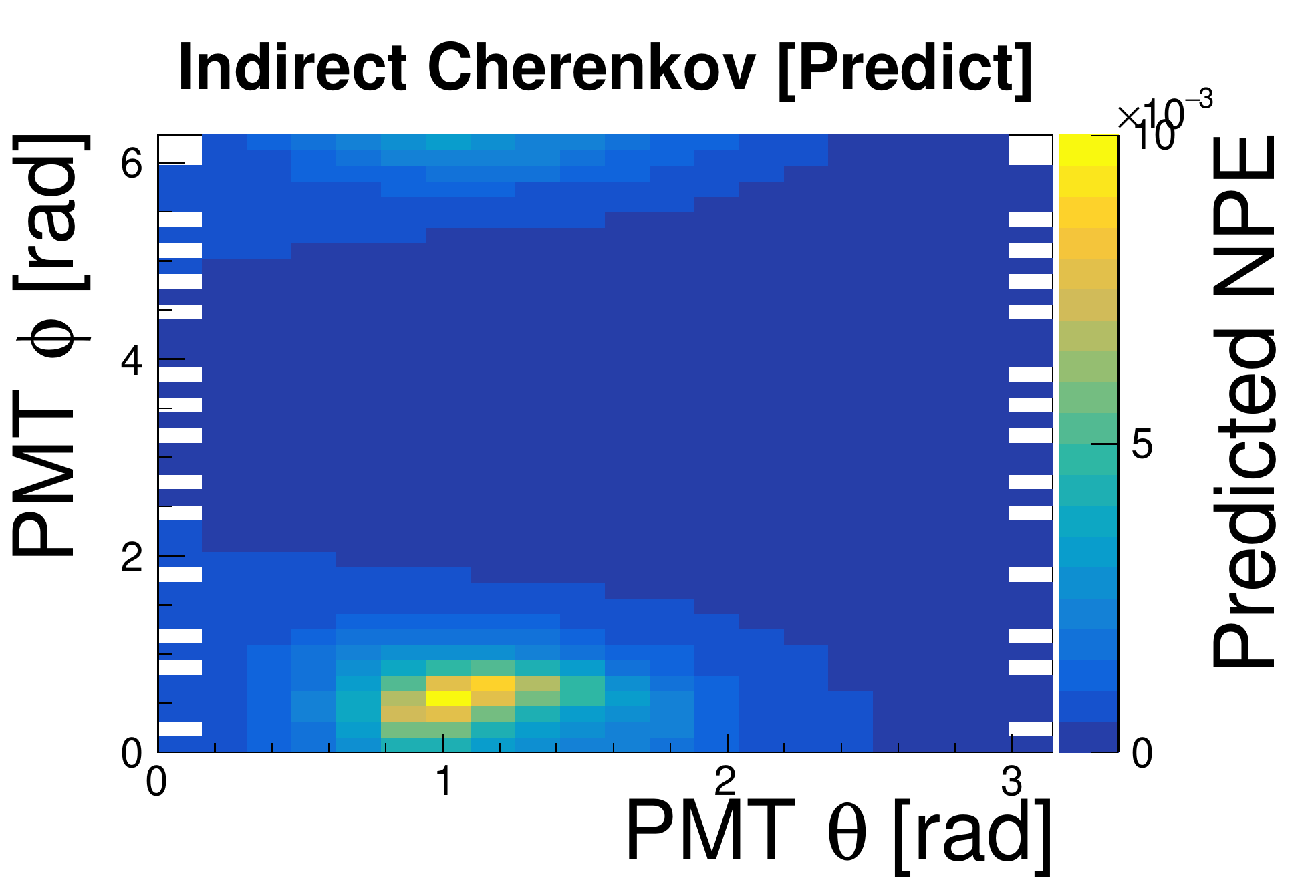}}
\subfigure{\includegraphics[scale=0.24]{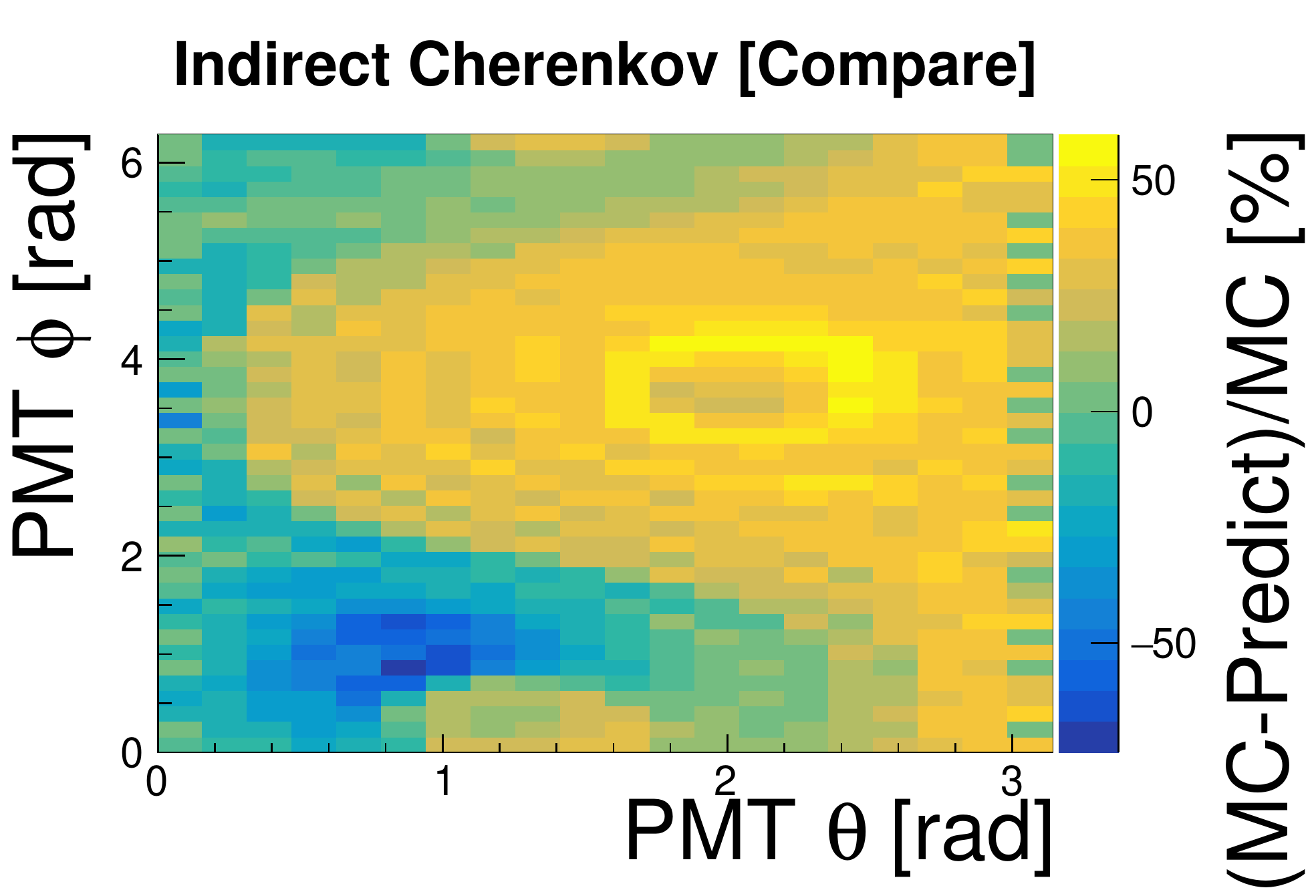}}
\caption{A total of 500,000 electrons are simulated in the direction of (1,~0,~0) at (\SI{3}{m},~\SI{2}{m},~\SI{2}{m}) with a kinetic energy of \SI{5}{MeV}. The number of the full MC-estimated PEs and the model-predicted PEs are compared. The comparison of the direct scintillation PEs is the first row. The comparison of the direct Cherenkov PEs is the second row. The comparison of the indirect scintillation row is the third row. The comparison of the indirect Cherenkov light is the fourth row.}
\label{fig: SD ComPared}
\end{figure}

The average number of PEs of a PMT is obtained by the full MC simulation. Moreover, we predict the number of PEs of this PMT by the model (section~\ref{Sec: ChargeProfile}). The number of the full MC-estimated PEs is compared with the number of the model-predicted PEs by using~$(\mathrm{MC}-\mathrm{Predict})/{\mathrm{MC}}$. To facilitate the comparison, $\theta$ and $\phi$ are used in the spherical coordinate system to represent the position of the PMTs. Figure~\ref{fig: SD ComPared2} and figure~\ref{fig: SD ComPared} display the comparison results. 

The results show that for the electrons close to the center, the difference between the MC-estimated PEs and the model-predicted PEs is better than that for the electrons at the boundary. The comparison of the four types of lights is discussed in detail below.

\paragraph{Direct Scintillation}

The first rows of figure~\ref{fig: SD ComPared2} and figure~\ref{fig: SD ComPared}, the direct scintillation PEs comparison is displayed. The differences between the predicted and the full MC estimated values for (\SI{1}{m},~\SI{1}{m},~0) and (\SI{3}{m},~\SI{2}{m},~\SI{2}{m}) are 1\% and 3\%, respectively.

\paragraph{Direct Cherenkov}

In the second rows of figure~\ref{fig: SD ComPared2} and figure~\ref{fig: SD ComPared}, the direct Cherenkov PEs comparison is displayed. The differences between the number of model-predicted and full MC-estimated PEs for (\SI{1}{m},~\SI{1}{m},~0) and (\SI{3}{m},~\SI{2}{m},~\SI{2}{m}) are less than 8\% and 25\%, respectively. In addition, in the critical zone, where the most Cherenkov PEs are collected, the discrepancies are less than 4\% and 8\%, respectively.

\paragraph{Indirect Scintillation}

In the third rows of figure~\ref{fig: SD ComPared2} and figure~\ref{fig: SD ComPared}, the comparison of the indirect scintillation PEs is displayed. The differences between the number of the model-predicted and the full MC-estimated PEs for (\SI{1}{m},~\SI{1}{m},~0) and (\SI{3}{m},~\SI{2}{m},~\SI{2}{m}) are 20\% and 40\%, respectively. 
In addition, in the critical zone, where the most indirect scintillation PEs are collected, the discrepancies are less than 20\% and 25\%, respectively.

\paragraph{Indirect Cherenkov}

In the fourth rows of figure~\ref{fig: SD ComPared2} and figure~\ref{fig: SD ComPared}, the indirect Cherenkov PEs comparison is displayed. The differences between the number of the model-predicted and the full MC-estimated PEs for (\SI{1}{m},~\SI{1}{m},~0) and (\SI{3}{m},~\SI{2}{m},~\SI{2}{m}) are 30\% and 80\%, respectively.
In addition, in the critical zone, where the most indirect Cherenkov PEs are collected, the discrepancies are less than 20\% and 40\%, respectively.

\section{Reconstruction}\label{Sec: Recon}

This section describes the reconstruction process, which includes details of the initial value calculation and the maximization of the likelihood function. The section also presents the reconstruction results, including the energy, position and angular resolution. Moreover, the PID results are also shown in this section.

\subsection{Calculation of Initial Values and Maximization}

The initial value is required for the maximum likelihood method for energy, position, event time, and direction. In the maximization process, the initial value needs to be obtained by a fast algorithm, which saves time. In addition, the maximization process can easily fall into a local extremum, so the reconstruction is iterated to finally obtain the reconstructed results.

\paragraph{Initial Value of Energy}
The total number of PEs of the event could be obtained by the results of waveform analysis, and the visible energy of the event $E_\mathrm{e}$ can be calculated as:
\begin{equation}
E_\mathrm{e} = \displaystyle\sum_{i=1}^{N_{\mathrm{hitPMT}}} n_{i}^{\mathrm{Obs}}/{E_{\mathrm{scale}}},
\end{equation}

\noindent where $E_{\mathrm{scale}}$ is the PE yield (energy scale), $N_{\mathrm{hitPMT}}$ is the total number of hit PMTs, and $n_{i}^{\mathrm{Obs}}$ is the number of PEs of the $i$th PMT.

\paragraph{Initial Value of Vertex Position}

The vertex position is calculated by using a weighted average approach from each PE of the PMT (Barycenter) and its position is shown below:
\begin{equation}
\vec{r}_{\mathrm{BC}} = A_c\cdot (\displaystyle\sum_{i=1}^{N_{\mathrm{hitPMT}}} {n_{i}^{\mathrm{Obs}}\vec{r}_i})/(\displaystyle\sum_{i=1}^{N_{\mathrm{hitPMT}}}n_{i}^{\mathrm{Obs}}),
\end{equation}

\noindent where $\vec{r}_{\mathrm{BC}}$ is the vertex position of the event, $A_c$ is a correction factor with a value of 0.93, and $\vec{r}_i$ is the position of the $i$th PMT.

\paragraph{Initial Value of Event Time}
The fit to vertex position, direction and time corresponds to changing the vertex position $(x,y,z)$, direction $\vec{d}_{\mathrm{Fit}}$, and event time $t_{\mathrm{event}}$ until the maximum number of PE time residual is concentrated at the peak of the time residual p.d.f.. Typically, $t_{\mathrm{event}}$ is between \SI{-5}{ns} and \SI{5}{ns}, so the initial value of the event time is set to 0.

\paragraph{Initial Value of Direction}

The initial values for the six standard directions are (1,~0,~0), (-1,~0,~0), (0,~1,~0), (0,~-1,~0), (0,~0,~1), and (0,~0,~-1).

\paragraph{Maximize the Likelihood Function}

The likelihood function is maximized by TMinuit~\citep{TMinuit}. 
We obtain the final reconstruction results in two steps. 

In the first step, the likelihood function is maximized six times with the six standard directions as mentioned above.

In the second step, the best-fit results of the energy, position, and event time of the first step are used as their corresponding initial values.
For the direction, it is easy to fall into a local maximum and a broader range of initial values is examined.
We further calculated 50 likelihood values with 50 directions which are uniformly distributed within a cone of 60 degrees around the best-fit direction of the first step. The three directions, with the highest likelihood values out of the 50, are chosen as the initial values of the second step fit, and the fits are carried out three times. The best fit result is selected out of the three fits. 

\subsection{Performance of Reconstruction}\label{Sec: ReconPerformance}

Electrons with 1 to \SI{9}{MeV} kinetic energy and uniform position and direction distributions are simulated for the twelve SlowLS samples. We reconstruct these events using the method described above. The results show that the larger the percentage of Cherenkov PEs (higher CSR), the simpler the reconstruction direction. We note that after-pulse has little impact on the reconstruction results of these SlowLS samples since it usually appears a few hundred nanoseconds after a signal~\citep{Zhao:2022gksAP}. In contrast, the angular reconstruction mainly depends on the forward Cherenkov light, usually in the first \SI{10}{ns}.

For the low light yield SlowLS sample reconstruction results, the samples with the fastest time profile ($\tau_r=\SI{0.5}{ns}$, $\tau_d=\SI{10}{ns}$) can reconstruct a satisfactory direction because of a higher CSR (>1: 1.3). Therefore, the reconstruction results of the slower time profile SlowLS samples are not provided. 

For the high light yield SlowLS sample reconstruction results, the samples with a quicker time profile ($\tau_r=\SI{0.5}{ns}$, $\tau_d=\SI{10}{ns}$ and $\tau_r=\SI{1.2}{ns}$, $\tau_d=\SI{27}{ns}$) are unable to reconstruct the direction. Because the fast rise time leads to a small CSR (<1: 6.1) in the first \SI{10}{ns}, the angular reconstruction is very challenging.

As a result, the $\mathrm{LS_{\mathrm{24~PE}}^{0.5~ns}}$, $\mathrm{LS_{\mathrm{35~PE}}^{0.5~ns}}$, $\mathrm{LS_{\mathrm{285~PE}}^{\mathrm{7.7~ns}}}$, and $\mathrm{LS_{\mathrm{376~PE}}^{\mathrm{7.7~ns}}}$ reconstruction results are displayed in detail. In addition, the direction of the reconstruction can be more accurate when the CSR in \SI{10}{ns} is greater than 1:~5. This is a guideline for the selection of the SlowLS candidates in the CSD.

\subsubsection{Energy Resolution}

\begin{figure}[!htbp]
\centering
\subfigure{\includegraphics[scale=0.37]{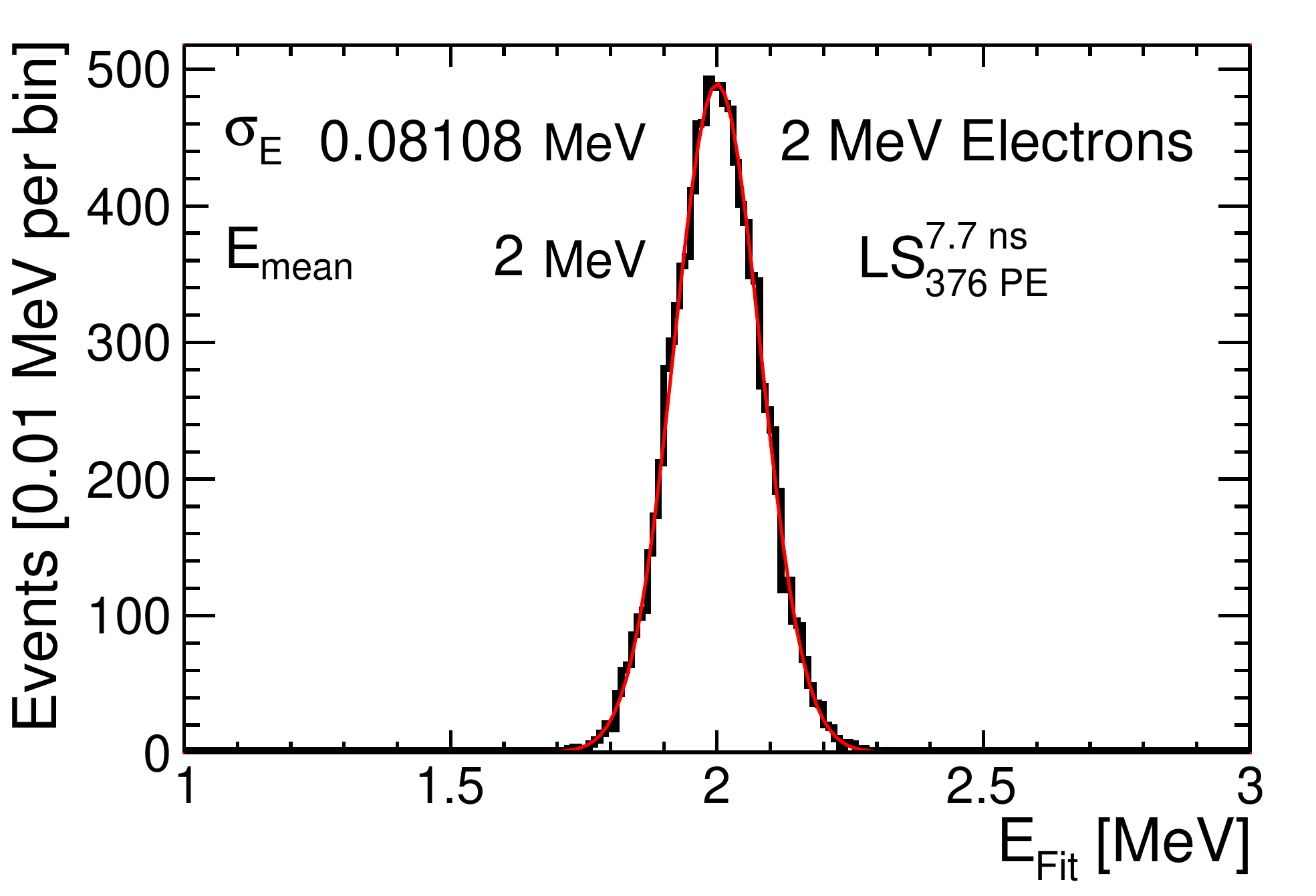}}
\subfigure{\includegraphics[scale=0.37]{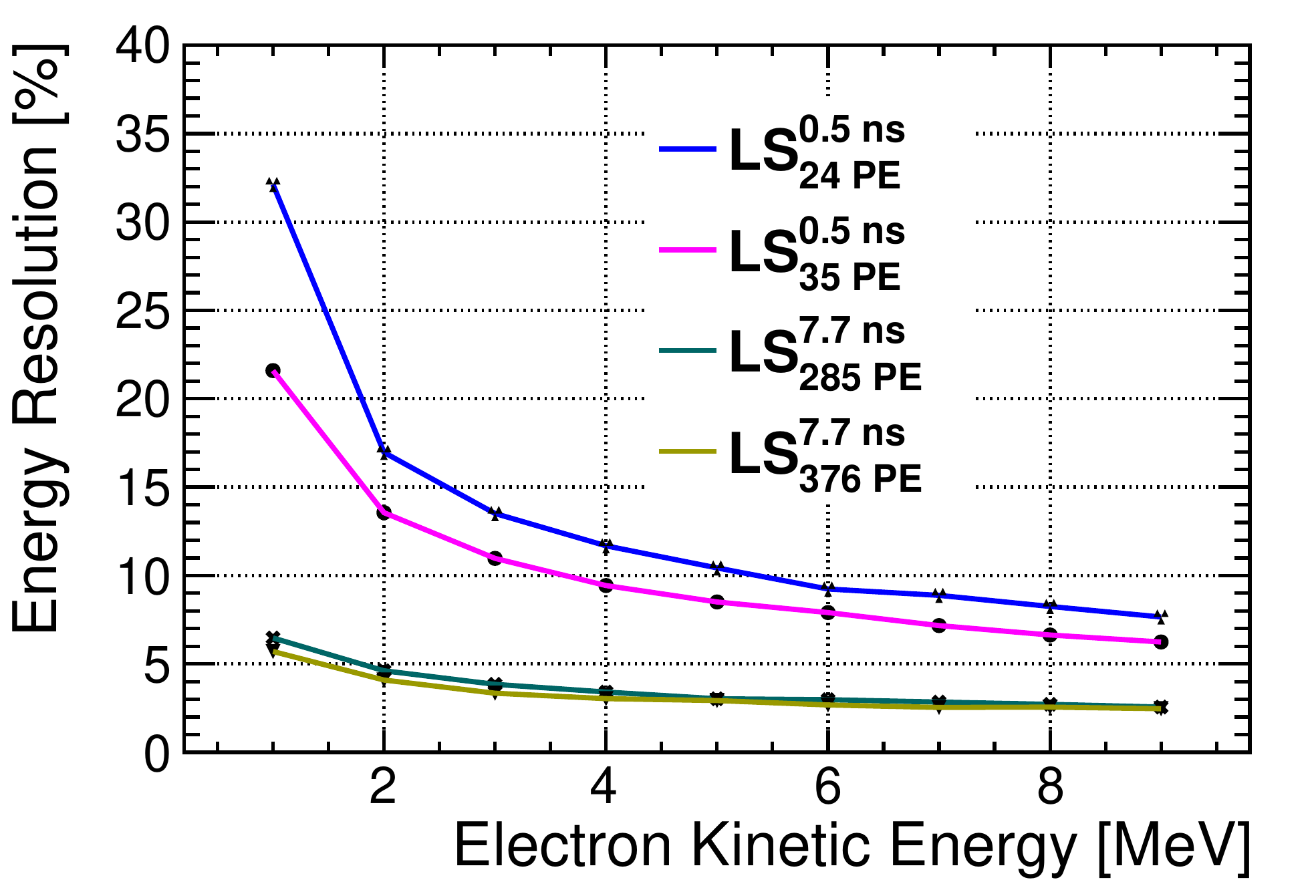}}
\caption{Left: An example of the distribution of reconstructed energy. For \SI{2}{MeV} electrons of the $\mathrm{LS_{\mathrm{376~PE}}^{\mathrm{7.7~ns}}}$, the energy resolution is approximately 4.1\%. Right: The distribution of energy resolution versus electron kinetic energy for four SlowLS samples is plotted.}
\label{Fig: ERecon}
\end{figure}

The energy distributions are obtained from the reconstructed result. The left side of figure~\ref{Fig: ERecon} is an example of a energy distribution with the $\mathrm{LS_{\mathrm{376~PE}}^{\mathrm{7.7~ns}}}$ for \SI{2}{MeV} electrons. A Gaussian function is used to fit this distribution. The definition of energy resolution is obtained by $\mathrm{\sigma_{E}/E_{mean}}$, where $\mathrm{\sigma_{E}}$ and $\mathrm{E_{mean}}$ are the standard deviation and central value in the Gaussian fit. As a result, the left panel of figure~\ref{Fig: ERecon} shows that the $\mathrm{LS_{\mathrm{376~PE}}^{\mathrm{7.7~ns}}}$ at \SI{2}{MeV} has an energy resolution of 4.1\%. 

As seen on the right side of figure~\ref{Fig: ERecon}, the energy resolution versus electron kinetic energy is plotted for four samples. It is evident that as the electron energy and scintillation light yield increase, the energy resolution improves. 

The energy of \SI{2}{MeV} is the region considered for $0 \nu \beta \beta$ decay research ($^{136}\rm{Xe}$ and $^{130}\rm{Te}$)~\citep{KamLAND-Zen2022,caccianiga2000neutrinoless,andreotti2011130te}. Therefore, the energy resolution of the \SI{2}{MeV} kinetic energy electrons with uniform position and direction distributions could be calculated by table~\ref{Tab: Conlusion}.

\begingroup
\setlength{\tabcolsep}{9pt} % Default value: 6pt
\renewcommand{\arraystretch}{1.35} % Default value: 1
\begin{table}[!htbp]
\centering
\begin{tabular}{@{}cccccc@{}}
\toprule
\multirow{2}{*}{\begin{tabular}[c]{@{}c@{}}\\SlowLS\\ \\Samples\end{tabular}}
&
\multirow{2}{*}{\begin{tabular}[c]{@{}c@{}}\\PE yield\\$E_{\mathrm{scale}}$\\(PE/MeV)\end{tabular}}
&
\multirow{2}{*}{\begin{tabular}[c]{@{}c@{}}\\CSR \\in the first \\\SI{10}{ns}\end{tabular}}
&
\multirow{2}{*}{\begin{tabular}[c]{@{}c@{}}\\Energy\\ \\Resolution\end{tabular} }
&
\multicolumn{2}{c}{Results with \SI{2}{MeV} Electrons} \\\cmidrule(l){5-6}
&
&  
 &
 & \begin{tabular}[c]{@{}c@{}}Position \\Resolution\\X-Projection\\(cm)\end{tabular} 
 & \begin{tabular}[c]{@{}c@{}}Angular \\Resolution\\ Containing 68\% Events\\(degrees)\end{tabular} \\ \midrule

 $\mathrm{LS_{\mathrm{24~PE}}^{0.5~ns}}$& 24 &\begin{tabular}[c]{@{}l@{}} 1:~0.7 \end{tabular} &$28.4\% /\sqrt{(E)}$ &15.3&51 \\ 
        
        $\mathrm{LS_{\mathrm{35~PE}}^{0.5~ns}}$ &35 & \begin{tabular}[c]{@{}l@{}} 1:~1.3 \end{tabular} & 19.6\%$/\sqrt{(E)}$ &16.6&57\\

         $\mathrm{LS_{\mathrm{285~PE}}^{\mathrm{7.7~ns}}}$ &285&\begin{tabular}[c]{@{}l@{}} 1:~2.4\end{tabular} & 6.5\%$/\sqrt{(E)}$ &10.0 &45\\ 
         
        $\mathrm{LS_{\mathrm{376~PE}}^{\mathrm{7.7~ns}}}$ &376 & \begin{tabular}[c]{@{}l@{}} 1:~3.2\end{tabular} & 5.7\%$/\sqrt{(E)}$ &9.2 &47\\ \bottomrule
\end{tabular}
\caption{The reconstruction results of the four SlowLS samples for \SI{2}{MeV} electrons with uniform position and direction distributions.}
\label{Tab: Conlusion}
\end{table}
\endgroup

\subsubsection{Vertex Position Resolution}

\begin{figure}[!htbp]
\centering
\subfigure{\includegraphics[scale=0.37]{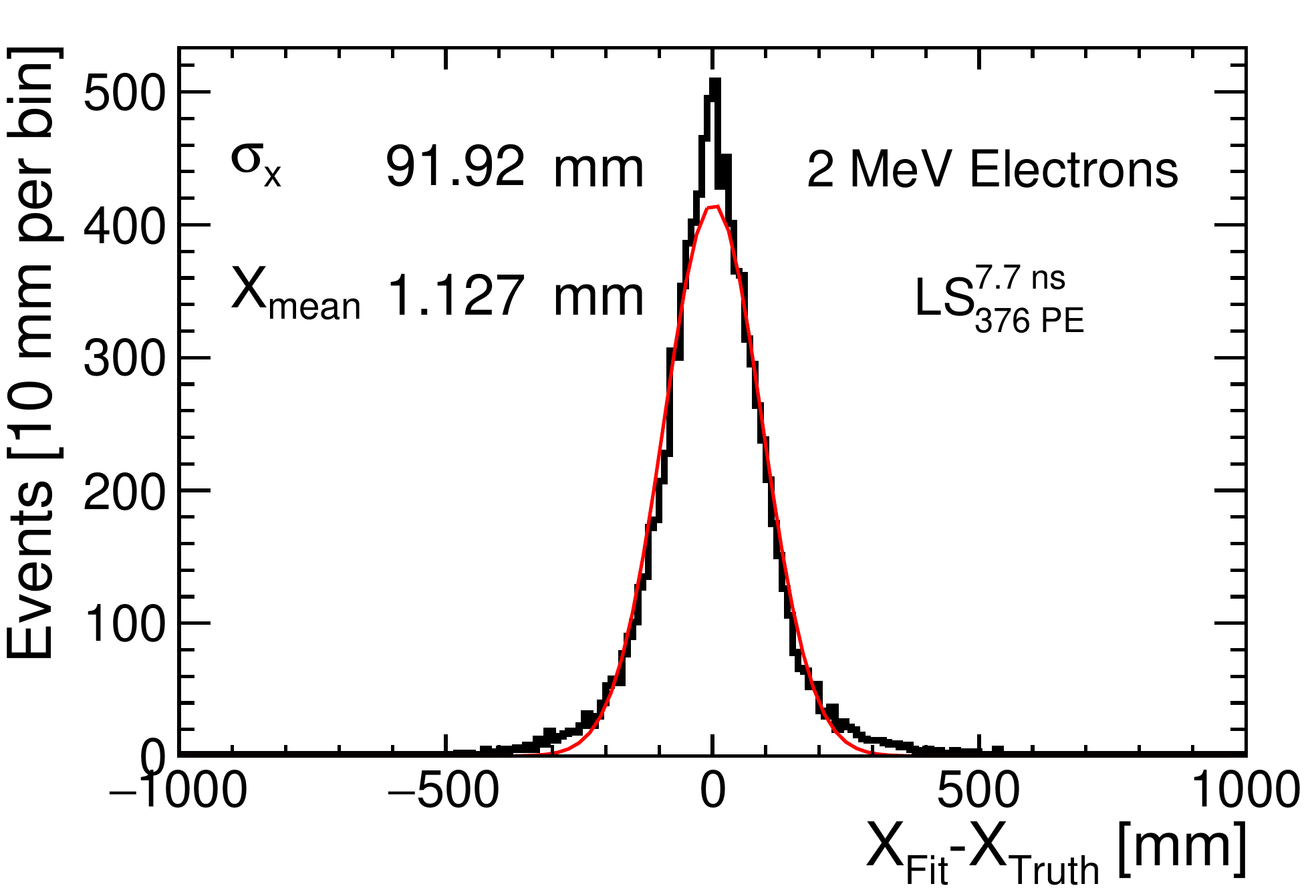}}
\subfigure{\includegraphics[scale=0.37]{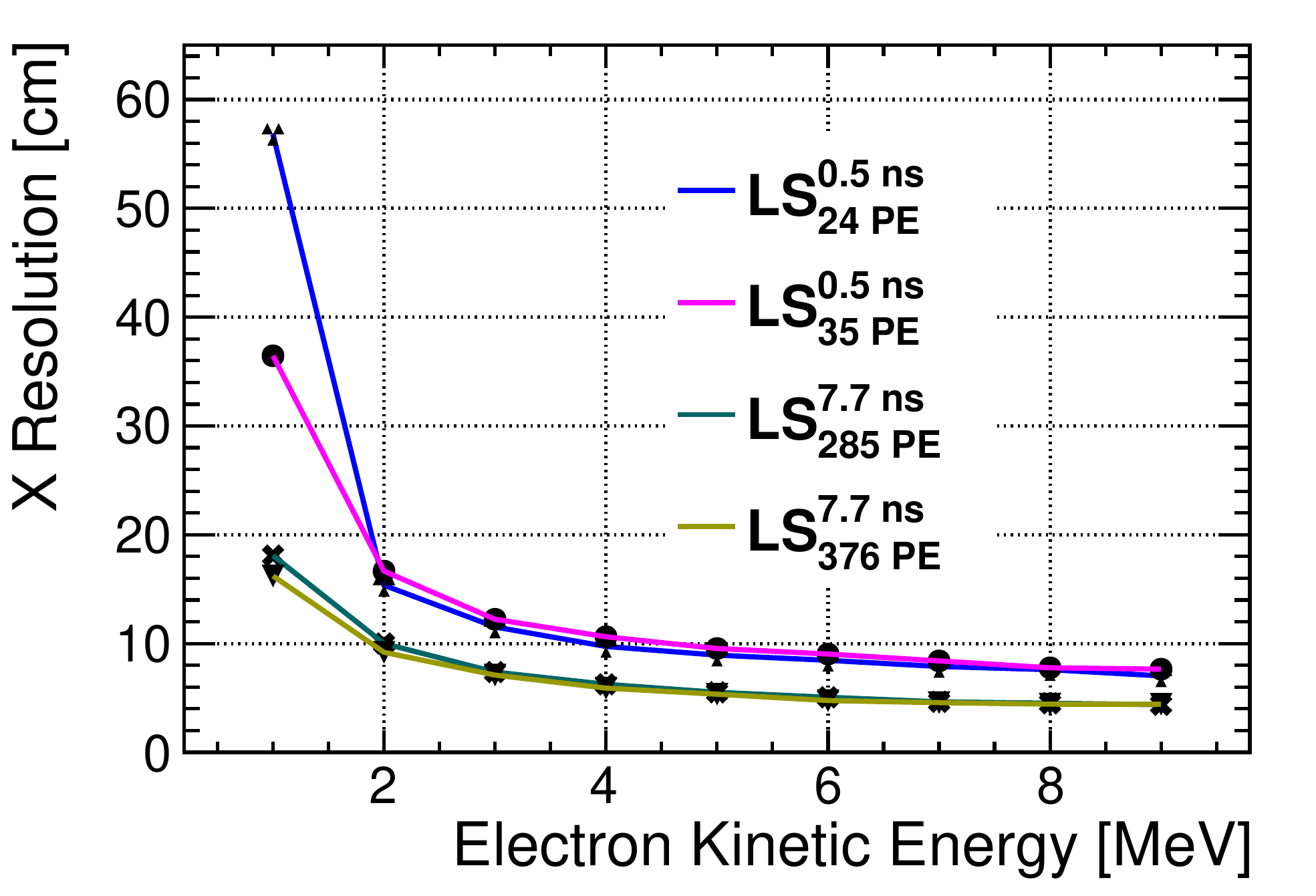}}
\caption{Left: A vertex position (X-projection) reconstruction example, which is for \SI{2}{MeV} electrons of $\mathrm{LS_{\mathrm{376~PE}}^{\mathrm{7.7~ns}}}$, and the position resolution is \SI{9.2}{cm}. Right: The resolution of the vertex position (X-projection) versus electron kinetic energy for four SlowLS samples.}
\label{Fig: XRecon}
\end{figure}

The vertex position distribution, e.g.,~X-projection, is obtained by subtracting the true position $X_{\mathrm{Truth}}$ from the reconstructed position $X_{\mathrm{Fit}}$. The left side of figure~\ref{Fig: XRecon} shows an example of a simulated sample. A Gaussian function is used to fit this distribution. The vertex position resolution is the fit standard deviation, $\sigma_x$, of the Gaussian function. As a result, the left panel of figure~\ref{Fig: XRecon} shows that the position resolution of the $\mathrm{LS_{\mathrm{376~PE}}^{\mathrm{7.7~ns}}}$ at \SI{2}{MeV} is \SI{9.2}{cm}.

The right side of figure~\ref{Fig: XRecon} shows the distribution of the vertex position versus electron kinetic energy of the four samples. As can be observed, the vertex position resolution improves as the electron energy increases, and similarly, the scintillation light yield increases as the vertex position resolution improves. 
However, the position resolution at \SI{1}{MeV} is poor for each of them. This is because of the high proportion of signal numbers due to the dark noise, which makes the reconstructed position resolution is poor.

In addition, $\mathrm{LS_{\mathrm{24~PE}}^{0.5~ns}}$ and $\mathrm{LS_{\mathrm{35~PE}}^{0.5~ns}}$ cross at \SI{2}{MeV},
because the better angular resolution of $\mathrm{LS_{\mathrm{24~PE}}^{0.5~ns}}$ (section~\ref{Sec: AngleRes}) improves the position resolution.

As a result, the vertex position resolution of the four samples is shown in table~\ref{Tab: Conlusion} for \SI{2}{MeV} kinetic energy electrons with uniform position and direction distributions, for the interest in $0 \nu \beta \beta$ decay research.

\subsubsection{Angular Resolution}\label{Sec: AngleRes}
\begin{figure}[!htbp]
\centering
\subfigure{\includegraphics[scale=0.37]{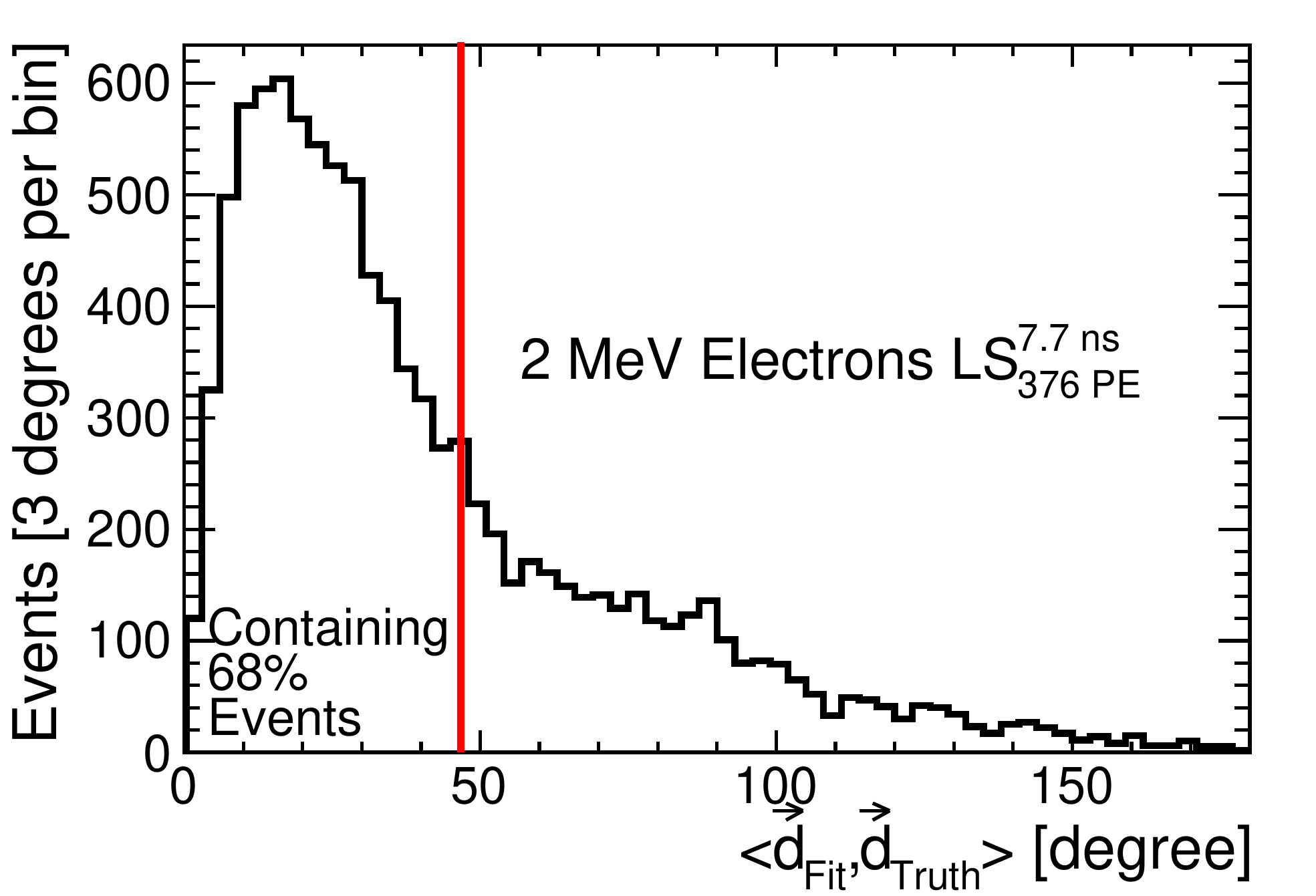}}
\subfigure{\includegraphics[scale=0.37]{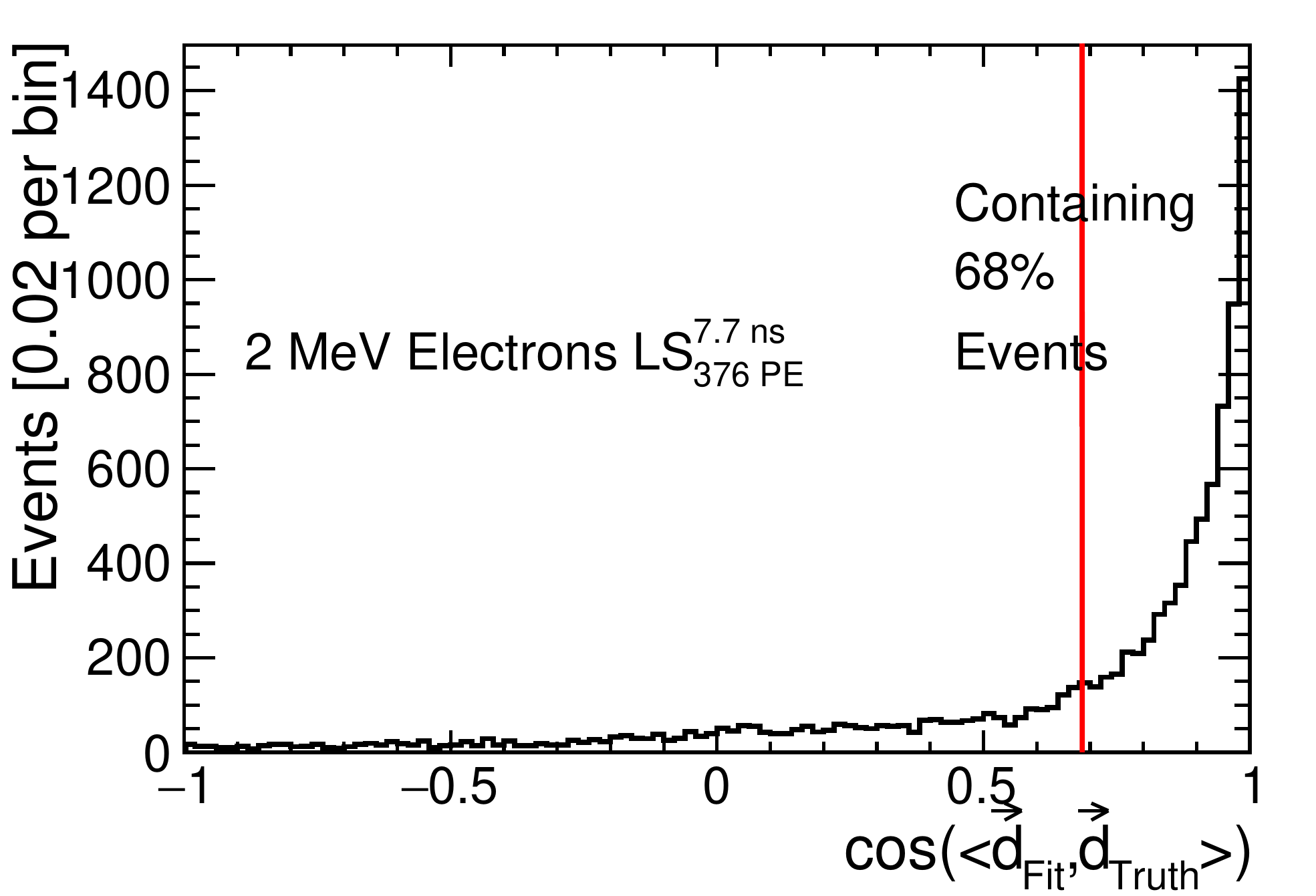}}

\subfigure{\includegraphics[scale=0.37]{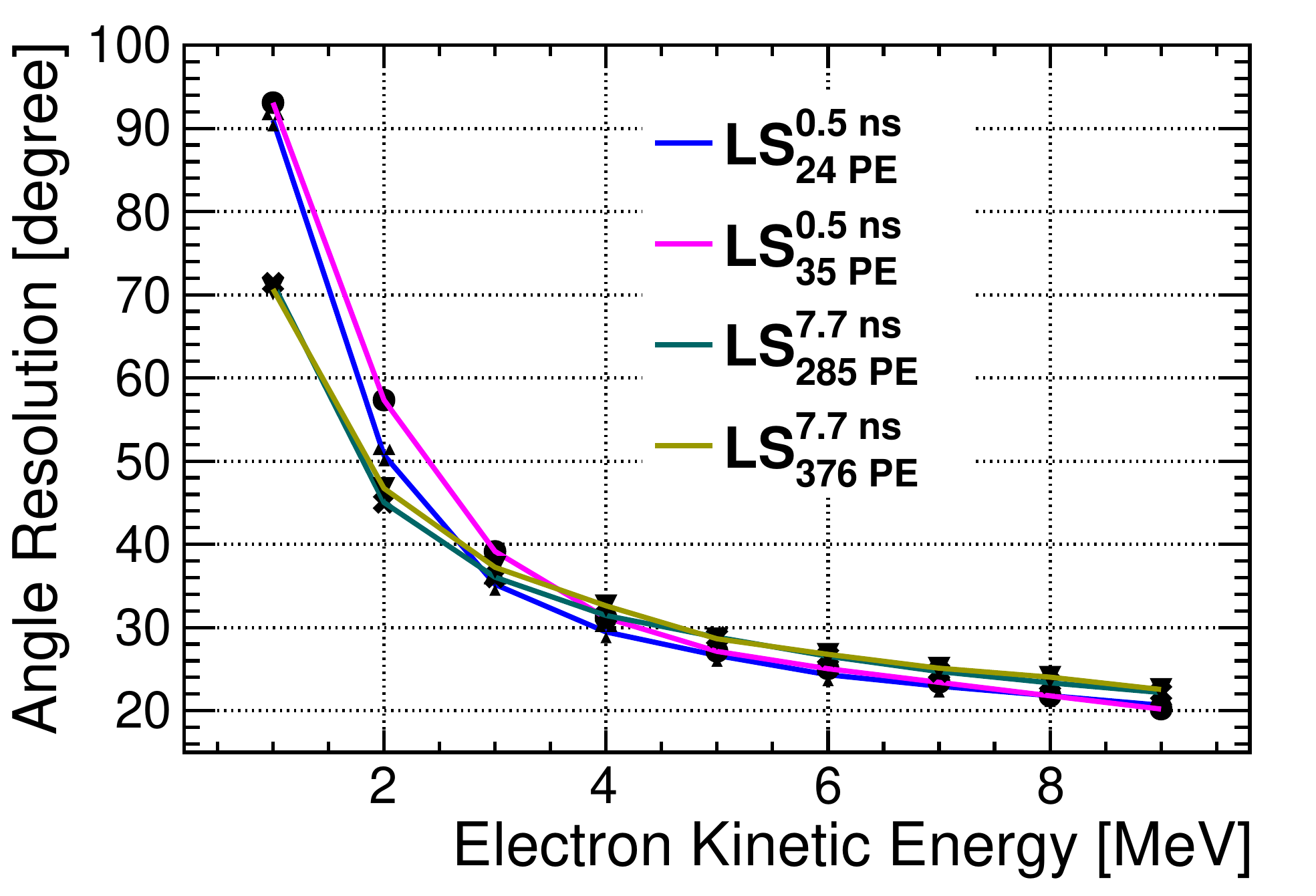}}
\caption{Top left: A sample of the reconstructed angular distribution. The X-axis displays the angle between the reconstructed direction, $\vec{d}_{\mathrm{Fit}}$, and the true direction, $\vec{d}_{\mathrm{Truth}}$. The angular resolution for \SI{2}{MeV} electrons of $\mathrm{LS_{\mathrm{376~PE}}^{\mathrm{7.7~ns}}}$ is approximately 47~degrees, which contains 68\% events. Top right: A sample of the reconstructed angular distribution with cosine angle. The X-axis displays the cosine of the angle between the $\vec{d}_{\mathrm{Fit}}$ and the $\vec{d}_{\mathrm{Truth}}$. Bottom: The distribution of angular resolution versus electron kinetic energy for four SlowLS samples.}
\label{Fig: AngleRecon}
\end{figure}

The angular distribution of the events can be obtained using the inner product of the reconstructed direction, $\vec{d}_{\mathrm{Fit}}$, and the true direction, $\vec{d}_{\mathrm{Truth}}$.
The top left panel of figure~\ref{Fig: AngleRecon} shows an angular distribution of the $\mathrm{LS_{\mathrm{376~PE}}^{\mathrm{7.7~ns}}}$ at \SI{2}{MeV}. The angular resolution is the angle that includes 68\% of the events. In addition, the top right panel of figure~\ref{Fig: AngleRecon} shows the cosine angle distribution of the $\mathrm{LS_{\mathrm{376~PE}}^{\mathrm{7.7~ns}}}$ at \SI{2}{MeV}. As a result, the bottom panel of figure~\ref{Fig: AngleRecon} shows that the angular resolution of the $\mathrm{LS_{\mathrm{376~PE}}^{\mathrm{7.7~ns}}}$ at \SI{2}{MeV} is 47~degrees.

The angular resolution of the different energies under the four samples is obtained using the above method, as shown in the right side of figure~\ref{Fig: AngleRecon}. The resolution of the angles increases with increasing energy. 

Additionally, the curves of the low light yield SlowLS samples and high light yield SlowLS samples in figure~\ref{Fig: AngleRecon} cross at \SI{4}{MeV}. There are two main reasons for this. First, at low energy, the dark noise has a significant effect on the angular reconstruction of the low light yield samples, making the angular resolution poor. Second, the high light yield samples have good position resolution at low energy, which improves the angular resolution.

The \SI{2}{MeV} electron angular reconstruction results are used for the interest in $0 \nu \beta \beta$ decay research. As a result, the angular resolution of the four samples is shown in table~\ref{Tab: Conlusion} for \SI{2}{MeV} kinetic energy electrons with uniform position and direction distributions. Furthermore, the angular resolution of  \SI{5}{MeV} electrons is in general agreement with some existing angular reconstructions~\citep{SNOAngle,land2021mev}.

\subsubsection{Particle Identification}

Events with visible energies (electron equivalent energy, unit is MeVee) of several-MeVee can be electrons, gamma rays, protons or alpha rays. Among them, protons and alpha rays are so massive that they do not produce directly Cherenkov photons for energies of a few MeV. In addition, the Cherenkov light produced by gamma rays relies mainly on the scattered electrons, which pass the Cherenkov threshold. Table~\ref{Tab: PID} lists the range of particle kinetic energy and the number of Cherenkov PE produced by the four particles when the visible energy is 2-\SI{3}{MeVee} in the full MC simulation of the SlowLS. Therefore, the difference in a particle's ability to emit Cherenkov radiation can be used to perform the PID.

\begingroup
\setlength{\tabcolsep}{9pt} % Default value: 6pt
\renewcommand{\arraystretch}{1.35} % Default value: 1
\begin{table}[!htbp]
\centering
\begin{tabular}{ccc}
\toprule
\begin{tabular}[c]{@{}c@{}}Particle Type\end{tabular} & \begin{tabular}[c]{@{}c@{}}Kinetic Energy of the Particle\\ with Visible Energy is 2-\SI{3}{MeVee}\\in the SlowLS\\(MeV) \end{tabular} & \begin{tabular}[c]{@{}c@{}} Number of Cherenkov PE\\with Visible Energy is 2-\SI{3}{MeVee}\\in the SlowLS\end{tabular} \\ \midrule
Electron                                                &               2-3                                                                                               &                10-40                                                  \\
Gamma                                                   &               1.7-3                                                                                            &                  5-35                                                \\
Proton                                                  &               13-19                                                                                             &                  0                                                \\
Alpha                                                   &               13-19                                                                                               &                 0                                                 \\ \bottomrule
\end{tabular}
\caption{The range of particle kinetic energies and the Cherenkov PE numbers produced by the particles are given for four different particles at visible energies of 2-\SI{3}{MeVee} in the full MC simulation of the SlowLS.}
\label{Tab: PID}
\end{table}
\endgroup

The four samples ($\mathrm{LS_{\mathrm{24~PE}}^{0.5~ns}}$, $\mathrm{LS_{\mathrm{35~PE}}^{0.5~ns}}$, $\mathrm{LS_{\mathrm{285~PE}}^{\mathrm{7.7~ns}}}$, and $\mathrm{LS_{\mathrm{376~PE}}^{\mathrm{7.7~ns}}}$) are studied in the following way. Electrons, gamma, and protons (alpha is similar to proton) with 1 to \SI{9}{MeVee} visible energy and uniform position and direction distributions are simulated for the four samples. Then the likelihood function is used to reconstruct these events. 

The number of Cherenkov PEs, $N^{\mathrm{C}}$, and scintillation PEs, $N^{\mathrm{S}}$, are obtained for each event from the reconstruction results, where $N^{\mathrm{C}}$ and $N^{\mathrm{S}}$ are shown as:
\begin{equation}
N^{\mathrm{C}}=\sum_{i}^{N_{\mathrm{hitPMT}}} N_{i}^{\mathrm{C}},
\end{equation}
\begin{equation}
N^{\mathrm{S}}=\sum_{i}^{N_{\mathrm{hitPMT}}}  N_{i}^{\mathrm{S}},
\end{equation}

\noindent where $N_{i}^{\mathrm{C}}$ is the fitted number of Cherenkov PEs of the $i$th PMT, and $N_{i}^{\mathrm{S}}$ is the fitted number of scintillation PEs of the $i$th PMT. The $N_{i}^{\mathrm{C}}$ and $N_{i}^{\mathrm{S}}$ could be fitted by a likelihood function as:
\begin{equation}
\mathcal{L}(n_i^{\mathrm{Obs}},t_{ij},E,x,y,z,\vec{d}_{\mathrm{Fit}}|N_i^{\mathrm{C}},N_i^{\mathrm{S}},N_i^{\mathrm{Dn}})=P_i^{\mathrm{C}}P_i^{\mathrm{S}}P_i^{\mathrm{Dn}},
\end{equation}

\noindent where $N_i^{\mathrm{Dn}}$ is the fitted number of dark noise of the $i$th PMT. $P_i^{\mathrm{C}}$, $P_i^{\mathrm{S}}$, and $P_i^{\mathrm{Dn}}$ are the Poisson distribution with the conditional expectation calculated in the observed sample of Cherenkov, scintillation, and dark noise, respectively. In addition, the sum of $N_i^{\mathrm{C}}$, $N_i^{\mathrm{S}}$ and $N_i^{\mathrm{Dn}}$ must be equal to $n_i^{\mathrm{Obs}}$.

\begin{figure}[!htbp]
\centering
\subfigure{\includegraphics[scale=0.37]{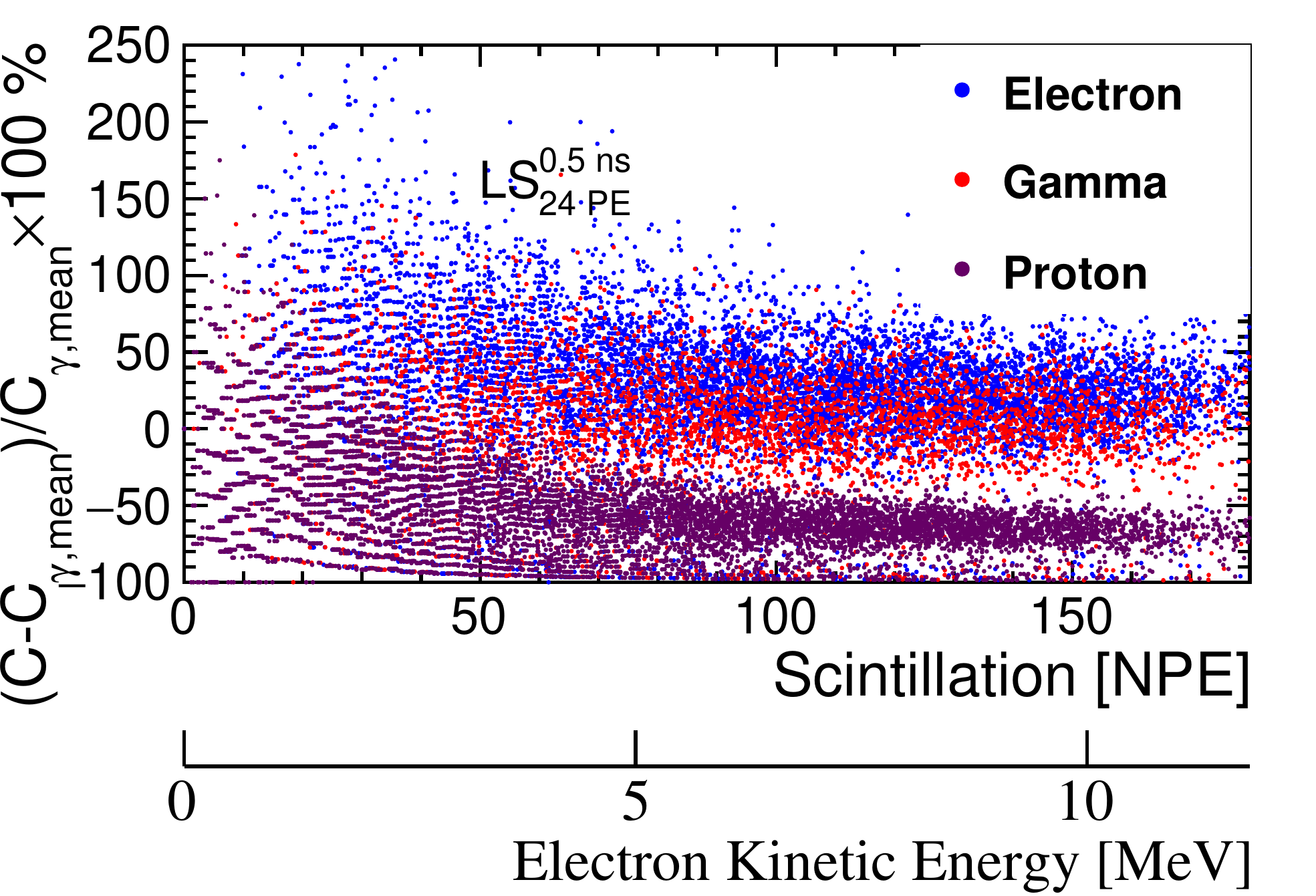}}
\subfigure{\includegraphics[scale=0.37]{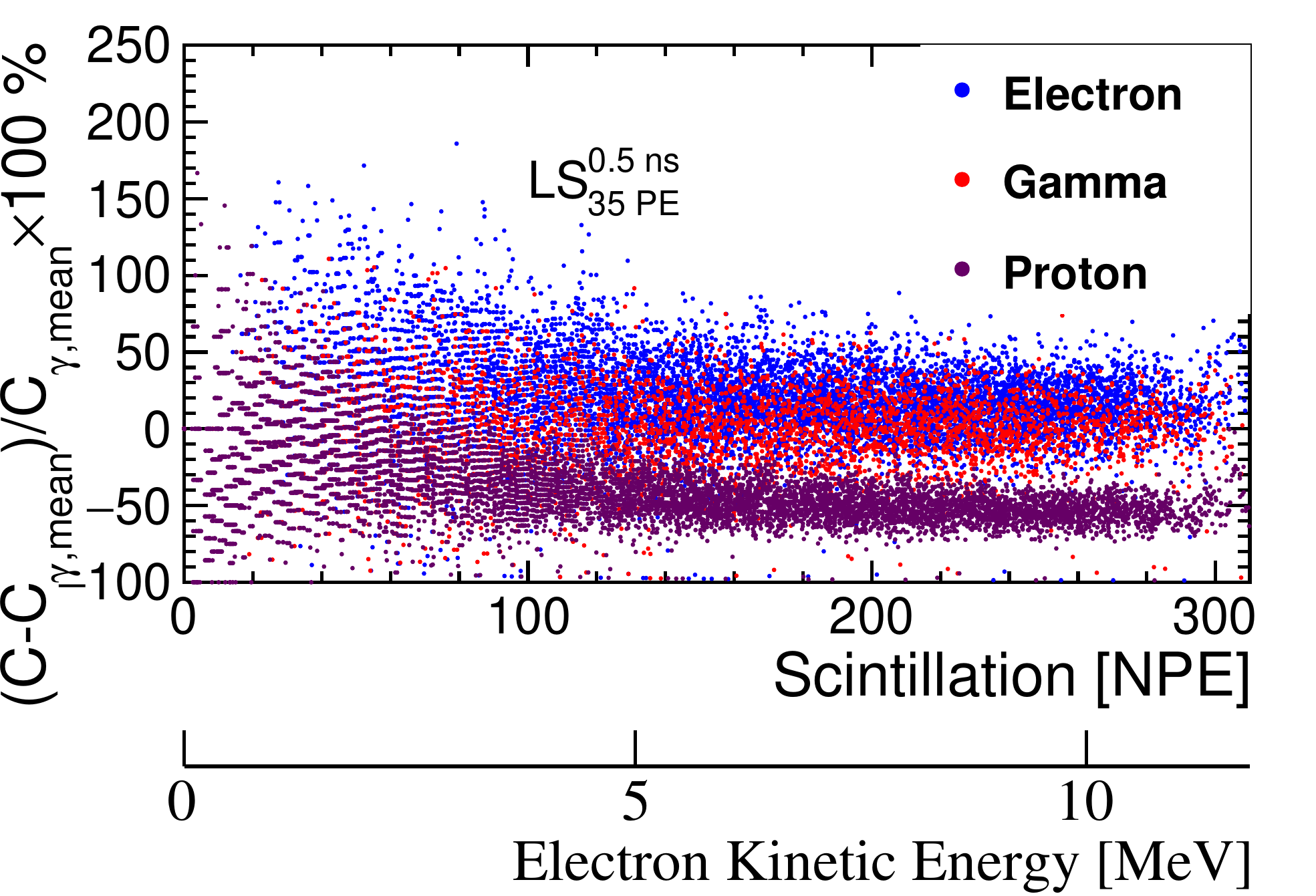}}
\subfigure{\includegraphics[scale=0.37]{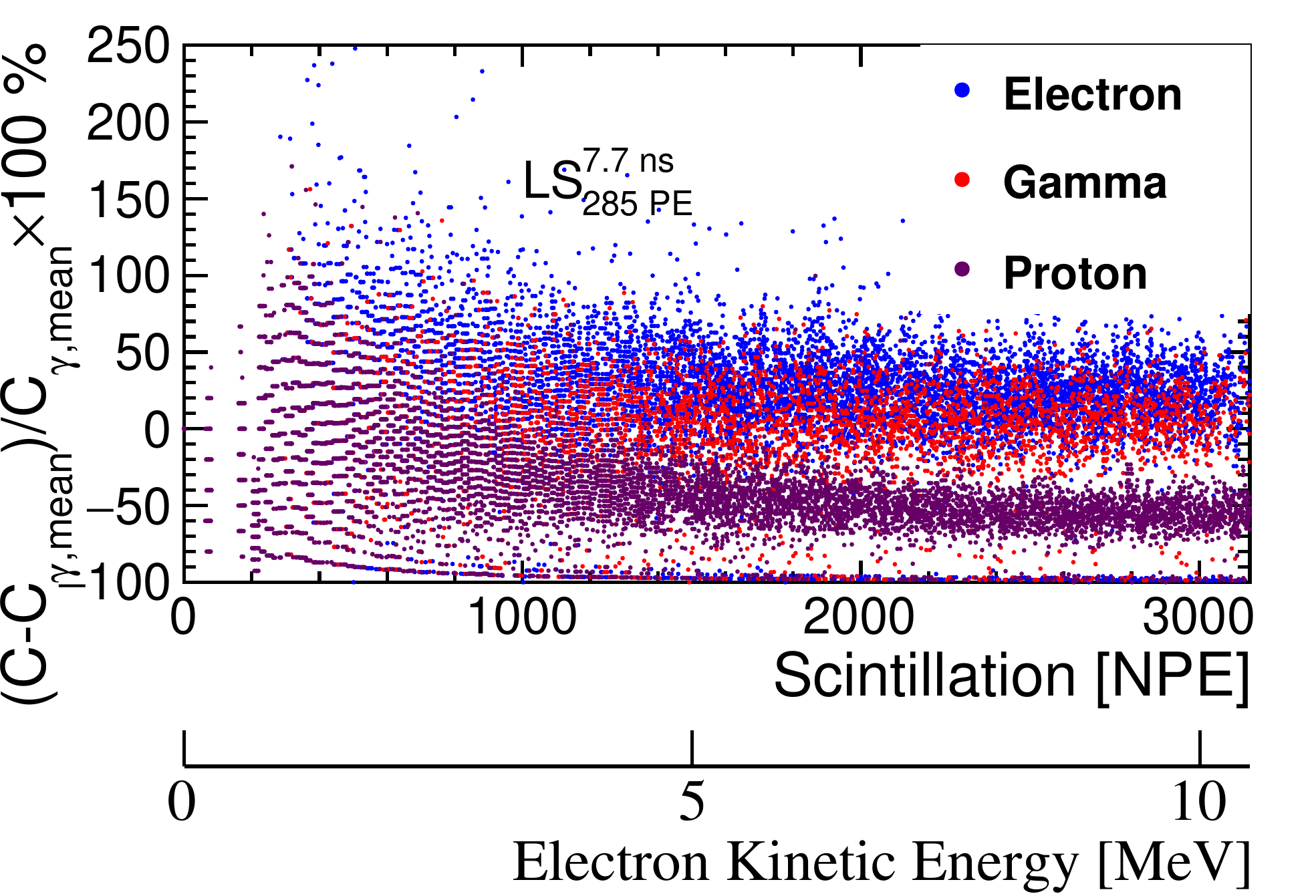}}
\subfigure{\includegraphics[scale=0.37]{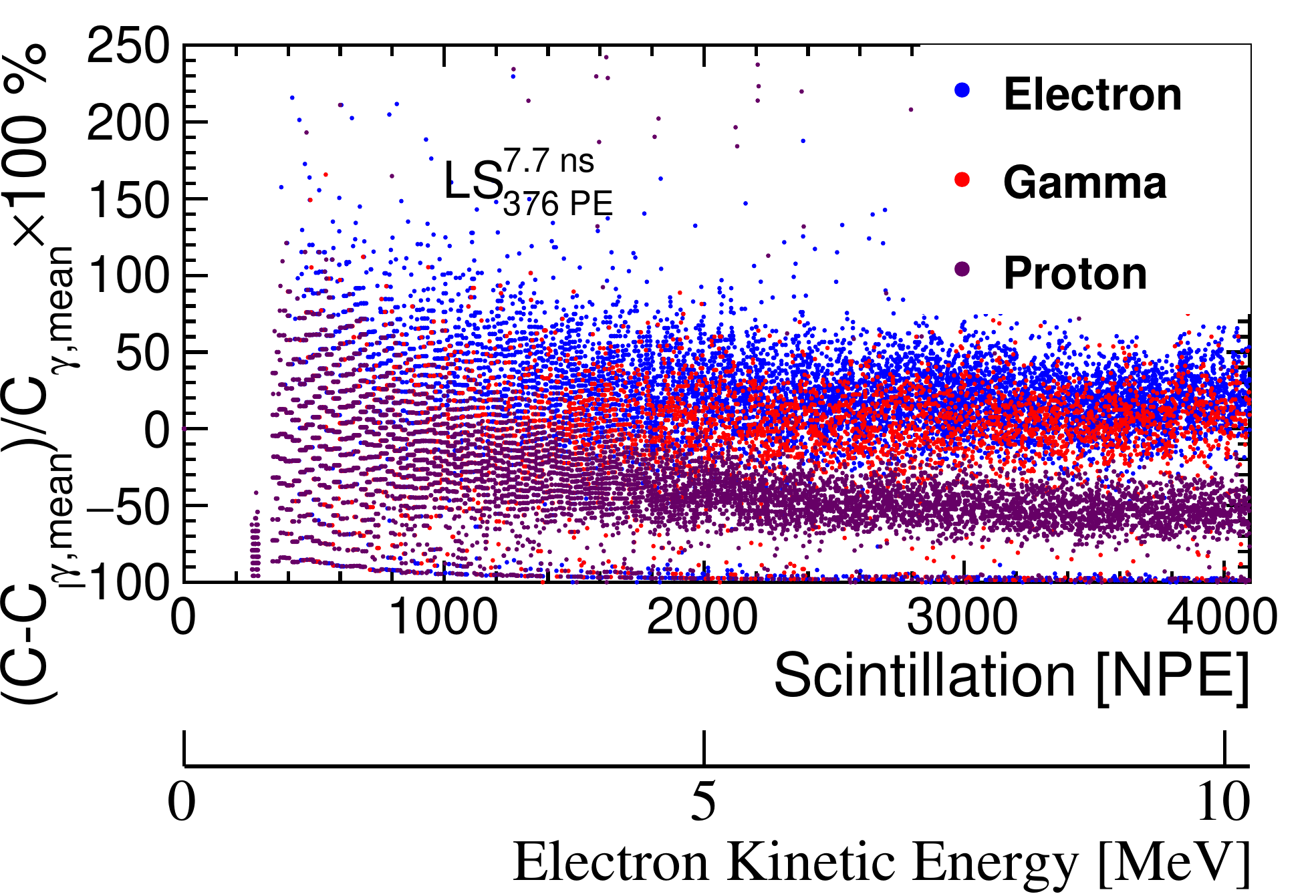}}

\caption{The reconstructions of the number of scintillation and Cherenkov PEs for four samples. Electrons are depicted by blue points, gamma by red points, and protons by magenta points. On the X-axis, the number of scintillation PEs and visible energy are shown. The Y-axis displays the fractional difference between the number of Cherenkov PEs, $\mathrm{C}$, to the average value of a gamma, $\mathrm{C}_{\gamma,\mathrm{mean}}$.}
\label{Fig: PIDGraph}
\end{figure}

To better demonstrate the difference in Cherenkov light produced by different particles, the average value of the number of Cherenkov PEs of gamma, $\mathrm{C}_{\gamma,\mathrm{mean}}$, is used as a reference and the difference from it is plotted. As shown in figure~\ref{Fig: PIDGraph}, the horizontal coordinates indicate the number of scintillation PEs, and the vertical coordinates indicate the fractional difference between the number of Cherenkov PEs, $\mathrm{C}$, to the $\mathrm{C}_{\gamma,\mathrm{mean}}$.

Figure~\ref{Fig: PIDGraph} shows that the Cherenkov light fraction of the electron is slightly higher than that of the gamma and much higher than that of the proton. Therefore, protons can be distinguished from electrons and gamma rays, while electrons and gamma rays are harder to separate. In addition, there is a discreteness in the bands for the lower scintillation PE number (NPE). Because the fitted PE numbers are integers, and the number of Cherenkov PE is only a few for the lower NPE.

\begin{figure}[!htbp]
\centering
\subfigure{\includegraphics[scale=0.37]{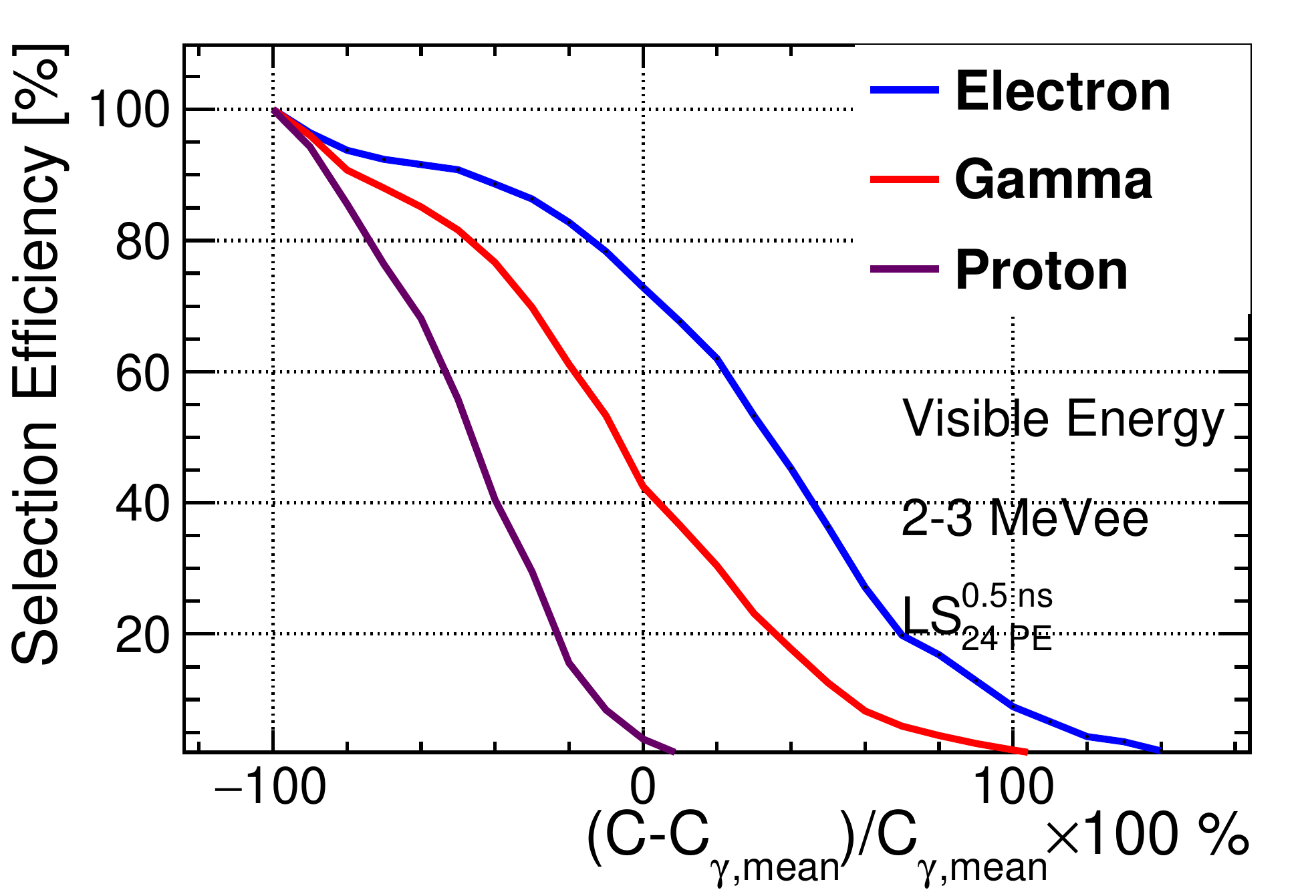}}
\subfigure{\includegraphics[scale=0.37]{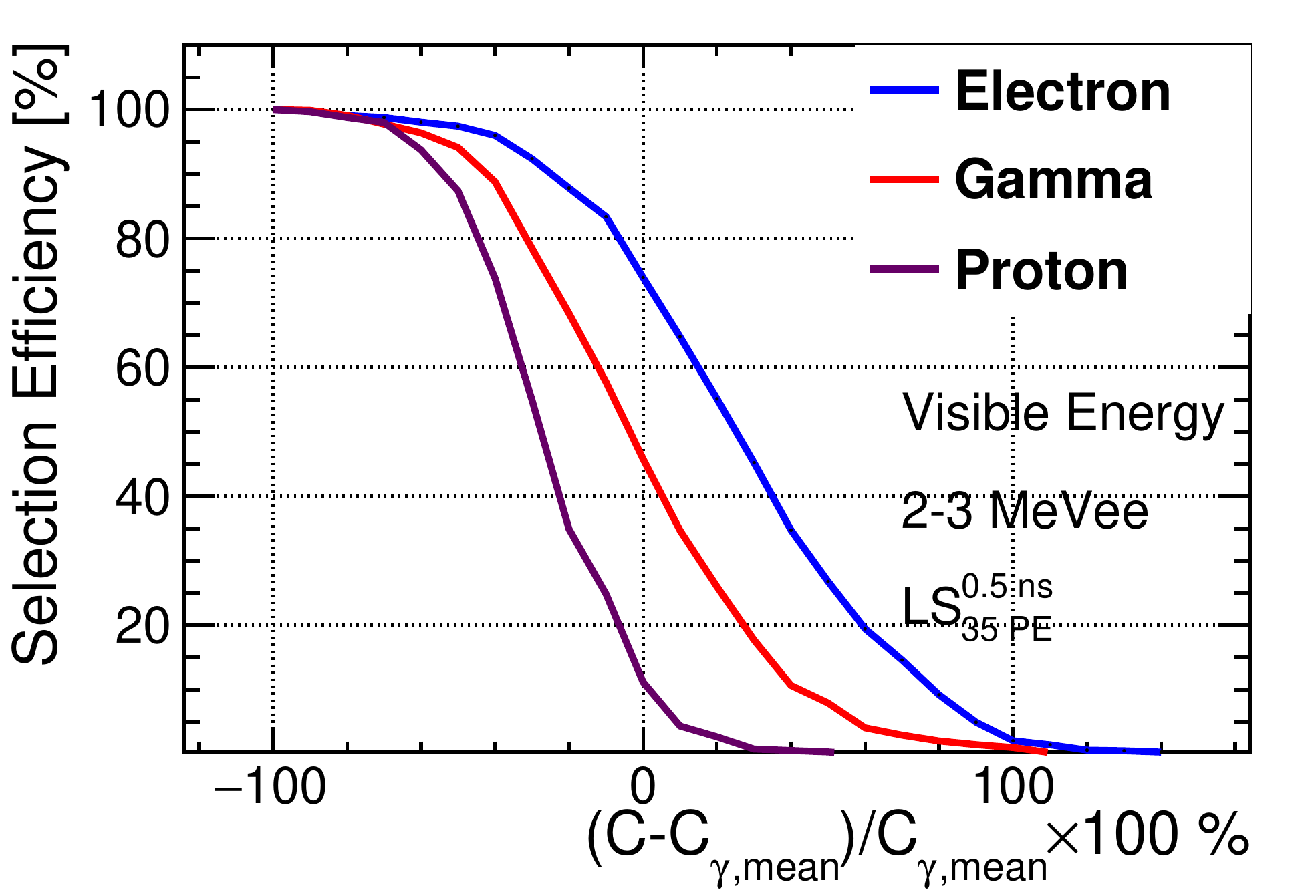}}
\subfigure{\includegraphics[scale=0.37]{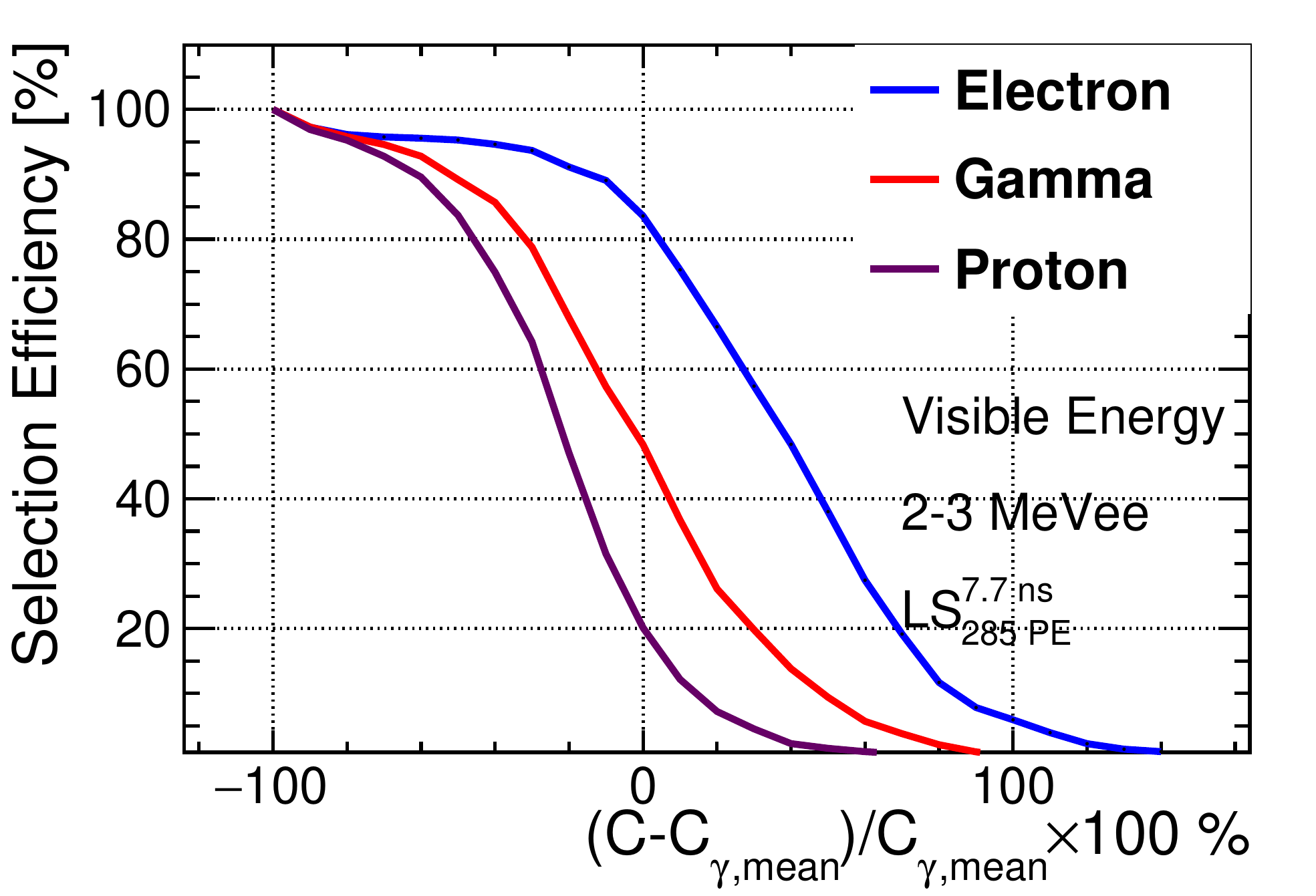}}
\subfigure{\includegraphics[scale=0.37]{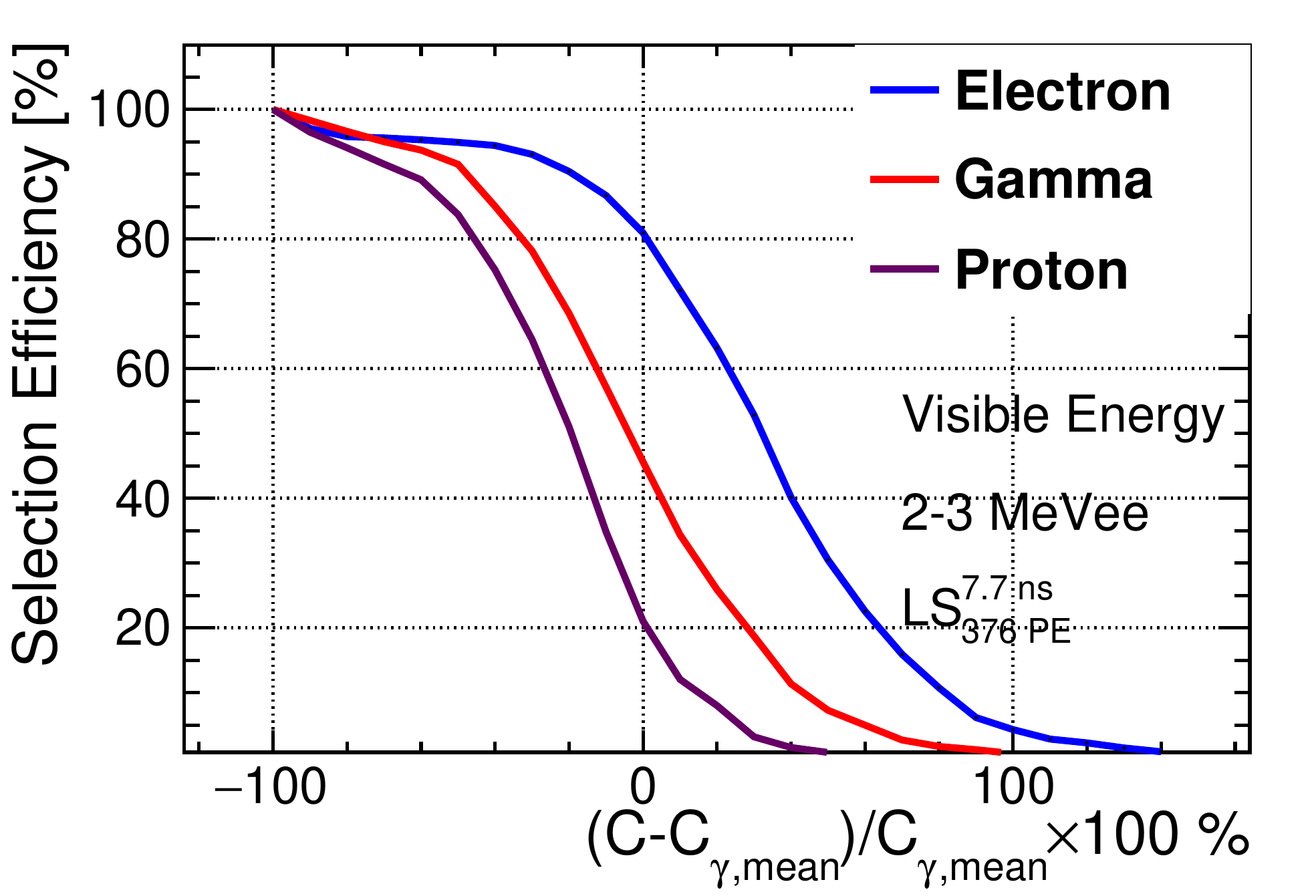}}
\caption{The distribution of the proportion of events retained by different particles versus the fractional difference between $\mathrm{C}$ and $\mathrm{C}_{\gamma,\mathrm{mean}}$ is obtained with 2-\SI{3}{MeVee} visible energy events for four samples. The electron is represented by blue lines, the gamma is represented by red lines, and the proton is represented by magenta lines.}
\label{Fig: PID0vbbRatio}
\end{figure}

The difference in figure~\ref{Fig: PIDGraph} can be further quantified  by projection onto the Y-axis. As shown in figure~\ref{Fig: PID0vbbRatio}, we select the events of 2-\SI{3}{MeVee} visible energy to obtain the distribution of the proportion of events retained versus the fractional difference between $\mathrm{C}$ and $\mathrm{C}_{\gamma,\mathrm{mean}}$ for the interest in $0 \nu \beta \beta$ decay research.

All four samples of SlowLS can identify particles with 2-\SI{3}{MeVee} visible energy, and the discrimination is comparable. As the light yield of scintillation increases, the electron-proton discrimination becomes progressively worsens. It can be seen that in the electron-gamma discrimination, gamma is reduced by 50\% and the electrons can be retained by 70-80\% for these samples; in the electron-proton discrimination, while retaining 70-80\% of the electrons, the protons are reduced by 95\%, 90\%, 80\%, and 75\% for the four samples ($\mathrm{LS_{\mathrm{24~PE}}^{0.5~ns}}$, $\mathrm{LS_{\mathrm{35~PE}}^{0.5~ns}}$, $\mathrm{LS_{\mathrm{285~PE}}^{\mathrm{7.7~ns}}}$, and $\mathrm{LS_{\mathrm{376~PE}}^{\mathrm{7.7~ns}}}$), respectively.

\section{Conclusion}\label{Sec: Conclusion}
In this paper, we focus on the reconstruction algorithm of a new CSD using SlowLS, expecting to reconstruct the energy, direction, position, and number of Cherenkov and scintillation PEs of a particle based on the full MC simulation. 

We conduct a comprehensive simulation study for a multi-hundred-ton spherical detector using the Geant4-based simulation software JSAP. 
The scintillation light yield and time profile of SlowLS are two key properties determining the performance of the CSD directional reconstruction. 
To study the effect, we select four typical light yields and three typical time profiles, forming twelve SlowLS samples. 
The twelve samples can be sorted according to the CSR, i.e.~Cherenkov light to scintillation light ratio in the first \SI{10}{ns}.

For the CSD reconstruction, we first analyze the waveform of the simulation output to obtain the number of PEs on each PMT and the time of the PEs. The number of fit PEs is consistent with the number of true PEs, and for 93\% of the PEs, the time difference between the fit result and the true value is less than \SI{0.5}{ns}.

Second, to understand the PE generation process, an exhaustive study of the signal generation process is conducted for the CSD using simulation. To save computation time, each signal process is represented with a simplified function. In addition, we introduce the concept of a virtual light source to solve the effect of the indirect photons in the reconstruction. 

Third, a likelihood function for charge and time is constructed and its performance is validated with the full simulation.
For the events near the detector boundary, the difference between the number of predicted direct scintillation PEs on PMTs and full simulation results is less than 3\%. The difference for the direct Cherenkov PEs on the Cherenkov ring is less than 5\%. Moreover, the closer the position to the center of the detector, the smaller the difference.

For the reconstruction process, the final reconstruction results are obtained in two steps to solve the local extremum problem.
Electrons with 1 to \SI{9}{MeV} kinetic energy and uniform position and direction distributions are simulated for the twelve SlowLS samples.
The reconstruction results show that the directional reconstruction is poor when the CSR is less than 1:~5. 

We demonstrate the reconstruction performance of four of the SlowLS samples with large CSR, $\mathrm{LS_{\mathrm{24~PE}}^{0.5~ns}}$, $\mathrm{LS_{\mathrm{35~PE}}^{0.5~ns}}$, $\mathrm{LS_{\mathrm{285~PE}}^{\mathrm{7.7~ns}}}$, and $\mathrm{LS_{\mathrm{376~PE}}^{\mathrm{7.7~ns}}}$ in detail, as shown in table~\ref{Tab: Conlusion}. 
The numbers of Cherenkov PEs and scintillation PEs are calculated with the reconstruction process, and their ratio is used for PID. 
Electrons and protons are easy to distinguish and electrons and gamma rays separation may also be possible. 
For four samples of 2-\SI{3}{MeVee} visible energy, it is possible to maintain 70-80\% selection efficiency for the electrons while reducing the gamma by 50\% in the case of electron-gamma discrimination, and in the electron-proton discrimination, more than 75\% of the protons are reduced and 70-80\% of the electrons remain.

The study based on simulation shows that the CSD could help the next generation of MeV-scale neutrino experiments, including solar-, geo-, and supernova neutrino and $0 \nu \beta \beta$ decay research, to fulfill the goal of achieving a better background suppression and precision measurement. The result is also a guideline for developing SlowLS candidates for a particular physics requirement on energy, angular resolution, and PID.

\acknowledgments

The authors would especially like to thank Ziyi Guo for useful discussions. This work was supported in part by the National Natural Science Foundation of China (No.~12127808, No.~12141503, No.~12221005 and No.~12175241), the Key Laboratory of Particle \& Radiation Imaging (Tsinghua University) and the Fundamental Research Funds for the Central Universities.

% We suggest to always provide author, title and journal data:
% in short all the informations that clearly identify a document

\bibliographystyle{JHEP}
\bibliography{output.bbl}

\end{document}